\documentclass[runningheads,a4paper,twocolumn]{article}

\usepackage{cmap}
\usepackage{latexsym}
\usepackage[cp1251]{inputenc}
\usepackage[unicode]{hyperref}
\usepackage{fancyhdr}
\usepackage{cmap}
\usepackage{amssymb}
\usepackage{amsmath}
\usepackage{tikz}
\usetikzlibrary{shapes,arrows, positioning}
\pgfmathtruncatemacro\distance{1}

\def\ral#1{\;\mathop{\longrightarrow}\limits^{\!#1}\;}

\def\blackbox{\;\vrule height 7pt width 7pt depth 0pt \;}

\def\be#1{\begin{equation}\label{#1}}
\def\ee{\end{equation}}

\def\re#1{(\ref{#1})}
\def\i{\item}

\def\bi{\begin{itemize}}
\def\ei{\end{itemize}}
\def\bn{\begin{enumerate}}
\def\en{\end{enumerate}}

\def\l#1{\langle #1 \rangle}

\def\c#1{\left\{\begin{array}{lllll}#1\end{array}\right\}}

\def\eqs#1{\mathop{\sim}\limits_{#1}}

\def\by{\begin{array}{llllllllllllll}}
\def\ey{\end{array}}

\def\ba{\left\{\begin{array}{llllllllll}}
\def\ea{\end{array} \right\}}
\def\bau{\left[\begin{array}{llllllllll}}
\def\eau{\end{array} \right]}
\def\beq{\begin{equation}}

\def\ra#1{\;\mathop{\to}\limits^{
\begin{picture}(0,5)
\put (-3,2){\makebox(0,1){$\scriptstyle #1$}}
\end{picture}
}\;}

\def\blackbox{\;\vrule height 7pt width 7pt depth 0pt \;}

\def\be#1{\begin{equation}\label{#1}}
\def\ee{\end{equation}}
\def\i{\item}

\def\re#1{(\ref{#1})}
\def\bn{\begin{enumerate}}
\def\en{\end{enumerate}}

\def\l#1{\langle #1 \rangle}
\def\c#1{\left\{\begin{array}{lllll}#1\end{array}\right\}}

\def\by{\begin{array}{llllllllllllll}}
\def\ey{\end{array}}
\def\bi{\begin{itemize}}
\def\ei{\end{itemize}}
\def\bn{\begin{enumerate}}
\def\en{\end{enumerate}}

\def\beenum{\begin{enumerate}}
\def\enenum{\end{enumerate}}
\def\ba{\left\{\begin{array}{llllllllll}}
\def\ea{\end{array} \right\}}
\def\bau{\left[\begin{array}{llllllllll}}
\def\eau{\end{array} \right]}
\def\beq{\begin{equation}}

\def\blackbox{\;\vrule height 7pt width 7pt depth 0pt \;}

\def\be#1{\begin{equation}\label{#1}}
\def\ee{\end{equation}}

\def\re#1{(\ref{#1})}
\def\i{\item}

\def\bi{\begin{itemize}}
\def\ei{\end{itemize}}
\def\bn{\begin{enumerate}}
\def\en{\end{enumerate}}

\def\l#1{\langle #1 \rangle}

\def\c#1{
{\def\arraystretch{0.7}
\left\{\begin{array}{lllll}#1\end{array}\right\}}}

\def\by{\begin{array}{llllllllllllll}}
\def\ey{\end{array}}

\def\ba{\left\{\begin{array}{llllllllll}}
\def\ea{\end{array} \right\}}
\def\bau{\left[\begin{array}{llllllllll}}
\def\eau{\end{array} \right]}
\def\beq{\begin{equation}}

\author{Andrew M. Mironov\\$\;$\\
Moscow State University,
Faculty of Mechanics and Mathematics,\\ Russia, 119991, Moscow, GSP-1, 1 Leninskiye Gory\\$\;$\\
Innopolis University, Leading Research Centre,\\
Russia, 420500, Innopolis, 1 Universitetskaya
}

\title{A Distributed Process Model 
of Cryptographic Protocols}

\date{
amironov66@gmail.com
}

\begin{document}

\maketitle

\newcounter{theorem}
\newcounter{lemma}
\newcounter{fig}

\newpage
$\;$
\newpage
$\;$
\newpage

\begin{abstract}

{\bf Cryptographic Protocols (CP)} are distributed algorithms intended for  secure communication 
in an insecure environment. They are used, for example, in electronic payments, electronic voting procedures, systems of confidential data  processing, etc. Errors in CPs can bring to great financial and social damage, therefore it is necessary to use mathematical methods to substantiate the correctness and safety of CPs. In this paper, a distributed process
model of CPs is presented, which allows one to formally describe CPs and their properties. It is shown how to solve the problems of verification of CPs on the base of this model.

\end{abstract}

\section{Introduction }

\subsection{A concept of a cryptographic protocol}

A {\bf cryptographic protocol (CP)} is a distributed algorithm that describes an order in which messages are exchanged between  communicating agents. Examples of such agents are computer systems, bank cards, people, etc.

To ensure  security properties of CPs (for example, 
integrity and confidentiality of transmitted data), cryptographic transformations  are used in CPs (encryption and decryption, hash functions, etc.). We assume that the cryptographic transformations used in  CPs are ideal, i.e. satisfy some axioms expressing, for example, the impossibility of extracting plain texts from ciphertexts without knowing the corresponding cryptographic keys. 

\subsection{Vulnerabilities in cryptographic protocols}

Many vulnerabilities in CPs are induced 
not by insufficient qualities of  cryptographic primitives used in CPs, but by logical errors in CPs. The most striking example of a vulnerability in a CP is a vulnerability in the Needham-Schroeder Public Key Authentication CP \cite{needhams}. This CP  was published in 1978, and 
a logical error in this CP was discovered  \cite{r2332}
more than 16 years after the start of the use of this CP.
This error was related with the possibility 
of  dishonest behavior of participants of this CP. The peculiarity of this error is that this CP is extremely simple, consisting of only three actions, and in a visual analysis of this CP, the absence of errors in  this CP did not raise any doubts. The error was discovered with use of an automated verification tool. 
Another example (taken from  \cite{mypapersverc}) is the following: in the CP, logging into the Google portal, which allows a user to identify himself only once, and then access various applications (such as Gmail or Google calendar), a logic error has been detected that allows a dishonest service provider to impersonate any of its users to another service provider. 
There are many other examples of CPs (see for example 
\cite{mypapers1},
\cite{mypapers3},
\cite{5}
\cite{mypaperskerberos}),
in which vulnerabilities of the following types were found:
\bi
\i   participants of these CPs can receive corrupted messages  as a result of interception and distortion of  transmitted messages by an adversary, which violates the integrity property of the transmitted messages,
\i an adversary can find out the secret information contained in the intercepted messages, as a result of that the confidentiality property of the transmitted messages is violated.
\ei

There are also examples of vulnerabilities in CPs used for authentication to mobile phone providers, for withdrawing money from  ATMs, for working with electronic passports, conducting electronic elections, etc.

All these examples justify  that in safety-critical systems there is not enough an informal analysis of  required security properties of CPs used in them, it is necessary to build
mathematical models of analyzed CPs and
prove that the  models satisfy (or do not satisfy) their properties, the procedure for constructing such proofs is called 
a verification of  CPs.

In this work, a 
distributed process
model of CPs is constructed.
In terms of this model it is possible to express  properties of CPs such  as  integrity and confidentiality of transmitted messages, and authentication of participants of CPs. 
We present new verification methods of CPs
based on this model. 

\subsection{An overview of methods
of
modeling and verification 
of cryptographic protocols}

The books \cite{C324M6} and \cite{mypa23persstrandlast} contain overviews of the most widely used methods for modeling and verification of CPs. The main classes of models of CPs
and approaches to  verification of CPs are the following.
\bn
\i {\bf Logical models}.

This class of models was the very first approach to modeling and verification of CPs. On the base of this class of models, the problem of  verification of CPs is reduced to the problem of constructing
proofs of theorems  (in some logical calculi) 
that CPs being analyzed have given properties. In the work \cite{mypapers5}, the first mathematical model of CPs (called BAN logic) was presented. This model has  limitations: it assumes that  participants of  analyzed CPs are honest, i.e. exactly fulfill their requirements. In addition, this model does not allow analyzing CPs with unlimited generation of sessions. The BAN logic approach was developed in works
\cite{CM87},
\cite{CM5},
\cite{CM126}
\cite{mypapers11},
\cite{CM99},
\cite{CM152},
\cite{CM151}, etc.
An important class of logical calculi for CP modeling and 
verification  
is the Protocol Composition Logic, see 
\cite{CM78},
\cite{CM66},
\cite{CM54},
\cite{CM67}, etc.

One of the classes of CP logical models is associated with logical programming. In these models, CPs are represented in the form of term rewriting rules, Horn clauses and constraint systems.
This approach is described in 
\cite{mypapershartog4},
\cite{mypapershartog3},
\cite{mypapersverc}, etc.

An important class of logical methods for  CP modeling and 
verification is the Paulson's inductive method:
\cite{CM130}, \cite{CM131},
\cite{mypapershartog5},
\cite {8}, etc.

\i Models based on {\bf process algebra}.

The source of this class of models is Milner's fundamental work \cite{milner} and its continuation  \cite{milnerpi}. 
In these works, a model of communicating processes is constructed, in which  processes are represented by terms. 
An observational equivalence  on these terms 
allows  to  express various properties of processes related to security (in particular, secrecy and anonymity). First work, which expounds a CP model based on  Milner's approach, is \cite{CM3}. Among other works related to this direction, it should be noted
\cite {82},
\cite{CM2},
\cite{mypapershartog10},
\cite{CM1},
\cite{mypapersrsm},
\cite{mypapersabf},
\cite{mypap32ersabf},
\cite{n2},
\cite{n1}.

\i Models based on {\bf CSP}.

CSP (Communicating Sequential Processes) is a mathematical approach developed by Hoare \cite{41} and intended for modeling and verification of distributed communicating processes. On the basis of this approach, a method for CP modeling and verification  is built, which is most fully described in the book \cite{CM1321}. Deductive CP verification method based on this approach uses the concept of a {\bf rank function}. Among the works related to this direction, it should be noted
\cite{88}, 
\cite{CM144}, 
\cite{61},
\cite{27},
\cite{86},
\cite{n324}.

\i Models based on {\bf strand spaces}.

Strand spaces allow to represent processes related to an execution of a CP in the form of graphical objects (called strands), which indicate  dependencies between actions of participants of the CP. Among the works related to  CP modeling and verification  based on strand spaces approach, it should be noted
\cite{CM153},
\cite {30},
\cite{mypapersstrand1},
\cite{CM154},
\cite {38},
\cite{CM91},
\cite{32CM21},
\cite{CM72},
\cite{CM74},
\cite{CM73},
\cite{mypapersstrandlast},
\cite{myre3papersverc},
\cite{mypapersae2bf}.
\en

\subsection{A comparison of the proposed 
model 
of cryptographic protocols
 with other models}

A distributed process
model  of CPs presented in this paper inherits the most essential qualities of  models from the above classes. \bi\i
In this model, CPs are represented in the form of distributed processes (DPs).
Each DP is a set of communicating sequential processes (SPs), where
SPs are models of participants of CPs
communicating by asynchronous message passing. 
\i Typically,  
these SPs are sequences of actions that can be graphically represented as strands, and an execution of the entire CP can be represented as a strand space,  where strands
are connected by edges  representing sending and receiving 
of messages, see for example 
\re{sdfg1dsgwer33r333},
\re{sdfgdsgwer33r333}.
\i
Properties  of CPs can be represented by logical formulas, which can be substantiated using  
algorithms
of logical inference. 
In section \ref{fdges5s5hysrhtgdf} we present an inference algorithm related to the introduced model of CPs.
\i
Some of properties of  CPs 
(for example, a property of anonymity in CPs of 
e-voting or e-commerce) can be  expressed as an 
observational equivalence between the corresponding DPs
(section \ref{agret54yywj75u6rt}), 
similarly to how it is done in CP models based on process algebra.
\ei

Main advantages of the proposed 
distributed process
model are the following.
\bn

\i Proofs of CP correctness properties based on this model are much shorter and simpler than  proofs of these properties based on other CP models. To substantiate this statement, we give examples of verification of two CPs: Yahalom \cite{CM1321} and a CP of message transmission with unlimited number of participants 
 taken from \cite{CM3}. Verification of these CPs in the above sources takes several dozen pages, while verification of the Yahalom CP based on the proposed model (items \ref{ratrete5y334rrrrre}, \ref{sfgdfsgsefds1}, \ref{fdsfgdstrhyh5tfddc}) takes less than 4 pages, and verification of the second CP (item \ref{fgsrh654sey5tg}) - less than 3 pages. In addition, the analysis of the proofs of the correctness of the CPs  shows that these proofs are produced according to a template method and can be generated automatically.

\i If a  CP consists of a finite number of components without cycles, then a verification of such CP can be carried out completely automatically. The method of verification of such CPs
is based on the concept of a transition graph (TG) introduced in section \ref{fds3w5y6udgdsff}. An application of this method
is shown on four examples of CPs (sections \ref{asfgdsgsrtr66yet3ttf}, 
\ref{54y6ewewrteryeuyut}, \ref{dfgfrw54h6yhytg}, \ref{dfgfrw54h6yhytg1}) taken from \cite{CM3}. 
A verification of the CPs in  \cite{CM3} is a non-trivial mathematical reasoning, while in the present work the CPs are verified by a standard
calculating of formulas related to nodes of   TGs. 

\i The language for describing DPs introduced in this paper makes it possible to construct such models of CPs that are substantially similar to the original CPs. This is essential when a flaw in a model of 
an analyzed CP is found, and it is necessary to correct the CP in such a way as to eliminate this flaw: for an elimination of the detected flaw \bi\i the model of this CP can first be corrected, \i which is then easily converted to a correction of the original CP. 
\ei
The language for describing DPs introduced in this paper can be considered as a new language for describing distributed algorithms. 
\en

\section{Sequential and distributed processes}

In this section, we introduce concepts of sequential and distributed processes. A sequential process is a model of a CP participant, and a distributed process is a model of the entire CP. The proposed model is a theoretical basis for  CP verification methods described in section \ref{dsfgstrhyytegrdfgs}.

\subsection{Auxiliary concepts }

\subsubsection{Types, constants, variables, function symbols}

We assume that there are given  sets $Types$, $Con$,
$ {\it {\it Var}}$  and  $Fun$,  elements of which are called
{\bf types}, {\bf constants}, {\bf variables}, and {\bf function symbols (FSs)}, respectively.
Each element
$x$ of $Con$, ${\it Var}$ and
$Fun$ is associated with a type
$\tau(x)\in Types$, and if 
$x\in Fun$, then 
$\tau(x)$ has the form
$$(\tau_1,\ldots,\tau_n)\to \tau,\quad\mbox{where  }
\tau_1,\ldots, \tau_n, \tau \in Types.$$

\subsubsection{Terms}

In this section we 
define a set of $ Tm $ of {\bf terms}, which are intended to describe messages sent during the execution of a CP.
The set $ Tm $ is defined inductively.
Each term $ e $ is associated with a type $ \tau (e) \in Types $.
The definition of a term is as follows: 
  \bi
  \i $\forall\,x\in  Con\cup {\it Var}\;\;x$
  is a term of the type
  $\tau(x)$,
\i 
if
$f\in Fun$,
$e_1,\ldots, e_n$ are terms, and
$$\tau(f)=
(\tau(e_1),\ldots, \tau(e_n))\to \tau,$$
then $f(e_1,\ldots, e_n)$  is a term of the type $\tau$.
\ei

We will use the following notation: 
$\forall\,e\in Tm$,
$\forall\,X\subseteq  {\it Var}$,
$\forall\,E\subseteq Tm$,
and 
$\forall\,\tau\in Types$
\bi
\i     ${\it Var}({e})=\{x\in Var\mid
     x\mbox{ occurrs in }e\}$,
\i $
Tm(X)=\{e\in Tm\mid
{\it Var}(e)\subseteq X\}$,
\i $E_{\bf X}=E\cap Var$, and
\i $E_\tau=\{e\in E\mid \tau(e)=\tau\}$.
\ei

Let $e,e'\in Tm$. Term
$e$ is said to be a  {\bf subterm} of the term $e'$,
if either $e=e'$, or $e'=f(e_1,\ldots, e_n)$, where $f\in Fun$,
and $\exists\,i\in\{1,\ldots, n\}$: $e$ is a subterm of
$e_i$.
Notation  $e\subseteq e'$, where $e,e'\in Tm$,
 means that $e$ is a subterm of $e'$.
Notation $e\subset e'$, where $e,e'\in Tm$,
 means that $e\subseteq e'$ and $e\neq e'$.
 
 By induction on the structure of a term $e\in Tm$
 it is easy to prove that
 \be{dzfgdfghxfhgfhde4e56}
\by
\mbox{if $e_1,e_2$ 
are different subterms of $e$,}\\
\mbox{then
  either $e_1\subset e_2$, or $e_2\subset e_1$, or
 $e_1$, $e_2$ }\\ 
\mbox{have no common components.}
  \ey
 \ee
 
 Notation $x\in e$, where $x\in Var, e\in Tm$
 denotes that $x\subseteq e$.

For each considered function
$f:E\to E'$, where $E,E'\subseteq Tm$,
we  assume that  $$\forall\,e\in E\;\;
\tau(f(e))=\tau(e)$$

\subsubsection{Examples of types }

We
assume that $ Types $ contains the following types:
\bi
\i  {\bf A}, terms of this type are called {\bf agents},
\i  {\bf C}, terms of this type are called {\bf channels}, they denote communication channels through which agents communicate with each other by  message passing,
\i  {\bf K}, terms of this type are called {\bf keys}, they denote cryptographic keys that agents can use to encrypt or decrypt messages,
\i  {\bf M}, terms of this type are called {\bf messages}, they denote messages that agents can send to each other during their execution,
\i {\bf N}, terms of this type are called {\bf nonces}, they denote variables with unique values ({\bf nonce} = ``{\bf n}umber only used {\bf once}''),
\i  {\bf P}, terms of this type are called {\bf processes}.
\ei

We  use the following notations and conventions:
\bi
\i  $ {\bf M} $ includes all other types from $ Types $, i.e. a term of any type is also a term of type $ {\bf M} $,
\i  $ Var_{\bf C} $ contains a variable denoted by $ \circ $ and called an {\bf open channel},
\i $ \forall \, n \geq 1, \forall \, \tau \in Types $ the set $ Types $ contains type $ \tau_n $, whose values are $n$--tuples consisting of terms of the type $ \tau $.
\ei

\subsubsection{Examples of function symbols }

We  assume that $ Fun $ contains the following FSs. 
\bi
\i FSs $encrypt$  and $decrypt$ of the type
$$({\bf K}, {\bf M})\to {\bf M}.$$
Terms $ encrypt (k, e) $ and $ decrypt (k, e) $ denote messages obtained by encrypting (and decrypting, respectively) the message $ e $ on the key $ k $. 
Term $ encrypt (k, e) $
is denoted by $ k (e) $, 
and is called an {\bf encrypted message (EM)}.

\i FSs $ shared \_key $ of the type $$ {\bf A}_n \to {\bf K} ,\mbox{ where } n \geq 2. $$
Term $ shared \_key (A_1, \ldots, A_n) $ is called a {\bf shared key} of agents $ A_1$, $\ldots$, $A_n $ and is denoted by $ k_{A_1 \ldots A_n} $.

\i FSs $ shared \_channel $ of the type 
$$ {\bf A}_n \to {\bf C} ,\mbox{ where } n \geq 2. $$
Term $ shared \_channel (A_1, \ldots, A_n) $ is called a {\bf shared channel} of agents $ A_1$, $\ldots$, $A_n $ and is 
denoted by $ c_{A_1 \ldots A_n} $. 

\ei

We will use the following notation: $ \forall \, e \in Tm $ 
$$
Keys(e)=
\{k\in Var_{\bf K}\mid \exists\,e'\in Tm: k(e')\subseteq e\}.$$

\subsubsection{Expressions }

In this section we
define a set $ Expr $ of {\bf expressions}, 
where each expression 
describes a  set of terms. Such a set can be a set of terms that are currently available to some process, or a set of messages that are currently in a channel.

An {\bf expression} is a notation of one of the following forms: 
\bi
\i $E$, where $E\subseteq Tm$,
\i $[P]$ and $[c]$, where $P\in Var_{\bf P}$, 
$c\in Tm_{\bf C}$, 
\i $k^{-1}(E)$, where $k\in Tm_{\bf K}$, and $E\in Expr$, 
\i $E\cap E'$,
$E\cup E'$,
where $E,E'\in Expr$.
\ei

$\forall\,E\in Expr\;\;{\it Var}({E})$
is a set of all variables 
occurred in $E$.

Expressions $k^{-1}([P])$
and $k^{-1}([c])$  are denoted by  
$k^{-1}[P]$
and $k^{-1}[c]$ respectively.
Expressions 
$\{e\}$, where $e\in Tm$, 
 are denoted without braces.


\subsubsection {Formulas}

In this section we define a set $Fm$ of
{\bf formulas}, which are intended to describe  properties of sets of terms. 
In this definition, a concept of {\bf elementary formula (EF)} is used, which is a notation of one of the following forms: 
\bn
\i $e\in E$,
$E= E'$, 
$E\subseteq  E'$, 
$E\supseteq E'$, 
where $e\in Tm$,
$E,E'\in Expr$,
\i 
$E\,\bot_{\bf C}\,P$,
$E\,\bot_{\bf K}\,P$,
where 
$E\subseteq Tm,
P\in Var_{\bf P}$,
\i $at_P=i$, where $P\in Var_{\bf P}$,
$i\in\{0,1,\ldots\}$.
\en

EFs express  properties of values of  expressions occurred in them. 
An example of a EF is
$$decrypt(k,k(e))=e,\;\mbox{where $ k\in {\it Tm}_{\bf K}, e\in
Tm.$}$$

A {\bf formula} is a set of EFs. Each formula 
$ \varphi = \{\varphi_i \mid i \in I \} $ expresses a statement that is a conjunction of statements expressed by EFs $ \varphi_i \; \; (i \in I) $.

$\forall\,\varphi \in Fm $ 
the set of all variables occurred in $ \varphi $
is denoted by 
 $ {\it Var} (\varphi) $.

$\forall\, \varphi_1 $, $ \ldots $, $ \varphi_n \in Fm $, the formula $ \varphi_1 \cup \ldots \cup \varphi_n $ is denoted by $ \{\varphi_1, \ldots, \varphi_n \} $.


A formula
$ \{E_0 \rho_1 E_1, E_1 \rho_2 E_2, \ldots, E_{n-1} \rho_n E_n \}$, where $ \rho_1 $, $ \ldots $, $ \rho_n $ are symbols from $ \{= , \subseteq \} $, is denoted by
$ E_0 \rho_1 E_1\rho_2 \ldots \rho_n E_n $.

\subsubsection {Bindings}

A
{\bf binding} is a function $ \theta: {\it Var} \to Tm $.
We  say that  $ \theta $ binds a variable $ x \in {\it Var} $ with  term $ \theta (x) $.

We  use the following notation:
\bi \i
the set of all bindings is denoted by $ \Theta $,
\i $ id $ denotes identical binding:
$$ \forall \, x \in {\it Var} \; \; id (x) = x,$$
\i $ \forall \, X \subseteq {\it Var}$ \\ 
$\Theta (X) =
\{\theta \in \Theta \mid \forall \, x \in {\it Var} \setminus X \; \; \theta (x) = x \}, $
\i
a binding $ \theta \in \Theta $ can be denoted by 
\be{sdfasdgfsdfdasfas} x \mapsto \theta (x) \;\mbox{or} \;
(\theta (x_1) / x_1, \ldots, \theta (x_n) / x_n), \ee
 second notation in \re{sdfasdgfsdfdasfas} is used when\\
$ \theta \in \Theta ({\{x_1, \ldots, x_n \}}) $,
\i
 $ \forall \, \theta \in \Theta, \; \forall \, e \in Tm $
  $ e ^ {\theta} $
denotes a term obtained from $ e $ by replacing $ \forall \, x \in {\it Var} (e) $  each occurrence of $ x $ in $ e $ by the term $ \theta (x) $,
  the term $ e $ is called a {\bf template} of $ e ^ \theta $ with respect to $ \theta $,
\i
 $ \forall \, \theta \in \Theta, \; \forall \, E \subseteq Tm $
  the set $ \{e ^ \theta \mid e \in E \} $ is denoted by
$ E ^ {\theta}$,
\i
$ \forall \, \theta, \theta '\in \Theta $
the binding $ x \mapsto (x ^ {\theta}) ^ {\theta '} $ is 
denoted by  $ \theta \theta' $.
\ei 

If $ X \subseteq X '\subseteq Var $,
$ \theta \in \Theta (X) $, $ \theta '\in \Theta (X') $,
and $ \forall \, x \in X \; \; \theta (x) = \theta '(x) $,
then $ \theta '$ is said to be an {\bf extension}
of $ \theta $.

\subsection {Sequential processes}

\subsubsection {Actions}

{\bf Actions} are notations of the following forms:
$$ \by
c! e, \; c? e, \; e: = e ', \;
\mbox{where } \;c \in Tm_{\bf C}, \; e, e '\in Tm, \ey $$
which is called 
a {\bf sending}  $ e $ to  $ c $,
a {\bf receiving}  $ e $ from  $ c $,
and an {\bf assignment}, respectively.

The set of all actions is denoted by $ Act $.

$ \forall \, \alpha \in {Act} $ the set of all variables occurred in $ \alpha $ is denoted by $ {\it Var} ({\alpha}) $.

If $ \theta \in \Theta $, $ \alpha \in Act $, then $ \alpha ^ {\theta} $ denotes the action 
$c^{\theta}!e^{\theta}$,
$c^{\theta}?e^{\theta}$ and
$e^{\theta}:=(e')^{\theta}$, if $\alpha=$
$c!e$,
$c?e$ and 
$e:=e'$, respectively.

Actions can be written in round  parentheses: \\$ (c! e) $, 
$ (c? e) $, $(e=e')$.

\subsubsection{A concept of a sequential process}

A {\bf sequential process (SP)} is a 4-tuple
$ (P, A, X, \bar X) $, whose components have the following meanings:
\bi \i $ P $ is a graph with a selected node (called an {\bf initial node}, and denoted by  $ Init (P) $), each edge of which is labeled by some action, 
\i $ A $ is an {\bf agent} associated with this SP,
\i $ X \subseteq {\it Var} $ is a set of {\bf initialized variables}, \i $ \bar X \subseteq X $ is a set {\bf hidden variables}, they denote secret keys, hidden channels, or nonces, 
these variables  are initialized with unique values.
\ei

A SP is a description of a behavior of a  system, 
a work of which consists of 
sending or receiving messages, and an initialization 
of uninitialized variables.
 
For each SP $ (P, A, X, \bar X) $
\bi
\i this SP can be denoted by the same notation $ P $ as its graph, the set of nodes of the graph $ P $ is also is denoted by $ P $,
\i  $ Agent (P) $, $ X (P) $, $ \bar X (P) $ denote  corresponding components of $ P $,
\i $ {\it Var} (P) $ is 
a set of all variables occurred in $ P $,
 \i $ \tilde X (P) =X (P) \setminus \bar X (P) $,\i
 $ \hat X (P) = Var (P) \setminus X (P) $.
  \ei

 Each SP is associated with a variable 
of the type {\bf P}, called a {\bf name} of this SP. 
 We will denote names of SPs by the same notations
  as the SPs.

If $ P $ has no edges and $ X (P) = \emptyset $, then $ P $ is denoted by {\bf 0}.

Actions of the form $ \circ! e $ and $ \circ? e $ will be abbreviated as $! e $ and $? e $ respectively. 

\subsubsection{A state of a sequential process}
\label{fgsdhg3}

A {\bf state} of a SP $ P $ is a 5-tuple 
$$s=(at,\alpha,[P],\theta,\{[c]\mid c\in Tm_{\bf C}\}),$$ 
where \bi\i $at\in P$ 
is a node of the graph $ P $ in  $ s $, 
\i $\alpha\in \{init\}\sqcup Act$ is an 
action before transition to $ s $, 
\i  $[P]\subseteq  {\it Var}$ is a set 
of initialized variables in  $s$, 
\i $\theta\in \Theta([P])$ is a binding in $s$, and
\i $\forall\,c\in Tm_{\bf C}\;\;[c]\subseteq Tm$ is a
content  of the channel $c$ in  $s$.
\ei

The components of $ s $ are denoted by  
 $at_s$, 
$\alpha_s$,
 $[P]_s$, 
  $\theta_s$, 
$[c]_s$
respectively. 

The set $Tm([P]_s)$
is denoted by
$\l{P}_s$.

A state of SP $P$ is said to be {\bf initial},
and is denoted by 
 $0_P$, if it has the form
$$(Init(P), 
init,
 X(P),
id,
\{\emptyset\mid c\in Tm_{\bf C}\}).$$

\subsubsection{Values of expressions and formulas in states}
\label{sdfsdf34wfg5ghstghdfs}

Let $P$ be a SP, $s$ be
a state of $ P $, $ E \in Expr$, 
and $ \varphi \in Fm$.

The notation $ E ^ s $ denotes a set of terms,
called a {\bf value} of $ E $ in $ s $, and defined as follows:
\bi
\i $\forall\,E\subseteq Tm\quad
E^s= \{e^{\theta_s} \mid e\in E\}$, \\
$\forall\,e\in Tm$
the set 
$\{e\}^s$, 
and the only element of this set, is  denoted by  $e^s$,
\i $[P]^s=([P]_s)^{s}$,
$\l{P}^s=(\l{P}_s)^{s}$,
 $[c]^s=[{c^{s}}]_s$,
 where $P\in Var_{\bf P}$,
 $c\in  Tm_{\bf C}$,
\i $k^{-1}(E)^s=\{e\in Tm\mid
\exists\,e'\in E^s:
k^{s}(e)\subseteq e'
\}$,

\i
$(E\cap E')^s=E^s\cap (E')^s$,
$(E\cup E')^s=E^s\cup (E')^s$.
\ei

The notation $s\models \varphi$ denotes the statement \begin{center}
{\bf  $\varphi$ holds at $s$}.
\end{center}
This statement is true if 
 $Var(\varphi)_{\bf P}\subseteq \{P\}$, and one of the 
 following conditions holds: 
\bi
\i 
$\varphi$ has the form $$(e\in E),
(E=E'), (E\subseteq E'), \mbox{ or }(E\supseteq E'),$$
where $e\in Tm$, 
$E,E'\in Expr$, and
$$e^s\in E^s,\;
E^s=(E')^{s},\; 
E^s\subseteq (E')^{s},\; 
E^s\supseteq (E')^{s},$$ 
respectively,
\i $\varphi=(E\,\bot_{\bf C}\,P),\;
\forall\,e\in E^s\;\;Agent(P)\not\in e$, and
\be{szdgdfhdg443e35gy6h5}
\left.\by\forall\,x\in E^s_{\bf X},
\forall\,y\in [P]_s\;\;
x\not\in y^s\\
\forall\,x\in E^s_{\bf X},
\forall\,c\in Tm_{\bf C}\;
\\ \hspace{3mm}
\mbox{ if }\exists\,e\in [c]_s: x\in e, 
\mbox{ then }c\in E^s
\ey\right\}\ee
\re{szdgdfhdg443e35gy6h5}
can be interpreted as the  statement: each variable from $ E^s_{\bf X} $ 
\bi\i is not occurred in terms available to  $ P $ in the state $ s $, and 
\i is occurred in terms from 
the content of only those channels that are not available for $ P $, 
\ei
\i 
$\varphi=(E\,\bot_{\bf K}\, P)$,
$\forall\,e\in E^s\;\;Agent(P)\not\in e$, and
\be{f443dgh1sdfsg4341}
\left.\by\forall\,x\in E^s_{\bf X},\,
\forall\,y\in [P]_s\;\; 
x\,\bot_{{\bf K},E}\, y^s\\
\forall\,x\in E^s_{\bf X},\,
\forall\,c\in Tm_{\bf C},
\forall\,e\in [c]_s\\\hspace{40mm}
x\,\bot_{{\bf K},E}\, e
\ey\right\}\ee
where  
$x\,\bot_{{\bf K},E}\, e$
means that 
\be{sdfdasgfadsfgrewgtwg}\by
\mbox{each occurrence of $x$ in $e$
}\\\mbox{is contained in a subterm }\\\mbox{$k(\ldots)\subseteq e$,
where $k\in E^s_{\bf K}$}\ey\ee

\re{f443dgh1sdfsg4341}
can be interpreted as the  statement: variables from 
$ E^s_{\bf X} $ are occurred
\bi
\i in  terms available to  $ P $ in the state $ s $, and
\i in terms from a content of any channel,\ei
 in a ``secure'' form, i.e. are occurred
 in subterms of the form $k(\ldots)$, where 
$k\in E^s_{\bf K}$,

\i $\varphi=(at_P=i)$, and $at_s=i$,

\i $\varphi=\{\varphi_i\mid i\in I\}$, and
$\forall\,i\in I\;\;s\models\varphi_i$. 

\ei

\subsubsection{An execution of a sequential process}
\label{hfdsklds}

An  {\bf execution} of a SP $P$ is a walk in the graph $P$, 
starting from $ Init (P)$, 
with an execution of actions that are labels of passed edges. Each step of the execution is associated with a state 
$s$ of $P$, called a {\bf current state} of $P$ 
at this step (a current state at first step is $0_P$).
If a  step of the execution  is not final, then 
the current state $ s $ is replaced by the state $ s' $, which will be a current state at the next step of the execution, 
for this \bn \i 
either an edge {\bf e}
outgoing  from $at_s$  is selected,
whose label $ \alpha $ has the following properties: 
\bi
\i if $\alpha^{\theta_s}$
contains a subterm $shared\_key(\ldots)$
or  $shared\_channel(\ldots)$, 
then $Agent(P)$ 
occurs in this subterm,
 \i one of the following conditions holds: 
\be{asdfsafgewe4444}
\!\!\!\!\!\!\!\!\!\!\!
\!\!\!\!\!\!\!\!\!\!\!
\left.\by
({\rm a}) & \alpha=c!e,\; 
c,e\in \l{P}_s\\
({\rm b})&\alpha=c?e,\; 
c\in \l{P}_{s},\\&\hspace{3mm}
Keys(e^s)\subseteq [P]_s,\\
&\hspace{3mm}
\exists\, \theta\in \Theta({{\it Var}(e)\setminus [P]_s}):
(e^{ \theta})^{s}\in [c]^s\\
({\rm c}) & \alpha=(e:=e'),\;  e'\in \l{P}_s, 
\\&\hspace{3mm}
Keys(e^s)\subseteq [P]_s,
\\
&\hspace{3mm}
\exists\, \theta\in \Theta({{\it Var}(e)\setminus [P]_s}):
e^{ \theta}=
e'
\ey\!\!\!\right\}
\ee\ei
and $s'$ is defined as follows:
$ at_{s'} $ is an end of the edge {\bf e}, 
$\alpha_{s'}=\alpha$,
and
\bi
\i if (a) in 
\re{asdfsafgewe4444} holds, then 
$$\by [P]_{s'}=[P]_{s},  
\theta_{s'}=\theta_{s}, \\
\,[c^s]_{s'}=[c^s]_s\cup
\{e^s\},\\
\forall\,c'\in Tm_{\bf C}\setminus \{c^s\}\;\;
[c']_{s'}=[c']_s,\ey$$
\i if (b) or (c) in 
\re{asdfsafgewe4444} holds, then 
$$\by
[P]_{s'}=[P]_{s}\cup 
{\it Var}(e), 
\theta_{s'}=\theta\theta_{s},\\ 
\forall\,c'\in Tm_{\bf C}\;\;
[c']_{s'}=[c']_s,\ey$$
(we say that %
each  $ x \in {\it Var} (e) \setminus [P]_s $ is initialized with the value $ x ^ {\theta_{s'}} $
in the transition from $ s $ to $ s '$),
 \ei
\i or all components of $ s '$, with the exception of the last component, 
are equal to  the corresponding components of  $ s $,
 and $ \forall \, c \in Tm_{\bf C} $ the set $ [c]_{s'} $ 
 either is equal to 
$ [c]_s $, or is obtained by adding a term to $ [c]_s $ as a result of an execution of a step by another SP.
\en

If  first (second) of the above situations takes place, then we say that $ s' $ is obtained by an {\bf active} ({\bf passive}, respectively) transition from $ s $, and  denote this  by
$ s \ra {P} s' $ ($ s \to s' $, respectively).

Variables in $ {\it Var} ({P}) $ have the following features:
$ \forall \, x \in {\it Var} ({P}) $
\bi \i
if $ x \in X(P)$
(or $x\in \hat X ({P})$), then $ x $ is  initialized (or not initialized,
respectively)
at the initial moment of each execution of  $ P $, 
 \i \label {sdafasgsdfgfds} if $ x \in \bar X ({P}) $, 
 then 
 $x$ is initialized by a {\bf unique value} 
 at the initial moment of each execution of $ P $.
\ei 

Conditions  (a), (b) and (c)  in \re{asdfsafgewe4444} have
the following meaning:
\bi
\i[(a)] is related to a sending a message $c!e$: \bi \i
a name $ c ^ s $ of a channel to which the message is sent
is available to $ P $ in the state $ s $, and \i the sent message $ e ^ s $ is a term whose components are  available to $ P $ in the state $ s $, \ei
\i[(b)] is related to a receiving a message $c?e$:
\bi
\i a name $ c ^ s $ of a channel from which the message is received, is available to $ P $ in the state $ s $,
\i each EM in the received message, which  
\bi \i must be decrypted during the receiving  this message,
and \i is not encrypted on a shared key, \ei
has the form $ k (\ldots) $, where $ k $ must be available to 
 $ P $ in the state $ s $, 
this property is expressed in second line of  \re{asdfsafgewe4444}(b),
\i  term $ e $ is a template of some term from $ [c] ^ s $ with respect to some extension of  $ \theta_s $, this property is expressed in  last line of  \re{asdfsafgewe4444}(b),
\ei

\i[(c)] is related to an assignment $e:=e'$:
\bi \i each component of $ (e ') ^ s $ 
must be available to $ P $ in the state  $ s $,
\i a meaning of  property in 
second line of 
\re{asdfsafgewe4444}(c) coincides with a meaning of  corresponding properties in \re{asdfsafgewe4444}(b): each 
EM in $ (e ') ^ s $, which must be decrypted during the assignment, has the form
$ k (\ldots) $ or $ Agent (P) (\ldots) $, and \bi \i
either $ k $ is a shared key, \i or $ k \in Var_{\bf K} $, and $ k $ is  available to $ P $ in the state $ s $, \ei
\i $ e $ is a template of $ e '$ with respect to some  $ \theta \in \Theta ({{\it Var} (e) \setminus [P]_s}) $.
\ei
\ei

\subsubsection {An adversary}
\label {sdfsfdgrawtrhyjtadfd}

An {\bf adversary} is a SP $ P_{\dagger}$ with the  properties:
\bi
\i the graph $ P_{\dagger} $ consists of a single node, 
\i  $\forall\,\tau\in Types$ the sets $ \bar X({P_{\dagger} })_\tau$
and $\hat X({P_{\dagger} })_\tau$
are countable,
\i $\forall\,\alpha\in Act\;\;{P_{\dagger} }$ 
has an edge labelled by $\alpha$.
\ei

We assume that $ P_{\dagger} $ is the only SP whose graph has cycles. 

\subsubsection{Renamings}
\label{fsdgdfgdsfsgdsgdass}

A {\bf renaming}
is an injective map 
$\eta:X\to X'$, where $X,X'\subseteq Var$.

For each renaming  $ \eta: X \to X '$, each $ e \in Tm $ and each SP $ P $,  notations $ e ^ {\eta} $ and $ P ^ \eta $ denote a term or a SP respectively, obtained from $ e $ or $ P $ 
respectively
by replacing $ \forall \, x \in X $  each occurrence of $ x $ with $ \eta (x) $.

If a renaming $ \eta $ has the form
$$\eta: 
\bar X(P)\cup \hat X(P)
\to
{\it Var}\setminus \tilde X(P),
$$
then SPs $ P $ and $ P ^ \eta $ are considered as equal. 

\subsection{Distributed processes}

In this section we introduce a concept 
of a distributed process (DP).
DPs are models of CPs. 
All CPs considered in the paper 
are represented as DPs.

\subsubsection{A concept of a distributed process}
\label{soglashe212nie}

A {\bf distributed process (DP)} is a family of SPs:
$
{\cal P}=\{P_i\mid i\in I\}
$.
Each DP is associated with a variable of the type {\bf P},
called a {\bf name} of this DP. 



A DP  is a model of a distributed algorithm,  components 
of which communicate with each other by message
passing through channels.

Let $ {\cal P} $ be a DP.
We will use the following notations and assumptions: 
\bi
\i ${\it Var}({\cal P})=\bigcup_{P\in {\cal P}}{\it Var}({P})$,
 the sets
$X({\cal P})$, $\bar X({\cal P})$, 
$\tilde X({\cal P})$, 
$\hat X({\cal P})$  are defined similarly,
\i we will assume that 
\be{dfgdsg3wt5wyg35}
\by
\mbox{components  of the family}\\
\{\bar X({P})\cup \hat X(P)\mid P\in {\cal P}\}\\
\mbox{are disjoint sets}\\\mbox{and do not intersect with  $\tilde X({\cal P})$}
\ey
\ee
(if this is not the case, then replace each $ P \in {\cal P} $ by an equal SP in the sense described at the end of  \ref{fsdgdfgdsfsgdsgdass}, so that  \re{dfgdsg3wt5wyg35} 
will be satisfied), 
\i a DP  ${\cal P}=\{P_i\mid i\in I\}$
can be denoted by \bi\i
$\{P_1,\ldots, P_n\}$,  
if $I=\{1,\ldots, n\}$
(in the case $ n = 1 $ the brackets can be omitted, i.e. the DP 
$ \{P_1 \} $ is denoted by $ P_1 $), and\i
$P^{*}$, if $I=\{1,2,\ldots\}$, and all SPs in $ {\cal P} $ are equal to $ P $,
\ei
\i the notation ${\cal P}_{\dagger} $   
 denotes the DP $\{{\cal P}, P_{\dagger}\}$,
\i if $\{{\cal P}_i\mid i\in I\}$ is a family of DPs, and for each
$i\in I$ ${\cal P}_i=\{P_{i'}\mid {i'\in I_i}
\}$, where sets of indices $I_i\;(i\in I)$ are disjoint (if this is not the case, then we replace them with the corresponding disjoint copies), then the notation $ \{{\cal P}_i \mid i \in I \} $  denotes also the DP 
$
\{P_{i'}\mid
{i'}\in \bigsqcup_{i\in I}{I_i}\}$.
\ei

We will use the following convention:
\bi \i if, in some reasoning related to a DP of the form $ P ^{*} $, some SP is the first of the considered SPs from $ P ^{*} $, then this SP and all its variables are denoted by the same notations
as  in  $ P $,
\i if, in addition to this SP, another SP from $P^{*}$
is considered, then it is denoted by 
$\grave P$, and in the notation of those of its variables that correspond to variables from $\bar X ({P}) \cup \hat X (P) $ are used backstrokes, etc.
\ei

\subsubsection{A concept of a state of a distributed process}

{\bf A state} of a DP $ {\cal P} $ is a family $ s = \{s_P \mid P \in {\cal P} \} $ of states of SPs occurred in ${\cal P }$ such that
$ \forall \, c \in Tm_{\bf C} $ all components of the family $ \{[c]_{s_P} \mid P \in {\cal P} \} $ are the same (we will denote them by $ [c ]_s $).

Let $ s = \{s_P \mid P \in {\cal P} \} $ be a state of a DP $ {\cal P} $. Then
\bi \i
$ s $ is said to be an {\bf initial} state of ${\cal P} $, and is denoted by $ 0_{\cal P} $ (or more briefly by 0,
if the DP ${\cal P}$ is clear from the context), if $ \forall \, P \in {\cal P}$ 
$s_P =0_P,$
\i
  $ at_s =\{at_{s_P} \mid P \in {\cal P} \} $,
 $ [{\cal P}]_s =
 \bigcup_{P \in {\cal P}} [P]_{s} $,
 $\l {{\cal P}}_s = Tm ([{\cal P}]_s) $,
\i $ \theta_s $
denotes a binding from $ \Theta ({[{\cal P}]_s}) $ such that
$$ \forall \, P \in {\cal P}, \;
\forall \, x \in [P]_{s} \quad
\theta_{s_P} (x) = \theta_s (x) $$
(an existence of such a binding follows from  assumption \re{dfgdsg3wt5wyg35}).
\ei

Concepts of values of expressions and formulas in states 
of DPs are defined similarly to the corresponding
  concepts for SPs. 

$\forall\, \varphi, \psi \in Fm $, the notation $ \varphi \leq \psi $ means that for each DP $ {\cal P} $ and each state
$ s$ of ${\cal P} $ the implication $ s \models \varphi \Rightarrow s \models \psi $ holds.

If  formulas $ \varphi, \psi \in Fm $ are such that $ \varphi \leq \psi $ and $ \psi \leq \varphi $, then we will consider such formulas as equal. If the formulas $ \varphi $ and $ \psi $ are equal, then we will denote this fact by  $ \varphi = \psi $.

Examples of equal formulas are the following: 
\bi\i
$\{f(e_1,\ldots, e_n)=
f(e'_1,\ldots, e'_n)\} \;(f\in Fun)$
and  $\{e_1=e'_1,\ldots, e_n=e'_n\}$,
\i $\{[c]=\{e\}, e'\in [c]\}$
and
$\{[c]=\{e\}, e=e'\}.$
\ei

\subsubsection{An execution  of a distributed process}
\label{vypraspr}

An {\bf execution} of a DP $ {\cal P} $ is a non-deterministic alternation of executions of
SPs  occurred in $ {\cal P} $. At each step of the execution 
only one SP from $ {\cal P} $ executes an active transition, and
other SPs from $ {\cal P} $ execute passive transitions. 

An execution of a DP $ {\cal P} $ can be understood as a
 generation of a sequence of states of $ {\cal P} $ (starting from $ 0_{\cal P} $), in which each pair $(s,s')$ of adjacent states 
 belongs to a
 {\bf transition relation}, which means the following: 
$\exists\,P\in {\cal P}$: 
\be{sdfagr33gr356ujyhfgd}
\!\!\!
\by
s_P\ra{P}s'_P,\;\;
\forall\,P'\in {\cal P}\setminus \{P\}\;\;
s_{P'}\to s'_{P'}\\
\mbox{where $s=\{s_P\mid P\in {\cal P}\}$, 
$s'=\{s'_P\mid P\in {\cal P}\}$.}
\ey\ee

The property \re{sdfagr33gr356ujyhfgd}
is denoted by $s\ral{\alpha_{P}}s'$,
where $\alpha=\alpha_{s_P'}$.

A set of  states of a DP $ {\cal P} $ can be considered
 as a graph in which there is an edge from $ s $ to $ s' $ labeled 
 by $ \alpha_{P} $ iff $ s \ral {\alpha_{P}} s' $. The designation $ P $ in the label $ \alpha_{P} $ can be omitted.

For each pair $ (s, s ')$ of states  of a DP
$ {\cal P} $ \bi\i $ s \to s' $ means that $( s,s') $ belongs to the transition relation, and \i $ s \Rightarrow s' $ means that there is a sequence  $ s_1, \ldots, s_n $ of states of ${\cal P}$ such that $ s_1 = s $, $ s_n = s' $, and $ \forall \, i = 1, \ldots, n-1 \; \; s_i \to s_{i + 1} $.
\ei

A state $ s $ of  $ {\cal P} $ is said to be {\bf reachable} if $ 0_{\cal P} \Rightarrow s $. The set of reachable states of $ {\cal P} $ is denoted by  $ \Sigma_{\cal P} $.
$\forall\,s,s'\in  \Sigma_{\cal P} $, 
the notation $ s \leq_\pi s' $ means that $\pi$ is a path 
such that $s,s'\in \pi$,
and either $s=s'$, or
$s$ is located on $\pi$ to the left of $s'$. The notation
$ s <_\pi s' $ means that $ s \leq_\pi s' $ and $ s \neq s' $. If 
$ \pi $ is clear from the context, then the designation 
of $\pi$ in $ \leq_\pi $ and $ <_\pi $ can be omitted.

\subsubsection{Observational equivalence
of distributed processes}
\label{agret54yywj75u6rt}

A concept of 
an observational  equivalence of DPs has the following  meaning: DPs $ {\cal P} $ and $ {\cal P} '$ are observationally equivalent if for each observer who analyzes an execution of $ {\cal P}_{\dagger} $ and ${\cal P} '_{\dagger} $ by observing 
the contents of $ \circ $, these DPs are indistinguishable.

Let $ {\cal P} $, $ {\cal P} '$ be DPs, 
$ s \in \Sigma_{{\cal P} _{\dagger}} $, $ s' \in \Sigma_{ {\cal P} '_{\dagger}} $, and $ \eta: X \to X'$ be a renaming.
The notation $s\eqs{\eta}s'$
 means that 
$ [\circ]_{s} \subseteq Tm (X)$
and
$[ \circ]_{s '} = \{e ^ \eta \mid e \in [\circ]_{s} \}$.

$ {\cal P} $ and $ {\cal P} '$ are said to be 
{\bf observationally equivalent} if there is a set $ \mu $ of triples 
$ (s, s', \eta) $, where $ s \in \Sigma_{{\cal P} _{\dagger}} $,
$ s' \in \Sigma_{{\cal P} '_{\dagger}} $, $s\eqs{\eta}s'$, such that
\bi\i
$(0_{{\cal P}_{\dagger}},0_{{\cal P}'_{\dagger}},\emptyset) \in \mu$ (where $ \emptyset $ is a function with empty domain), and 
\i $ \forall \, (s, s', \eta) \in \mu $
if $ s \to \tilde s $ or $ s' \to \tilde s' $, then $ \exists \, (\tilde s, \tilde s', \tilde \eta) \in \mu $: $ \tilde \eta $ is an extension 
of $ \eta $, and $ s '\Rightarrow \tilde s' $ or $ s \Rightarrow \tilde s $, respectively.
\ei

Note that
the above definition is not the only possible definition of an observational equivalence, and can be modified depending on the problem being solved. In some problems, a more appropriate definition of an observational equivalence is a coarsening of the equivalence defined above, such that,
for example, the DPs 
${\cal P}=\{P\}$ and
${\cal P}'=\{P'\}$
are
equivalent, where
$ P =
 \bullet \! \! \! \ral {\! \! ! k (e)} \! \! \! \bullet$,
$P' = \bullet \! \! \! \ral {! k '(e')}  \! \! \! \bullet$,
$ k \in \bar X ({P})_{\bf K},$
 $k '\in \bar X ({P}')_{\bf K}. $

 \subsection{Preservation theorems for values of formulas}
\label{sadfaaerg}

In this section, we formulate and prove  theorems
about a preservation of  values  of some formulas under transitions of DPs.
These theorems are used for verification of  DPs. In  examples of applications of these theorems below, \bi \i
SP $P$ mentioned in these theorems is $ P_{\dagger} $, and
\i informally speaking, these theorems state that \bi \i if  names 
of some channels are secure with respect to $P_{\dagger}$, then  contents of these channels cannot be changed by $P_{\dagger}$, and \i if some  keys are secure with respect to $P_{\dagger}$, then  contents of  EMs encrypted with these keys are inaccessible to $P_{\dagger}$.
\ei
\ei 

\subsubsection{Secure channel theorems }

First theorem is related to a preservation of values  of formulas of the form \be{zfgfdgdsfgdsrr} E \, \bot_{\bf C} \, P,
\quad
\mbox{where $ E \subseteq Tm $, and $ P $ is a SP}
\ee
under transitions of DPs. This theorem can be interpreted as the following statement: if 
${\cal P}$ is a DP, $P\in {\cal P}$, 
 there is no messages in $E$ which are available for $P$, and
 in a current state of ${\cal P}$ there are no messages from $E$ in channels available to $P$,
then no own activity of $P$ will lead to availability  for $P$
messages from $ E $.
Channels whose names are occurred in $ E $ can be interpreted as channels {\bf secure}  with respect to $P$. \\

\refstepcounter{theorem}
{\bf Theorem \arabic{theorem}\label{doredtusl2e3nab33222eq}}.

Let $ {\cal P} $ be a
DP, $ P \in {\cal P} $, 
and $ s, s'$ be states of ${\cal P}$
 such that $ s \ral {\! \alpha_{P}} s' $.

Then $ \forall \, E \subseteq \l {{\cal P}}_0 $ the following implication holds:
$$s\models E\,\bot_{\bf C}\,P\;\;\Rightarrow 
\;\;s'\models E\,\bot_{\bf C}\,P.$$

{\bf Proof}.

$s\models E\,\bot_{\bf C}\,P$  means that
$\forall\,e\in E\;\;Agent(P)\not\in e$, and
$\forall\,x\in E_{\bf X}$,
$\forall\,c\in Tm_{\bf C}$
\be{szdgdfhdge35gy6h5}
\left.\by
\forall\,y\in [P]_s\;\;
x\not\in y^s\\
\mbox{if }\exists\,e\in [c]_s: x\in e, 
\mbox{ then }c\in E
\ey\right\}\ee

Prove that  
\re{szdgdfhdge35gy6h5} implies that
$s'\models E\,\bot_{\bf C}\,P$, i.e.
$\forall\,x\in E_{\bf X}$,
$\forall\,c\in Tm_{\bf C}$
\be{s32z32dgdfhdge35gy6h5}
\left.\by
\forall\,y\in [P]_{s'},\;\;
x\not\in y^{s'}\\
\mbox{if }\exists\,e\in [c]_{s'}: x\in e, 
\mbox{ then }c\in E
\ey\right\}\ee
\bn
\i If first statement in \re{s32z32dgdfhdge35gy6h5} is false, then
first statement in \re{szdgdfhdge35gy6h5} implies  $ [P]_s \neq [P]_{s'} $, and one of the following two cases holds.
\bi
\i First case: 
\be{1asfdgagdsfgfsg34}
\!\!\!\!\!\!\left.\by 
\mbox{$\alpha = c?e$, where }\\
c\in \l{P}_s, e^{s'}\in [c^s]_s,\\
\,[P]_{s'} = [P]_s\cup 
 {\it Var}(e), \\
 \exists\,x\in E_{\bf X},
 \exists\,y\in Var(e):
x\in y^{s'}.
\ey\right\}
\ee
The statement 
$ x \in y ^ {s '} \subseteq e ^ {s'} \in [c ^ s]_s $ and 
second statement in \re{szdgdfhdge35gy6h5} imply 
  $ c ^ s \in E $.

If $ c ^ s \not \in E_{\bf X} $, then $ c $ has the form $ shared \_channel (\ldots) $, and in this case,  the definition of 
an execution  of a SP in \ref{hfdsklds} implies 
that $ Agent (P) \in c ^ s $. But this fact 
and the statement $ c ^ s \in E $ contradict the assumption $ \forall \, e \in E \; \; Agent (P) \not \in e $.

Thus, $ c ^ s \in E_{\bf X} $,
that implies $ c \in [P]_s $.
According to  first statement in \re{szdgdfhdge35gy6h5} (in which we take $ c ^ s $ and $ c $ as $ x $ and $ y $, respectively), the statement $ c ^ s \not \in c ^ s $ holds, but this is false.

\i Second case: 
\be{1asfdgagdsfgfsg333243}
\!\!\!\!\!\!
\left.\by 
\mbox{$\alpha = (e:=e')$,
where} \\
e'\in \l{P}_s,
e^{s'}=(e')^s,\\
\,[P]_{s'} = [P]_s\cup 
 {\it Var}(e), \\
\exists\,x\in E_{\bf X},
 \exists\,y\in {\it Var}(e):
x\in y^{s'}
\ey\right\}
\ee
Since
$x\in y^{s'}\subseteq e^{s'}=(e')^s\in \l{P}^s$, then
\be{fgfdsgesgsdfgds}
\exists\,z\in [P]_s:x\in z^s.\ee
 \re{fgfdsgesgsdfgds} contradicts  first statement in \re{szdgdfhdge35gy6h5}. 
\ei
\i If second statement in \re{s32z32dgdfhdge35gy6h5}
is false,
i.e. $$\by\exists\,x\in E_{\bf X},
\exists\,c'\in Tm_{\bf C},
\exists\,e'\in [c']_{s'}: \\x\in e', 
\mbox{ but }c'\not\in E\ey$$
then second statement in \re{szdgdfhdge35gy6h5} implies
that $ [c ']_{s'} \neq [c']_s $, and
$$
\mbox{$ \alpha = c! e $, where $ c, e \in \l {P}_s, x \in e '= e ^ s $}.
$$
From $ x \in e ^ s $ it follows that $ \exists \, y \in [P]_s: x \in y ^ s $, which contradicts  first statement in \re{szdgdfhdge35gy6h5}. 
$\blackbox$
\en

The following theorem is a strengthening of  theorem \ref{doredtusl2e3nab33222eq}. It states that under  conditions of  theorem \ref{doredtusl2e3nab33222eq} the lower and upper bounds on the contents of secure channels do not change when 
actions  of
$ P $ are executed.
\\

\refstepcounter{theorem}
{\bf Theorem \arabic{theorem}\label{doredtuslna32b23322eq}}.

Let $ {\cal P} $ be a
DP, $ P \in {\cal P} $, 
and $ s, s'$ be states of ${\cal P}$
 such that $ s \ral {\! \alpha_{P}} s' $.

Then $ \forall \, E \subseteq \l {{\cal P}}_0 $, 
$ E ', 
E''\subseteq Tm $, 
$c \in E_{\bf C } $ the 
following implication holds:
$$\by
s\models \varphi\Rightarrow
s'\models \varphi,\\
\mbox{ where }\varphi=
\{E\,\bot_{\bf C}\,P, 
E'\subseteq [c]\subseteq E''\}.
\ey$$

{\bf Proof}.

According to  theorem \ref{doredtusl2e3nab33222eq}, $ s \models E \, \bot_{\bf C} \, P $ implies $ s' \models E \, \bot_{\bf C} \, P $.
In addition, $ [c]_s \subseteq [c]_{s'} $.
Thus, to prove the theorem, it suffices to prove  implication 
\be{sfddsgdshgfdsghdsf}
s\models \varphi
\;\;\Rightarrow\;\;
s'\models  [c]\subseteq E''.\ee

If the conclusion of  implication \re{sfddsgdshgfdsghdsf} 
does not hold, then
$ [c]_s \neq [c]_{s'} $.

By assumption $ c \in E_{\bf C} \subseteq \l {{\cal P}}_0 $, so $ [c]_s \neq [c]_{s'} $ is possible only if $ \alpha = c '! e $, where $ (c') ^ s = c $.

If $ c $ is not a variable, then $ c $ is a shared channel, and by definition of an execution the action  $ c '! e $, in this case the condition $ Agent (P) \in c $ holds, which contradicts the assumption $ s \models E \, \bot_{\bf C} \, P $ (because, in particular, $ Agent (P) $ has no occurrences in terms from $ E $). Thus, $ c \in Var $,  $ c '\in Var $,  $ c' \in [P]_s $.

Since $ c \in E_{\bf X} $ and $ c '\in [P]_s $, then 
by  assumption  $ s \models E \, \bot_{\bf C} \, P $ 
we have
$ c \not \in c $, which is false.
 $ \blackbox $
 
 \subsubsection {Secure key theorems}

In this subsection, we prove theorems similar to theorems \ref{doredtusl2e3nab33222eq} and \ref{doredtuslna32b23322eq}. Now we consider secure keys instead of secure channels.

First theorem is related to a preservation of  values  of formulas 
\be{sadfgdg54ehrtge}
E\,\bot_{\bf K}\, P, 
\quad
\mbox {where $E\subseteq Tm$, and $P$ is a SP}
\ee
under transitions of DPs. 
This theorem can be interpreted as the following statement: if 
${\cal P}$ is a DP, $P\in {\cal P}$, 
 there is no messages in $E$ which are available for $P$, and
 in a current state of ${\cal P}$ 
\re{sadfgdg54ehrtge} holds,
then no own activity of $P$ will lead to availability for $P$
keys from $ E $. 
These keys can be interpreted as {\bf secure} keys 
with respect to $P$. \\

\refstepcounter{theorem}
{\bf Theorem \arabic{theorem}\label{doredtusl2nab33222eq}}.

Let $ {\cal P} $ be a
DP, $ P \in {\cal P} $, 
and $ s, s'$ be states of ${\cal P}$
 such that $ s \ral {\! \alpha_{P}} s' $.

Then $ \forall \, E \subseteq \l {{\cal P}}_0 $ the following
implication holds:
$$s\models E\,\bot_{\bf K}\, P\Rightarrow s'\models E\,\bot_{\bf K}\, P.$$

{\bf Proof}.

$ s \models E \, \bot_{\bf K} \, P $ means that $ \forall \, e \in E \; \; Agent (P) \not \in e $, and $\forall\,x\in E_{\bf X}$,
$\forall\,c\in Tm_{\bf C}$
\be{fdgh1sdfsg4341}\!\!\!\!\!\!\!\!
\left.\by
\forall\,y\in [P]_s\;\; 
x\,\bot_{{\bf K},E}\, y^s\\
\forall\,e\in [c]_s\;\; 
x\,\bot_{{\bf K},E}\, e
\ey\right\}\ee

Prove that \re{fdgh1sdfsg4341} implies 
that $ s' \models E \, \bot_{\bf K} \, e $, i.e. 
$\forall\,x\in E_{\bf X}$,
$\forall\,c\in Tm_{\bf C}$
\be{fdg1h3sdfsg432341}\!\!\!\left.
\by
\forall\,y\in [P]_{s'}\;\; 
s'\models x\,\bot_{{\bf K},E}\, y^{s'}
\\
\forall\,e\in [c]_{s'}\;\; 
s'\models x\,\bot_{{\bf K},E}\, e
\ey\right\}\ee
\bn\i
If  first statement in \re{fdg1h3sdfsg432341} is wrong, 
then first statement in \re{fdgh1sdfsg4341} implies 
$ [P]_s \neq [P]_{s'} $, and one of two cases does hold:
\bi
\i First case: the following statement holds
\be{asfdgagdsfgfsg34}
\left. \by
\mbox{$ \alpha = c? e $},
c \in \l {P}_{s},
e ^ {s'} \in [c ^ s]_s, 
\\
\,[P]_{s'} = [P]_s \cup
 {\it Var} (e),\\
\mbox{$\exists \, y \in {\it Var} (e) $},\\
\mbox{$ \exists $ an occurrence
of $ x $ in $ y ^ {s '} $}\\
\mbox{that is not contained }\\
\mbox{in any subterm
of the form 
}\\\mbox{$ k (\ldots) \subseteq y ^ {s'} $, 
where $ k \in E_{\bf K} $.}
\ey \right \}
\ee

Since the occurrence of $ x $ mentioned in \re{asfdgagdsfgfsg34} occurs in the term $ y ^ {s '} \subseteq e ^ {s'} \in [c ^ s]_s $, then second statement in \re{fdgh1sdfsg4341} implies 
that this occurrence of $ x $ occurs in the subterm $ k (\tilde e) \subseteq e ^ {s'} $, where $ k \in E_{\bf K} $.

 \re{asfdgagdsfgfsg34} 
implies that $ k (\tilde e) $ is not a subterm of $ y ^ {s'} $. Since the terms $ k (\tilde e) $ and $ y ^ {s'} $ have a non-empty intersection (both contain the above occurrence of $ x $), then  \re{dzfgdfghxfhgfhde4e56} implies that
$ y ^ {s'} \subset k (\tilde e) $.
Thus, \be{112rfgrsfgrtt} y ^ {s'} \subset
k (\tilde e)
\subseteq e ^ {s'}. \ee

Prove by induction on the structure of $ e $ that \re{112rfgrsfgrtt} implies 
\be{zfdgdsghsdd} \exists \, z \in Var (e): \; \;
k (\tilde e) \subseteq z ^ {s '} \subseteq e ^ {s'}. \ee

If $ e \in Con \cup {\it Var} $, then  
\re{zfdgdsghsdd} holds. 

If $ e = f (e_1, \ldots, e_n) $, where $ f \in Fun $, then \bi \i if $ f = encrypt $, i.e. $ e = k_1 (e_1) $, then $ k_1 \in Keys (e) $, and \re{asdfsafgewe4444}(b) implies  
the inclusion $ Keys (e) \subseteq [P]_s $, 
thus $ k_1 \in [P]_s $, and the following cases are possible:
\bi
\i $ k (\tilde e) = e ^ {s '} = k_1 ^ {s'} (e_1 ^ {s '}) $, in this case $ k = k_1 ^ {s'} = k_1 ^ s \in [P] ^ s $, but since $ k \in E_{\bf K} $, then, according to  first statement in \re{fdgh1sdfsg4341}, the occurrence of $ k $ in $ k $ occurs in a subterm of the form
$ k '(\ldots) \subseteq k $, which is impossible,
\i $ k (\tilde e) \subseteq k_1 ^ {s'} $, this case is impossible by the definition of terms of the type $ {\bf K} $,
\i $ k (\tilde e) \subseteq e_1 ^ {s'} $, in this case  statement \re{zfdgdsghsdd} follows from the inductive hypothesis,
\ei
\i if $ f \neq encrypt $, then $ \exists \, i \in \{1, \ldots, n \}: k (\tilde e) \subseteq e_i ^ {s'} $, and statement \re{zfdgdsghsdd} follows from inductive hypothesis.
\ei

From \re{112rfgrsfgrtt} and \re{zfdgdsghsdd} it follows that
\be{sdafgadsgewwewtwryehree}
y ^ {s'} \subset k (\tilde e) \subseteq z ^ {s'} \subseteq e ^ {s'}. \ee
Thus, the term $ e $ contains occurrences of the variables $ y $ and $ z $ with the following property: $ y ^ {s '} \subset z ^ {s'} $, whence for these occurrences the inclusion $ y \subset z $
holds, which is impossible.

\i Second case: the following statement holds
\be{asfdgagdsfgfsg333243}
\left. \by
\mbox{$ \alpha =( e: = e ')$,
} \\\mbox{
 $ e '\in \l {P}_s ,
e ^ {s '} = (e') ^ s$},\\ 
\mbox{$[P]_{s'} = [P]_s \cup
 {\it Var} (e)$,}\\
\mbox{$
 \exists \, y \in {\it Var} (e)$, 
 } \\\mbox{$ \exists $ an occurrence of $ x $ in $ y ^ {s '} $} \\
\mbox{that is not contained}\\
\mbox{in any subterm of the form 
} \\\mbox{$ k (\ldots) \subseteq y ^ {s'} $, where $ k \in E_{\bf K} $.}
\ey \right \}
\ee

Since the occurrence of $ x $ mentioned in \re{asfdgagdsfgfsg333243} is contained in 
the term $ y ^ {s'} \subseteq e ^ {s'} = (e ') ^ s $, then this occurrence of $ x $ is contained in the subterm $ ( z ') ^ s \subseteq (e') ^ s $, where $ z '\in Var (e') $.

By assumption, $ e '\in \l {P}_s $, therefore $ Var (e') \subseteq [P]_s $, so $ z '\in [P]_s $. From  first statement in \re{fdgh1sdfsg4341} it follows that the occurrence of $ x $ in $ (z ') ^ s $ mentioned in \re{asfdgagdsfgfsg333243} is contained in some subterm $ k (\tilde e) \subseteq (z') ^ s $, where $ k \in E_{\bf K} $.

From \re{asfdgagdsfgfsg333243} it follows that $ k (\tilde e) $ 
is not a subterm of $ y ^ {s'} $.

Since the terms $ k (\tilde e) $ and $ y ^ {s'} $ have a non-empty intersection (both contain the occurrence of $ x $ mentioned in \re{asfdgagdsfgfsg333243}), then from \re{dzfgdfghxfhgfhde4e56} it follows that $ y ^ {s'} \subset k (\tilde e) $.

The equality $ e '= e ^ \theta $ implies
that
$\exists\,z \in Var (e) $: the above occurrence of $ z' $ in $ e '$ occurs in the subterm $ z ^ \theta \subseteq e ^ \theta = e '$.
Therefore, $ (z ') ^ s \subseteq (z ^ \theta) ^ s = z ^ {s'} \subseteq e ^ {s'} $.

So we get:
\be{sadfadsw3w4t5y365}
y ^ {s'} \subset
k (\tilde e)
\subseteq (z ') ^ s \subseteq z ^ {s'}
\subseteq e ^ {s'}.
\ee
As in the previous section, on the reason of 
\re{sadfadsw3w4t5y365}, we conclude that the term $ e $ contains occurrences of the variables $ y $ and $ z $ with the following property: $ y ^ {s '} \subset z ^ {s'} $, whence the inclusion of $ y \subset z $ holds for these occurrences, which is impossible.

\ei
\i
If  second statement in \re{fdg1h3sdfsg432341} is not true, then from  second statement in \re{fdgh1sdfsg4341} it follows that
\be{asfdgag32dsfgfsg34}
\left. \by
\mbox{$ \alpha = c! e $, where $ e \in \l {P}_s $}, \\
\mbox{$ \exists $ occurrence of $ x $ in $ e ^ {s} $}\\\mbox{that is not contained} \\\mbox{in any subterm of the form 
}\\\mbox{$ k (\ldots) \subseteq e ^ s $, where $ k \in E_{\bf K} $.}
\ey \right \}
\ee
Since $ e \in \l {P}_s $, then the occurrence of $ x $ in $ e ^ s $ mentioned in \re{asfdgag32dsfgfsg34} is contained in a subterm of the form $ y ^ s $ of the term $ e ^ s $, where $ y $ is some variable from $ [P]_s $.
According to  first statement in \re{fdgh1sdfsg4341}, this occurrence of $ x $ in $ y ^ s $ is contained in a subterm of the form $ k (\ldots)
\subseteq y ^ s \subseteq e ^ s $, where $ k \in E_{\bf K} $.
But this contradicts to \re{asfdgag32dsfgfsg34}.
$ \blackbox $ 
\en 

\refstepcounter{theorem}
{\bf Theorem \arabic{theorem}\label{dor1edtusln32a32b322eq}}.

Let ${\cal P}$ be a DP such that $ Var ({\cal P})_{\bf C} = \{\circ \} $,  $ P \in {\cal P} $,  $ s, s '$ be states from
$\Sigma_{\cal P} $ 
such that $ s \ral {\! \alpha_{P}} s' $, and $ E \subseteq \l {{\cal P} }_0 $.

Then $ \forall \, k \in E_{\bf K}$ 
the following implication holds:
 $$ s \models \varphi \Rightarrow
s' \models \varphi$$
where $\varphi =
\{
E \, \bot_{\bf K} \, P,
k ^ {- 1} [{P}] \subseteq k ^ {- 1} [\circ] \}.
$\\

{\bf Proof}.

By theorem \ref{doredtusl2nab33222eq},  
the statement $ s \models E \, \bot_{\bf K} \, P $ implies 
$ s' \models E \, \bot_{\bf K} \, P $. Thus, to prove theorem \ref{dor1edtusln32a32b322eq}, it suffices to prove the implication
\be{1sfddsgdshgfdsghdsf21221}
s \models \varphi \; \; \Rightarrow \; \;
s' \models k ^ {- 1} [{P}] \subseteq k ^ {- 1} [\circ]. \ee

If \re{1sfddsgdshgfdsghdsf21221} does not hold,
then
 $ \alpha $ is not a sending, $ [P]_{s '} = [P]_s \cup {\it Var} ({e'}) $, and $\exists\, e \in k ^ {- 1} [P] ^ {s'} $ such that
\be{1adsfadsfsfgewrgfgwf3343}\by
\mbox{$ e \not \in k ^ {- 1} [\circ]_{s} $, i.e. 
$\exists \, x \in [P]_{s '}:$}\\\mbox{$ k (e) \subseteq x ^ {{s'}}, \forall \, \dot e \in [\circ]_s \; \; k (e) \not \subseteq \dot e. $}
\ey\ee

The assumption $ s \models \varphi $ and \re{1adsfadsfsfgewrgfgwf3343} imply that $ x \in {\it Var} ({e '}) $, whence we get: $ k (e) \subseteq x ^ { {s'}} \subseteq (e ') ^ {{s'}} $.

Consider separately each of the two possible types of $ \alpha $.
\bn
\i \label {adsgfdsghsdfgdsa} $ \alpha =? e '$, in this case $ k (e) \subseteq (e') ^ {{s'}} \in [\circ]_s $. Setting in \re{1adsfadsfsfgewrgfgwf3343} the term $ \dot e $ be equal to $ (e ') ^ {{s'}} $, we get a contradiction.

\i \label {adsgfdsghsdfgdsa32}

$ \alpha = (e ': = e'') $, in this case $ e''\in \l {P}_s $, $ (e') ^ {{s'}} = (e'') ^ {s} $, so \be{1s3fddsgdshgfdsg3hdsf1} k (e) \subseteq (e'') ^ {s}. \ee
Prove that $ k \not \in e'' $ and $ k \in E_{\bf X} $.

Suppose $ k \in e'' $. 
If $ k = shared \_key (\ldots) $, then, according to the definition of SP execution in section \ref{hfdsklds}, the statement $ Agent (P) \in k \in E_{\bf K} $ holds, which contradicts  first condition of the property $ s \models E \, \bot_{\bf K} \, P $.
Recall that this property has the form:
$ \forall \, \tilde e \in E $ $ Agent (P) \not \in \tilde e $, and
\be{1f332dgh1s43dfsg4343321} 
\left. \by \forall \, x \in E_{\bf X}, \,
\forall \, y \in [P]_s \; \;
x \, \bot_{{\bf K}, E} \, y ^ s \\
\forall \, x \in E_{\bf X}, \,
\forall \, \tilde e \in [\circ]_s \; \;
x \, \bot_{{\bf K}, E} \, \tilde e
\ey \right \} \ee
Therefore, $ k \in E_{\bf X} $, so  
$ k \in e'' \in \l {P}_s $ implies
$ k \in [P]_s $. However, assuming in  first statement in \re{1f332dgh1s43dfsg4343321} $ x $ and $ y $ be equal to
 $ k $, we get $ k \, \bot_{{\bf K}, E} \, k $, which is false by definition
 \re{sdfdasgfadsfgrewgtwg}.

 Similarly to the proof of the implication \re{112rfgrsfgrtt} $ \Rightarrow $ \re{zfdgdsghsdd} in theorem \ref{doredtusl2nab33222eq}, we  prove 
 that  the statements $ k \not \in e'' $ and 
  \re{1s3fddsgdshgfdsg3hdsf1}
 imply
 \be{1dsafsf3w3w5t5367ui8} \exists \, y \in {\it Var} ({e''}) \subseteq [P]_s: k (e) \subseteq y ^ {{s}}. \ee

By assumption, $ s \models k ^ {- 1} [{P}] \subseteq k ^ {- 1} [\circ] $, i.e.
$ k ^ {- 1} [{P}] ^ s \subseteq k ^ {- 1} [\circ]_s $. From \re{1dsafsf3w3w5t5367ui8} it follows that $ e \in k ^ {- 1} [P] ^ s $. Therefore, $ e \in k ^ {- 1} [\circ]_s $, which contradicts the assumption \re{1adsfadsfsfgewrgfgwf3343}. $ \blackbox $
\en

The following theorem is a strengthening of theorem \ref{dor1edtusln32a32b322eq}. It states that under the conditions of  theorem \ref{dor1edtusln32a32b322eq} the lower and upper bounds for the set of EMs contained in  open channel and encrypted with secure keys do not change when  actions of the SP $ P $ are executed. \\

\refstepcounter {theorem}
{\bf Theorem \arabic {theorem} \label {doredtuslna32b322eq}}.

Let $ {\cal P} $ be a DP
such that $ Var ({\cal P})_{\bf C} = \{\circ \} $, $ P \in {\cal P} $, $ s, s '$ be states from $\Sigma_{\cal P} $ such that $ s \ral {\! \alpha_{P}} s' $, and $ E \subseteq \l {{\cal P} }_0 $, $ E ', E''\subseteq Tm $.

Then $ \forall \, k \in E_{\bf K}$ 
the following implication holds:
 \be{sfdgfdgdsfhgsdghsdg}
\by
s \models \varphi \Rightarrow
s' \models \varphi,
\mbox{where }\\ \varphi =
\c {
E \, \bot_{\bf K} \, P, k ^ {- 1} [{P}] \subseteq k ^ {- 1} [\circ] \\E '\subseteq k ^ {- 1} [ \circ] \subseteq E''}.
\ey\ee

{\bf Proof}.

By theorem \ref{dor1edtusln32a32b322eq}, to prove \re{sfdgfdgdsfhgsdghsdg} it is enough to prove that \be{sfddsgdshgfdsghdsf21221} s\models
 \varphi  \Rightarrow  s '\models \{E' \subseteq k ^ {- 1} [\circ], k ^ {- 1} [\circ] \subseteq E'' \}. \ee

\bi \i
The statement $ s' \models E '\subseteq k ^ {- 1} [\circ] $ follows from $ [\circ]_s \subseteq [\circ]_{s'} $.
\i Prove the statement $ s' \models k ^ {- 1} [\circ] \subseteq E'' $. If it is not true, then $ [\circ]_s \neq [\circ]_{s'} $. This is possible only if $ \alpha =! e $, where $ e \in \l {P}_s $, and \be{dsfasgfagregt54} 
\by
[\circ]_{s'} = [\circ]_s \cup \{e ^ s \} ,\\ 
\exists \, e '\not \in (E'') ^ {s} \supseteq k ^ {- 1} [\circ]_s:
k (e') \subseteq e ^ s .\ey
\ee

As in theorem \ref{dor1edtusln32a32b322eq},
we prove that $ k \in E_{\bf X} $, and 
$ e \in \l {P}_s $ implies
 $ k \not \in e $.
 
 Similarly to the proof of  implication \re{112rfgrsfgrtt} $ \Rightarrow $ \re{zfdgdsghsdd} in theorem \ref{doredtusl2nab33222eq}, we can prove that 
 the statements
 $ k \not \in e $ and $ k (e ') \subseteq e ^ s $ imply $$ \exists \, x \in {\it Var} ({e}) \subseteq [P]_s: \; k (e ') \subseteq x ^ {{s}}, $$ therefore $ e' \in k ^ {- 1} [P] ^ s $. Hence, using the assumption $ s \models \varphi $, which results to the inclusion $ k ^ {- 1} [{P}] ^ s \subseteq k ^ {- 1} [\circ]_s $, we get: $ e '\in k ^ {- 1} [\circ]_s $, which contradicts \re{dsfasgfagregt54}. $ \blackbox $
\ei

\subsection {$ \! \! \! \! \! $
Theorem for proving a correspondence property}

A theorem stated in this section can be used to prove a {\bf correspondence property} of authentication protocols, which has the following meaning: if one of  participants of an authentication protocol, after executing this protocol, has come to the conclusion that  declared name and parameters of 
other participant of this protocol are authentic, then
then this is indeed the case. A theorem proved below is used to prove that if \bi \i
a DP $ {\cal P} $ uses only open channel $ \circ $ for communication, and
\i in some state $ s \in \Sigma_{\cal P} $ this channel contains a message containing a subterm $ k (e) $, where the key $ k $ is secure 
in $s$ with respect to 
$ P \in {\cal P} $, \ei
then in some state $ s '<_\pi s $ another SP $ P'\in {\cal P}\setminus\{P\}$ sent a message to $ \circ $, containing the  subterm $ k (e) $.

In sections \ref{sdfdsgerre434343456} and \ref{sdfgstrghw54y46ywh} we consider examples of applying this theorem to verification of the Yahalom 
CP,
and  to verification of the CP of EMs passing
between several agents.
\\

\refstepcounter{theorem}
{\bf Theorem \arabic{theorem}\label{doredtuslna32b322e3323q}}.


Let $ {\cal P} $ be a DP, $ Var ({\cal P})_{\bf C} = \{\circ \} $, $ P \in {\cal P} $, 
$E$ be a subset of $\l {{\cal P}}_0 $ 
and $ s\in \Sigma_{\cal P}$ 
be a state 
such that
\bi \i $ s \models E \, \bot_{\bf K} \, P $, and
\i $ [\circ]_s $ contains a term with a subterm $ k (e) $, where $ k \in E_{\bf K} $.
\ei

Then, for each path $ \pi $ from $0_{\cal P}$  to $ s $, 
there is a SP $ P '\in {\cal P} \setminus \{P \} $ such that $ \pi $ has an edge of the form
\be{fgfdsgdsgsdfgfdsgh} {\dot s} \ral {(! \dot e)_{P '}} s', \quad \mbox{where $ k (e) \subseteq \dot e ^ {\dot s} $.} \ee

{\bf Proof}.

Let $ s' $ be a first state on $ \pi $ such that $ [\circ]_{s'} $ has a term $ e '$ with the subterm $ k (e) $. Since $ [\circ]_0 = \emptyset $, then $ s' \neq 0 $.

Let $ \dot s \ra {\alpha_{P '}} s' $ be 
an edge on $ \pi $ ending at $ s' $. 
Since $ e '\not \in [\circ]_{\dot s} $, then $ \alpha =! \Dot e $, where $ \dot e ^ {\dot s} = e' $. If $ P '\neq P $, then the theorem is proven.

Prove that another possible case ($ P '= P $) is impossible.

Suppose $ P '= P $, i.e. $ \dot s \ral {(! \dot e)_{P}} s' $.

Prove that $ k \in E_{\bf X} $.
If $ k \not\in E_{\bf X} $, 
i.e. $ k $ is a shared key, then, by the definition of an 
execution  of a
SP in \ref{hfdsklds}, $ Agent (P) \in k \in E_{\bf K} $, which contradicts the assumption $ \forall \, \tilde e \in E \; \; Agent (P) \not \in \tilde e $.

 $ s \models E \, \bot_{\bf K} \, P $ implies 
 $ \dot s \models E \, \bot_{\bf K} \, P $, whence we get $ k \not \in [P]_{\dot s} $, because if $ k \in [P]_{\dot s} $, then, according to \re{f443dgh1sdfsg4341}, the occurrence of $ k $ in $ k ^ {\dot s} = k $ must be contained in a subterm of the form $ k '( \ldots) $, which is impossible.

The statement
$ k \not \in [P]_{\dot s} $ and condition $ \dot e \in \l {P}_{\dot s} $ (which is true according to \re{asdfsafgewe4444}(a)) imply
$ k \not \in \dot e $.

Similarly to the proof of the implication \re{112rfgrsfgrtt} $ \Rightarrow $ \re{zfdgdsghsdd} in theorem \ref{doredtusl2nab33222eq}, we can prove that 
 statements $ k (e) \subseteq e '= \dot e ^ {\dot s} $ and $ k \not \in \dot e $ imply
\be{asdfsga34t53} \exists \, x \in Var (\dot e) \subseteq [P]_{\dot s}: k (e) \subseteq x ^ {\dot s} \in [P] ^ { \dot s}. \ee

Let $ s'' $ be a first state on  $ \pi $ such that $ [P] ^ {s''} $ has a term with the subterm $ k (e) $, that is, \be{dfgdsgdsfg335} \exists \, x \in [P]_{s''}: k (e) \subseteq x ^ {s''}. \ee 

\re{asdfsga34t53} implies that $ s''$ is  
to the left of $ s' $ on $\pi$.
It is clear that
$ s'' \neq 0 $, so there is an edge 
$ \ddot s \ral {\; \alpha_{P''}} s'' $
on $ \pi $. 
From the choice of $ s'' $ it follows that $ x \not \in [P]_{\ddot s} $, thus  $ P'' = P $, and two cases are possible:
\bn
\i $ \alpha =? \ddot e, x \in Var (\ddot e),
\ddot e ^ {s''} \in [\circ]_{\ddot s} $,

since $ k (e) \subseteq x ^ {s''} \subseteq \ddot e ^ {s''} \in [\circ]_{\ddot s} $, then we get a contradiction with the choice of $ s' $ as the  first state on $ \pi $ such that $ [\circ]_{s '} $ contains the term $ e' $ with the subterm $ k (e) $: the state $ \ddot s $ has the same property, and 
is located to the left of $ s' $,
\i
$ \alpha = (\ddot e: = \tilde e),
x \in Var (\ddot e),
\tilde e \in \l {P}_{\ddot s},
\ddot e ^ {s''} = \tilde e ^ {\ddot s} $,

since \bi \i
$ k (e) \subseteq x ^ {s''} \subseteq \ddot e ^ {s''} =
\tilde e ^ {\ddot s} $ and \i $ \tilde e $ does not contain $ k $, because  it was proven above that $ k \not \in [P]_{\dot s} $, therefore, taking into account the property $ \ddot s \leq \dot s $, which implies the inclusion $ [P]_{\ddot s } \subseteq
[P]_{\dot s} $, we get: $ k \not \in [P]_{\ddot s} $, and therefore the term $ \tilde e \in \l {P}_{\ddot s} $ also does not contain $ k $, \ei then, similarly to the proof of the implication \re{112rfgrsfgrtt} $ \Rightarrow $ \re{zfdgdsghsdd} in  theorem \ref{doredtusl2nab33222eq}, we can prove that
$$ \exists \, y \in [P]_{\ddot s}:
k (e) \subseteq y ^ {\ddot s}, $$
which contradicts the choice of $ s'' $ as a  first state on  $ \pi $ with the property \re{dfgdsgdsfg335}: $ \ddot s $ has the same property and is located to the left of $ s'' $. $ \blackbox $
\en

\subsection {Diagrams of distributed processes}
\label {fadsgadsg3w54ggfd}

\subsubsection {Prefix sequential processes}

A SP $ P $ is said to be a {\bf prefix SP} if \be{afdsga23sd342t34t5ytwe}\by
\begin{picture}(20,15)
\put(-70,0){\circle*{4}}
\put(-25,0){\circle*{4}}
\put(25,0){\circle*{4}}
\put(75,0){\circle*{4}}
\put(-68,0){\vector(1,0){41}}
\put(27,0){\vector(1,0){46}}
\put(-23,0){\line(1,0){10}}
\put(13,0){\vector(1,0){10}}
\put(0,0){\makebox(0,0){$\ldots$}}
\put(-90,0){\makebox(0,0){$P=$}}
\put(-70,5){\makebox(0,0)[b]{$0$}}
\put(-25,5){\makebox(0,0)[b]{$1$}}
\put(25,5){\makebox(0,0)[b]{$n-1$}}
\put(75,5){\makebox(0,0)[b]{$n$}}
\put(-48,2){\makebox(0,0)[b]{$ \alpha_{1}$}}
\put(50,2){\makebox(0,0)[b]{$ \alpha_{n}$}}
\put(80,0){\makebox(0,0)[l]{$P'$}}
\put(80,3){\oval(40,22)}
\end{picture}\ey
\ee
i.e. $ P $ contains nodes numbered by natural numbers $ 0,1, \ldots, n \; (n \geq 1) $, and $ Init (P) = 0 $, $ \forall \, i = 0, \ldots , n-1 $ there is exactly one outgoing edge from  node $ i $ with the end $ i + 1 $ and labeled by $ \alpha_i $. The 
notation $ P '$ in \re{afdsga23sd342t34t5ytwe} denotes a subgraph of the graph $ P $, consisting of the nodes and edges of the graph $ P $, with the 
exception of  the nodes $ 0, \ldots, n-1 $ and   edges associated with these nodes.

The
subgraphs $ 0 \ra {\alpha_1} 1 \ra {\alpha_2} \ldots
\ra {\alpha_n} n $ and $ P '$ of the graph \re{afdsga23sd342t34t5ytwe} are called 
a {\bf prefix} and a {\bf postfix} of the SP $ P $, respectively, and are denoted by  $ {\it Pref} (P) $ and $ {\it Post} (P) $ respectively. The last node of $ {\it Pref} (P) $ is called a {\bf final} node of this prefix. If $ {\it Post} (P) $ consists of one node, then it is denoted by {\bf 0}.

If a SP $ P $ has the form \re{afdsga23sd342t34t5ytwe}, then we will denote this fact by 
\be{sadfdasgfdgfadse3rrtge}
P = \alpha_1; \ldots; \alpha_n; P'. \ee

\subsubsection {A concept of a diagram of 
a distributed process}

Let $ {\cal P} $ be a DP, such that each
$P \in {\cal P}$ is  a prefix SP, 
and for each sending (each receivng) in 
$ {\it Pref} ( P) $  it is assumed that 
\bi \i 
an intended receiver (an intended sender) of the message that is sent (received) when performing this action
is some $ P '\in  {\cal P} $, and
\i an action of the SP $ P '$ corresponding to the receiveng (sending) of this message is in $ {\it Pref} (P') $. \ei

These dependencies between the actions can be expressed 
by a {\bf diagram} of the  DP ${\cal P}$, which has the following form:
\bi
\i each SP $ P \in {\cal P} $ is represented 
by a {\bf thread}
in this diagram, i.e. by a vertical line on which  points corresponding to  nodes of $ {\it Pref} (P) $ are marked, the upper point corresponds to $ Init (P) $, and
\bi
\i each point has the number of the corresponding node,
\i a name $P$ of the SP $P$ is indicated near the upper point,
\i if
$ {\it Post} (P) =P'\neq {\bf 0} $, then 
$P'$ is indicated at the bottom point,
\i near to each segment $ l $ connecting adjacent points
on the thread,
there is a label $ \alpha_l $ of an edge from $ {\it Pref} (P_i) $ corresponding to $ l $,
\ei
\i for each segment $ l $ connecting adjacent points
of the thread, if $ \alpha_l $ is a sending, 
$ \alpha_{l '} $ is 
an  intended receiving for $ \alpha_l $,
then the diagram contains an arrow, 
a start of which lies on $ l $, and an end 
of which lies on $ l '$.
\ei 
For example if $ P_i = \alpha_{1}; \; \ldots \; \alpha_{n}; P'_i $, where $ \alpha_{1} $ is a sending, and $ \alpha_{n} $ is a receiving, then  SP $ P_i $ corresponds to a thread \be{afdsgasd342t34t5ytwe}\by
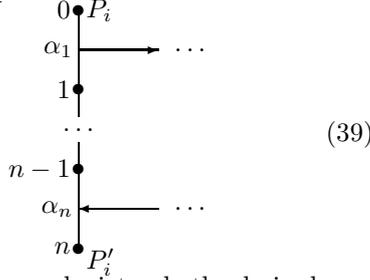
\begin{picture}(0,90)
\put(0,90){\circle*{4}}
\put(0,60){\circle*{4}}
\put(0,30){\circle*{4}}
\put(0,0){\circle*{4}}
\put(0,90){\line(0,-1){40}}
\put(0,0){\line(0,1){40}}
\put(0,45){\makebox(0,0){$\ldots$}}
\put(-2,75){\makebox(0,0)[r]{$ \alpha_{1}$}}
\put(-2,15){\makebox(0,0)[r]{$ \alpha_{n}$}}
\put(-3,90){\makebox(0,0)[r]{$0$}}
\put(-3,60){\makebox(0,0)[r]{$1$}}
\put(-3,30){\makebox(0,0)[r]{${n-1}$}}
\put(-3,0){\makebox(0,0)[r]{$n$}}
\put(3,-5){\makebox(0,0)[l]{$P_i'$}}
\put(3,90){\makebox(0,0)[l]{$P_i$}}
\put(0,75){\vector(1,0){30}}
\put(30,15){\vector(-1,0){30}}
\put(36,75){\makebox(0,0)[l]{$\ldots$}}
\put(36,15){\makebox(0,0)[l]{$\ldots$}}
\end{picture}\ey
\ee

Note that the arrows depict only the desired connection between   sendings and receivings, but they have no relation with real communication: it is possible that the sent message will be received by a SP which is  different from a SP to which it was intended.

For the sake of greater clarity, we will use the following convention in the notation of variables:
\bi
\i we will indicate a horizontal bar above a designation of a variable $ x $ if $x\in \bar X ({\cal P}) $ (i.e. this variable is denoted by $ \bar x $),
\i if $ P $ is a SP of the form \re{sadfdasgfdgfadse3rrtge},  
$ x \in \hat X (P) $,
and $ i $ is the first index such that $x\in \alpha_i $ 
 (i.e. $ \forall \, i '= 1, \ldots, i-1 $ $ x \not\in \alpha_{i '} $), 
then  occurrences of $x$ 
in the label $ \alpha_i $ of 
the $ i $--th segment of the thread of $ P $
 are denoted by 
$ \hat x $.
\ei
These variable designations also will be used in notations
    of the form \re{sadfdasgfdgfadse3rrtge}.

\subsubsection {Examples of diagrams of distributed processes }
\label {pkergnmf43564654632}

\bn \i
First example is  DP $ {\cal P}_1 = \{A, B \} $, which is a model of a transmission from $ A $ to $ B $ 
a message $ x $ through a  channel $ c_{AB} $, where only $ A $ and $ B $ know the name of this channel, i.e. $$ c_{AB} = shared \_channel (A, B) .$$

This DP works as follows:
\bi \i
 $ A $ sends the message $ x $ to   $ c_{AB} $,
\i
$ B $ receives this message from $ c_{AB} $ and writes it to the variable $ y $, after which it behaves like the SP $ P $.
\ei 

SPs $ A $ and $ B $ are defined as follows:
$$A = c_{AB}! x; {\bf 0},\;\;
B = c_{AB}? \hat y; P.$$

A
diagram of $ {\cal P}_1 $ has the following form: 
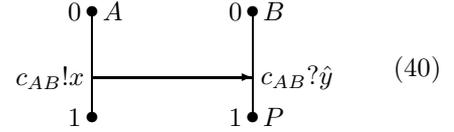
\be{primerdfadsljfk1}\by
\begin{picture}(60,40)
\put(0,15){\vector(1,0){60}}
\put(0,0){\line(0,1){40}}
\put(60,0){\line(0,1){40}}
\put(4,40){\makebox(0,0)[l]{$A$}}
\put(64,40){\makebox(0,0)[l]{$B$}}
\put(0,40){\circle*{4}}
\put(60,40){\circle*{4}}
\put(0,0){\circle*{4}}
\put(60,0){\circle*{4}}
\put(-4,40){\makebox(0,0)[r]{$0$}}
\put(-4,0){\makebox(0,0)[r]{$1$}}
\put(56,40){\makebox(0,0)[r]{$0$}}
\put(56,0){\makebox(0,0)[r]{$1$}}
\put(64,0){\makebox(0,0)[l]{$P$}}
\put(-3,15){\makebox(0,0)[r]{$ c_{AB}!x$}}
\put(63,15){\makebox(0,0)[l]{$ c_{AB}?\hat y$}}
\end{picture}\ey
\ee

\i
Second example is  DP $ {\cal P}_2 = \{A, B \} $, which is a model of transmission from $ A $ to $ B $  EM $ k_{AB} (x) $ through an open channel $ \circ $.
It is assumed that $ A $ and $ B $ have a shared secret key $ k_{AB} $, on which they can encrypt and decrypt messages using a symmetric encryption system, and only $ A $ and $ B $ know   $ k_{AB } $, i.e. $$ k_{AB} = shared \_key (A, B). $$

This DP works as follows:
\bi
\i $ A $ sends   EM $k_{AB} (x) $ to
 $ \circ $,
\i $ B $ receives EM $ k_{AB} (x) $ from 
 $ \circ $,
decrypts it, writes the extracted message $ x $ to  variable $ y $, after which it behaves like SP $ P $.
\ei

SPs $ A $ and $ B $ are defined as follows:
$$A =! k_{AB} (x); {\bf 0},\;\;
B =? k_{AB} (\hat y); P.$$

A  diagram of  $ {\cal P}_2 $ has the following form: 
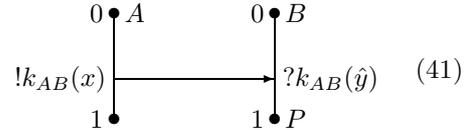
\be{primerdfadsljfk2}\by
\begin{picture}(60,40)
\put(0,40){\circle*{4}}
\put(60,40){\circle*{4}}
\put(0,0){\circle*{4}}
\put(60,0){\circle*{4}}
\put(-4,40){\makebox(0,0)[r]{$0$}}
\put(-4,0){\makebox(0,0)[r]{$1$}}
\put(56,40){\makebox(0,0)[r]{$0$}}
\put(56,0){\makebox(0,0)[r]{$1$}}
\put(0,15){\vector(1,0){60}}
\put(0,0){\line(0,1){40}}
\put(60,0){\line(0,1){40}}
\put(4,40){\makebox(0,0)[l]{$A$}}
\put(64,40){\makebox(0,0)[l]{$B$}}
\put(64,0){\makebox(0,0)[l]{$P$}}
\put(-3,15){\makebox(0,0)[r]{$! k_{AB}(x)$}}
\put(63,15){\makebox(0,0)[l]{$? k_{AB}(\hat y)$}}
\end{picture}\ey
\ee

\i

Third example is  DP $ {\cal P}_3 = \{A, B, J \} $, which is a model of transmission from $ A $ to $ B $ a message $ x $ over  secret channel $ \bar c $ with use of  a {\bf trusted intermediary} $ J $, where $ A $ and $ J $ ($ B $ and $ J $) interact via 
channel $ c_{AJ} $ ($ c_{BJ} $), and only $ A $ and $ J $ ($ B $ and $ J $) know the name $ c_{AJ} $ ($ c_{BJ} $), i.e.
$$\by c_{AJ} = shared \_channel (A, J) ,\\
 c_{BJ} = shared \_channel (B, J). \ey$$

This DP works as follows:
\bi
\i
 $ A $ sends $ J $ the name of  secret channel $ \bar c $ (which at first only $A$ knows) through the channel $ c_{AJ} $,
\i $ J $ sends $ B $ the received channel name $ \bar c $ 
through   channel $ c_{BJ} $,
\i
 $ A $ sends to channel $ \bar c $ the message $ x $,
\i
$ B $ receives from  channel $ \bar c $
the message $x$
 and writes it to the variable $ y $, after which it behaves like SP $ P $.
\ei

SPs $ A $, $ B $ and $ J $ are defined as follows: 
\be{sdfgdsfgdsfgw3rt4ge46h5urjhyr}\!\!\!\!
\by
A=  \alpha_1;
 \alpha_2;
{\bf 0},\;  \alpha_1= c_{AJ}! \bar c,\;
 \alpha_2= \bar c!  x,\\
J=  j_1;
 j_2;{\bf 0},\;
 j_1=
  c_{AJ}?\hat u,\;
 j_2= c_{BJ}! u,\\
B=  \beta_1;
 \beta_2;P,\;
 \beta_1=  c_{BJ}?\hat v,\;
 \beta_2=v?\hat y.
\ey
\ee

A diagram of  $ {\cal P}_3 $ has the following form: 
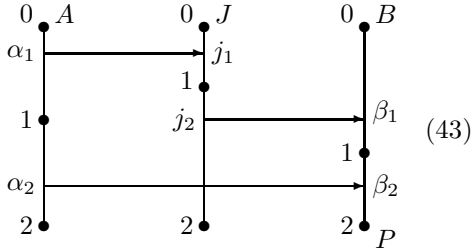
\be{primerdfadsljfk3}\by
\begin{picture}(0,85)
\put(-60,80){\circle*{4}}
\put(0,80){\circle*{4}}
\put(60,80){\circle*{4}}

\put(-60,45){\circle*{4}}
\put(0,57.5){\circle*{4}}
\put(60,32.5){\circle*{4}}

\put(-60,5){\circle*{4}}
\put(0,5){\circle*{4}}
\put(60,5){\circle*{4}}

\put(-64,85){\makebox(0,0)[r]{$0$}}
\put(-64,45){\makebox(0,0)[r]{$1$}}
\put(-64,5){\makebox(0,0)[r]{$2$}}

\put(-4,85){\makebox(0,0)[r]{$0$}}
\put(-4,60){\makebox(0,0)[r]{$1$}}
\put(-4,5){\makebox(0,0)[r]{$2$}}

\put(56,85){\makebox(0,0)[r]{$0$}}
\put(56,32.5){\makebox(0,0)[r]{$1$}}
\put(56,5){\makebox(0,0)[r]{$2$}}

\put(-60,70){\vector(1,0){60}}
\put(0,45){\vector(1,0){60}}
\put(-60,20){\vector(1,0){120}}
\put(60,5){\line(0,1){75}}
\put(0,5){\line(0,1){75}}
\put(-60,5){\line(0,1){75}}

\put(-56,85){\makebox(0,0)[l]{$A$}}
\put(4,85){\makebox(0,0)[l]{$J$}}
\put(64,85){\makebox(0,0)[l]{$B$}}
\put(64,0){\makebox(0,0)[l]{$P$}}
\put(-63,70){\makebox(0,0)[r]{$ \alpha_1$}}
\put(-3,45){\makebox(0,0)[r]{$ j_2$}}
\put(3,70){\makebox(0,0)[l]{$ j_1$}}
\put(-63,20){\makebox(0,0)[r]{$ \alpha_2$}}
\put(63,20){\makebox(0,0)[l]{$ \beta_2$}}
\put(63,48){\makebox(0,0)[l]{$ \beta_1$}}

\end{picture}\ey
\ee

\i
Fourth example is $ {\cal P}_4 = \{A, B, J \} $ (called  {\bf 
Wide-Mouth Frog (WMF)} protocol), which is a model 
of a transmission
from $ A $ to $ B $   EM $ \bar k (x) $ through  open channel $ \circ $ with use of a trusted intermediary $ J $, with whom $ A $ and $ B $ interact through $\circ$. $ A $ creates a secret key $ \bar k $, sends $ J $ this encrypted key for $B$, 
and then sends $ B $  EM $ \bar k (x) $.

It is assumed that $ A $ and $ J $ ($ B $ and $ J $) have a shared secret key $ k_{AJ} $ ($ k_{BJ} $), on which they can encrypt and decrypt messages using a symmetric encryption system, and only $ A $ and $ J $ ($ B $ and $ J $) know $ k_{AJ} $ ($ k_{BJ} $), i.e.
$$\by k_{AJ} = shared \_key (A, J) ,\\
 k_{BJ} = shared \_key (B, J). \ey$$

This DP works as follows.

\bi
\i $ A $ creates a secret key $ \bar k $ (at first only $ A $ knows this key) and sends $J$  EM $ k_{AJ} (\bar k) $  through $ \circ $, then $ A $ sends $B$  EM $ \bar k (x) $,
\i $ J $ receives a message from $ A $, decrypts it,
then encrypts the extracted key $ \bar k $ with the key $ k_{BJ} $, and sends $ B $ the EM
 $ k_{BJ} (\bar k) $,

\i $ B $ extracts the key $ \bar k $ from the received message from $ J $, and then uses this key to extract the message $ x $ from the received message from $ A $, writes it to the variable $ y $, and then behaves like  SP $ P $.
\ei

SPs $ A $, $ B $ and $ J $ are defined as follows: 
\be{sdfgdsfgdsfgw3rt4ge46h5urjhyr1}
\!\!\!\!
\!\!
\!\!
\by
A= \alpha_1;
\alpha_2;
{\bf 0},\; \alpha_1=! k_{AJ}(\bar  k),\;
\alpha_2=! \bar k(x),\\
J= j_1;
j_2;{\bf 0},\; j_1=?  k_{AJ}(\hat u),\;
j_2=!{ k_{BJ}(u)},\\
B= \beta_1;
\beta_2;P,\; \beta_1=?  k_{BJ}(\hat v),\;
\beta_2=? v(\hat y).
\ey
\ee

A diagram of $ {\cal P}_4 $ has the form 
\re{primerdfadsljfk3}.
\en

\subsection{Transition graphs of distributed processes}
\label{fds3w5y6udgdsff}

In this section, we consider DPs consisting of a finite number of SPs. For a visual representation of an execution of such DPs, the concept of {\bf transition graph} of a DP is introduced. An execution of a DP can be presented as a walk in a GP corresponding to this DP.

Below in this section, the symbol $ {\cal P} $ denotes a DP consisting of a finite number of SPs, each of which is different from
$ P_{\dagger} $.

\subsubsection{A concept of a transition graph of a distributed process }
\label{dsfasfdsagfg}

Let $ {\cal P}$ be a DP of the form
$\{P_1, \ldots, P_n \} $.
A {\bf transition graph (TG)} of $ {\cal P} 
$ is a graph $ G_{\cal P} $, where
\bi
\i each node of $G_{\cal P}$ is a list $ at = (at_1, \ldots, at_n)$, where
$\forall \, i = 1, \ldots, n \;\;  at_i \in P_i, $
\i each edge of $G_{\cal P} $ has the form
$$(at_1, \ldots, at_n)
\ral {\! \alpha_{P_i}}
(at'_1, \ldots, at'_n),$$
where $i\in \{1,\dots, n\}$,
$ P_i $ has  edge $ at_i \ra {\alpha} at'_i $, and $ at_{i '} = at'_{i '} $ for $ i' \neq i $.
\ei

A node $Init (G_{\cal P})= (Init (P_1), \ldots, Init (P_n)) $ of 
$ G_{\cal P} $ is said to be
{\bf initial}.

$\forall\, s \in \Sigma_{\cal P} $, the component $ at_s $ of  
$ s $ can be considered as a node of $ G_{\cal P} $.

Let there is given 
an execution of a DP $ {\cal P} $,
and $ s_0, s_1, \ldots, s_n $ be 
a sequence of states generated in this execution.
From the definition of a transition relation
\re{sdfagr33gr356ujyhfgd} it follows that this sequence corresponds to a walk in TG $ G_{\cal P} $, which can be 
considered as a  representation 
of the execution of $ {\cal P} $:
$$ \!\!\!\by
Init (G_{\cal P}) =
at_{s_0} \!\ral{(\alpha_1)_{P_{i_1}}}\!
at_{s_1} 
\!\ral{(\alpha_2)_{P_{i_2}}}\!
\ldots
\!\ral{(\alpha_n)_{P_{i_n}}} \!
at_{s_n} \ey $$

Recall that ${\cal P} _{\dagger}=\{{\cal P}, P_{\dagger}\}$.
The TG $ G_{{\cal P} _{\dagger}} $ can be considered as 
a graph obtained from $ G_{{\cal P}} $
by adding 
 edges $ at \ral {\alpha_{P_{\dagger}}} at $, where $ at \in G_{{\cal P}} $, $ \alpha \in Act $.

A node $ at \in G_{{\cal P}} $ is said to be 
{\bf reachable} if 
for some
$s \in \Sigma_{{\cal P} _{\dagger}}\;\;
 at = at_{s} $.

An edge of $ G_{\cal P} $ is said to be  {\bf realizable}, 
if it lies on a path corresponding to some execution of 
 $ {\cal P} _{\dagger} $.

The following conventions will be used in graphical
representation of TGs:
\bi
\i each node $ at = (at_1, \ldots, at_n) $ of a TG 
is represented by an oval, with a list 
$ at_1 \ldots at_n $ of components of  $ at $
inside this oval,
\i an initial node is represented by a double oval,
\i a black circle on an edge of a TG means that this edge is unrealizable (this unrealizability 
should be justified  by special reasoning),
\i in order to abbreviate notations, a label of an edge $ at \ral {\alpha_{P}} at '$  of a 
TG can be denoted simply by the action $ \alpha $ in this label (without specifying the SP $ P $ performing the action $ \alpha $ on this transition).
\ei

\subsubsection{Examples of transition graphs of distributed processes}
\label{dfgdshghdfghdsfsd4}

In this section we present
TGs for DPs defined
in  \ref{pkergnmf43564654632}. We  use the following convention: if $ A $ is a name of a SP occurred in some of these DPs, 
and $ i $ is a number of a point on a thread corresponding to this SP, then the node of the graph $ A $ corresponding to this point is denoted by  $ A ^ i $.

\bn \i  TGs for  $ {\cal P}_1 $,
$ {\cal P}_2$ 
described by 
\re{primerdfadsljfk1}, 
\re{primerdfadsljfk2},
have the form
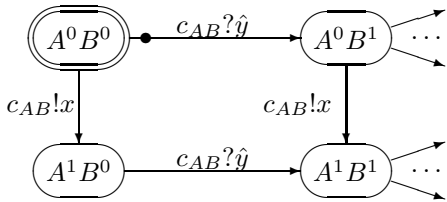
\be{zfdsas3453y46354y325221}
\by
\begin{picture}(150,65)
\put(25,50){\circle*{4}}
\put(0,50){\oval(34,20)}
\put(0,50){\makebox(0,0)[c]{${ A^0B^0}$}}
\put(100,50){\oval(34,20)}
\put(100,50){\makebox(0,0)[c]{${ A^0B^1}$}}
\put(0,0){\oval(34,20)}
\put(0,50){\oval(38,24)}
\put(0,0){\makebox(0,0)[c]{${A^1B^0}$}}
\put(100,0){\oval(34,20)}
\put(100,0){\makebox(0,0)[c]{${A^1B^1}$}}

\put(0,38){\vector(0,-1){28}}
\put(100,40){\vector(0,-1){30}}

\put(17,0){\vector(1,0){66}}
\put(19,50){\vector(1,0){64}}

\put(2,25){\makebox(0,0)[r]{
$c_{AB}! x$
}}
\put(98,25){\makebox(0,0)[r]{
$c_{AB}! x$
}}

\put(50,0){\makebox(0,0)[b]{
$c_{AB}? \hat y$
}}

\put(50,50){\makebox(0,0)[b]{
$c_{AB}? \hat y$
}}

\put(117,54){\vector(3,1){20}}
\put(117,46){\vector(3,-1){20}}
\put(130,50){\makebox(0,0){
$\ldots$
}}

\put(117,4){\vector(3,1){20}}
\put(117,-4){\vector(3,-1){20}}
\put(130,0){\makebox(0,0){
$\ldots$
}}
\end{picture}\ey
\ee

\be{zfdsas3453y46354y3252211}
\by
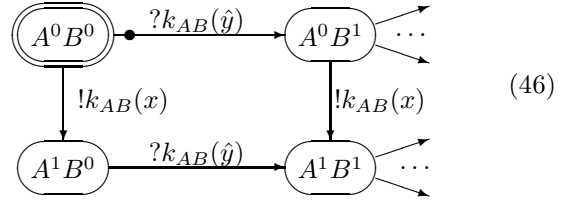
\begin{picture}(150,65)
\put(25,50){\circle*{4}}
\put(0,50){\oval(34,20)}
\put(0,50){\makebox(0,0)[c]{${ A^0B^0}$}}
\put(100,50){\oval(34,20)}
\put(100,50){\makebox(0,0)[c]{${ A^0B^1}$}}

\put(0,0){\oval(34,20)}
\put(0,50){\oval(38,24)}
\put(0,0){\makebox(0,0)[c]{${A^1B^0}$}}
\put(100,0){\oval(34,20)}
\put(100,0){\makebox(0,0)[c]{${A^1B^1}$}}

\put(0,38){\vector(0,-1){28}}
\put(100,40){\vector(0,-1){30}}

\put(17,0){\vector(1,0){66}}
\put(19,50){\vector(1,0){64}}

\put(2,25){\makebox(0,0)[l]{
$! k_{AB}(x)$
}}
\put(98,25){\makebox(0,0)[l]{
$! k_{AB}(x)$
}}

\put(50,0){\makebox(0,0)[b]{
$? k_{AB}(\hat y)$
}}

\put(50,50){\makebox(0,0)[b]{
$? k_{AB}(\hat y)$
}}

\put(117,54){\vector(3,1){20}}
\put(117,46){\vector(3,-1){20}}
\put(130,50){\makebox(0,0){
$\ldots$
}}
\put(117,4){\vector(3,1){20}}
\put(117,-4){\vector(3,-1){20}}
\put(130,0){\makebox(0,0){
$\ldots$}}

\end{picture}

\ey\vspace{4mm}
\ee
 where slant arrows indicate edges of  the TGs outgoing 
 from the corresponding nodes, as well as  parts of the TGs that are reachable after traversing those edges that are not represented in this graphical representation, this convention will also be used in other TG examples.
 

\i A TG for DPs $ {\cal P}_3 $ and $ {\cal P}_4 $, described by 
 \re{primerdfadsljfk3}
 has the form
 \re{zfdsas3453y46354y51}.

\begin{figure*}
{\def\arraystretch{1}
{\small
\be{zfdsas3453y46354y51}
\by
\begin{picture}(150,270)
\put(-75,200){\circle*{4}}
\put(-89,216){\circle*{4}}

\put(117,4){\vector(3,1){20}}
\put(117,-4){\vector(3,-1){20}}
\put(130,0){\makebox(0,0){$\ldots$}}
\put(147,34){\vector(3,1){20}}
\put(147,26){\vector(3,-1){20}}
\put(160,30){\makebox(0,0){$\ldots$}}
\put(177,64){\vector(3,1){20}}
\put(177,56){\vector(3,-1){20}}
\put(190,60){\makebox(0,0){$\ldots$}}

\put(117,104){\vector(3,1){20}}
\put(117,96){\vector(3,-1){20}}
\put(130,100){\makebox(0,0){$\ldots$}}
\put(147,134){\vector(3,1){20}}
\put(147,126){\vector(3,-1){20}}
\put(160,130){\makebox(0,0){$\ldots$}}
\put(177,164){\vector(3,1){20}}
\put(177,156){\vector(3,-1){20}}
\put(190,160){\makebox(0,0){$\ldots$}}

\put(117,204){\vector(3,1){20}}
\put(117,196){\vector(3,-1){20}}
\put(130,200){\makebox(0,0){$\ldots$}}
\put(147,234){\vector(3,1){20}}
\put(147,226){\vector(3,-1){20}}
\put(160,230){\makebox(0,0){$\ldots$}}
\put(177,264){\vector(3,1){20}}
\put(177,256){\vector(3,-1){20}}
\put(190,260){\makebox(0,0){$\ldots$}}

\put(-100,200){\oval(34,20)}
\put(-100,200){\oval(38,24)}
\put(-100,200){\makebox(0,0)[c]{${\scriptstyle A^0J^0B^0}$}}
\put(0,200){\oval(34,20)}
\put(0,200){\makebox(0,0)[c]{${\scriptstyle A^0J^0B^1}$}}
\put(100,200){\oval(34,20)}
\put(100,200){\makebox(0,0)[c]{${\scriptstyle A^0J^0B^2}$}}

\put(-70,230){\oval(34,20)}
\put(-70,230){\makebox(0,0)[c]{${\scriptstyle A^0J^1B^0}$}}
\put(30,230){\oval(34,20)}
\put(30,230){\makebox(0,0)[c]{${\scriptstyle A^0J^1B^1}$}}
\put(130,230){\oval(34,20)}
\put(130,230){\makebox(0,0)[c]{${\scriptstyle A^0J^1B^2}$}}

\put(-40,260){\oval(34,20)}
\put(-40,260){\makebox(0,0)[c]{${\scriptstyle A^0J^2B^0}$}}
\put(60,260){\oval(34,20)}
\put(60,260){\makebox(0,0)[c]{${\scriptstyle A^0J^2B^1}$}}
\put(160,260){\oval(34,20)}
\put(160,260){\makebox(0,0)[c]{${\scriptstyle A^0J^2B^2}$}}

\put(-100,100){\oval(34,20)}
\put(-100,100){\makebox(0,0)[c]{${\scriptstyle A^1J^0B^0}$}}
\put(0,100){\oval(34,20)}
\put(0,100){\makebox(0,0)[c]{${\scriptstyle A^1J^0B^1}$}}
\put(100,100){\oval(34,20)}
\put(100,100){\makebox(0,0)[c]{${\scriptstyle A^1J^0B^2}$}}

\put(-70,130){\oval(34,20)}
\put(-70,130){\makebox(0,0)[c]{${\scriptstyle A^1J^1B^0}$}}
\put(30,130){\oval(34,20)}
\put(30,130){\makebox(0,0)[c]{${\scriptstyle A^1J^1B^1}$}}
\put(130,130){\oval(34,20)}
\put(130,130){\makebox(0,0)[c]{${\scriptstyle A^1J^1B^2}$}}

\put(-40,160){\oval(34,20)}
\put(-40,160){\makebox(0,0)[c]{${\scriptstyle A^1J^2B^0}$}}
\put(60,160){\oval(34,20)}
\put(60,160){\makebox(0,0)[c]{${\scriptstyle A^1J^2B^1}$}}
\put(160,160){\oval(34,20)}
\put(160,160){\makebox(0,0)[c]{${\scriptstyle A^1J^2B^2}$}}

\put(-100,0){\oval(34,20)}
\put(-100,0){\makebox(0,0)[c]{${\scriptstyle A^2J^0B^0}$}}
\put(0,0){\oval(34,20)}
\put(0,0){\makebox(0,0)[c]{${\scriptstyle A^2J^0B^1}$}}
\put(100,0){\oval(34,20)}
\put(100,0){\makebox(0,0)[c]{${\scriptstyle A^2J^0B^2}$}}

\put(-70,30){\oval(34,20)}
\put(-70,30){\makebox(0,0)[c]{${\scriptstyle A^2J^1B^0}$}}
\put(30,30){\oval(34,20)}
\put(30,30){\makebox(0,0)[c]{${\scriptstyle A^2J^1B^1}$}}
\put(130,30){\oval(34,20)}
\put(130,30){\makebox(0,0)[c]{${\scriptstyle A^2J^1B^2}$}}

\put(-40,60){\oval(34,20)}
\put(-40,60){\makebox(0,0)[c]{${\scriptstyle A^2J^2B^0}$}}
\put(60,60){\oval(34,20)}
\put(60,60){\makebox(0,0)[c]{${\scriptstyle A^2J^2B^1}$}}
\put(160,60){\oval(34,20)}
\put(160,60){\makebox(0,0)[c]{${\scriptstyle A^2J^2B^2}$}}

\put(-100,188){\vector(0,-1){78}}
\put(0,190){\vector(0,-1){80}}
\put(100,190){\vector(0,-1){80}}
\put(-100,90){\vector(0,-1){80}}
\put(0,90){\vector(0,-1){80}}
\put(100,90){\vector(0,-1){80}}

\put(-70,220){\vector(0,-1){80}}
\put(30,220){\vector(0,-1){80}}
\put(130,220){\vector(0,-1){80}}
\put(-70,120){\vector(0,-1){80}}
\put(30,120){\vector(0,-1){80}}
\put(130,120){\vector(0,-1){80}}

\put(-40,250){\vector(0,-1){80}}
\put(60,250){\vector(0,-1){80}}
\put(160,250){\vector(0,-1){80}}
\put(-40,150){\vector(0,-1){80}}
\put(60,150){\vector(0,-1){80}}
\put(160,150){\vector(0,-1){80}}

\put(-81,200){\vector(1,0){64}}
\put(17,200){\vector(1,0){66}}
\put(-83,100){\vector(1,0){66}}
\put(17,100){\vector(1,0){66}}
\put(-83,0){\vector(1,0){66}}
\put(17,0){\vector(1,0){66}}

\put(-53,230){\vector(1,0){66}}
\put(47,230){\vector(1,0){66}}
\put(-53,130){\vector(1,0){66}}
\put(47,130){\vector(1,0){66}}
\put(-53,30){\vector(1,0){66}}
\put(47,30){\vector(1,0){66}}

\put(-23,260){\vector(1,0){66}}
\put(77,260){\vector(1,0){66}}
\put(-23,160){\vector(1,0){66}}
\put(77,160){\vector(1,0){66}}
\put(-23,60){\vector(1,0){66}}
\put(77,60){\vector(1,0){66}}

\put(-93,212){\vector(1,1){10}}
\put(10,210){\vector(1,1){10}}
\put(110,210){\vector(1,1){10}}
\put(-60,240){\vector(1,1){10}}
\put(40,240){\vector(1,1){10}}
\put(140,240){\vector(1,1){10}}

\put(-90,110){\vector(1,1){10}}
\put(10,110){\vector(1,1){10}}
\put(110,110){\vector(1,1){10}}
\put(-60,140){\vector(1,1){10}}
\put(40,140){\vector(1,1){10}}
\put(140,140){\vector(1,1){10}}

\put(-90,10){\vector(1,1){10}}
\put(10,10){\vector(1,1){10}}
\put(110,10){\vector(1,1){10}}
\put(-60,40){\vector(1,1){10}}
\put(40,40){\vector(1,1){10}}
\put(140,40){\vector(1,1){10}}

\put(-97,150){\makebox(0,0)[r]{
$\alpha_1$
}}
\put(3,150){\makebox(0,0)[r]{
$\alpha_1$
}}
\put(103,150){\makebox(0,0)[r]{
$\alpha_1$
}}

\put(-67,180){\makebox(0,0)[r]{
$\alpha_1$
}}
\put(33,180){\makebox(0,0)[r]{
$\alpha_1$
}}
\put(133,180){\makebox(0,0)[r]{
$\alpha_1$
}}

\put(-42,210){\makebox(0,0)[l]{
$\alpha_1$
}}
\put(58,210){\makebox(0,0)[l]{
$\alpha_1$
}}
\put(158,210){\makebox(0,0)[l]{
$\alpha_1$
}}

\put(-97,50){\makebox(0,0)[r]{
$\alpha_2$
}}
\put(3,50){\makebox(0,0)[r]{
$\alpha_2$
}}
\put(103,50){\makebox(0,0)[r]{
$\alpha_2$
}}

\put(-67,80){\makebox(0,0)[r]{
$\alpha_2$
}}
\put(33,80){\makebox(0,0)[r]{
$\alpha_2$
}}
\put(133,80){\makebox(0,0)[r]{
$\alpha_2$
}}

\put(-42,110){\makebox(0,0)[l]{
$\alpha_2$
}}
\put(58,110){\makebox(0,0)[l]{
$\alpha_2$
}}
\put(158,110){\makebox(0,0)[l]{
$\alpha_2$
}}

\put(-50,0){\makebox(0,0)[b]{
$\beta_1$
}}
\put(50,0){\makebox(0,0)[b]{
$\beta_2$
}}
\put(-20,30){\makebox(0,0)[b]{
$\beta_1$
}}
\put(70,30){\makebox(0,0)[b]{
$\beta_2$
}}
\put(10,60){\makebox(0,0)[b]{
$\beta_1$
}}
\put(110,60){\makebox(0,0)[b]{
$\beta_2$
}}

\put(-50,100){\makebox(0,0)[b]{
$\beta_1$
}}
\put(50,100){\makebox(0,0)[b]{
$\beta_2$
}}
\put(-20,130){\makebox(0,0)[b]{
$\beta_1$
}}
\put(70,130){\makebox(0,0)[b]{
$\beta_2$
}}
\put(10,160){\makebox(0,0)[b]{
$\beta_1$
}}
\put(110,160){\makebox(0,0)[b]{
$\beta_2$
}}

\put(-50,200){\makebox(0,0)[b]{
$\beta_1$
}}
\put(50,200){\makebox(0,0)[b]{
$\beta_2$
}}
\put(-20,230){\makebox(0,0)[b]{
$\beta_1$
}}
\put(70,230){\makebox(0,0)[b]{
$\beta_2$
}}
\put(10,260){\makebox(0,0)[b]{
$\beta_1$
}}
\put(110,260){\makebox(0,0)[b]{
$\beta_2$
}}


\put(-82,17){\makebox(0,0)[r]{
$j_1$
}}
\put(18,17){\makebox(0,0)[r]{
$j_1$
}}
\put(118,17){\makebox(0,0)[r]{
$j_1$
}}

\put(-82,117){\makebox(0,0)[r]{
$j_1$
}}
\put(18,117){\makebox(0,0)[r]{
$j_1$
}}
\put(118,117){\makebox(0,0)[r]{
$j_1$
}}

\put(-87,221){\makebox(0,0)[r]{
$j_1$
}}
\put(18,217){\makebox(0,0)[r]{
$j_1$
}}
\put(118,217){\makebox(0,0)[r]{
$j_1$
}}

\put(-52,47){\makebox(0,0)[r]{
$j_2$
}}
\put(48,47){\makebox(0,0)[r]{
$j_2$
}}
\put(148,47){\makebox(0,0)[r]{
$j_2$
}}

\put(-52,147){\makebox(0,0)[r]{
$j_2$
}}
\put(48,147){\makebox(0,0)[r]{
$j_2$
}}
\put(148,147){\makebox(0,0)[r]{
$j_2$
}}

\put(-52,247){\makebox(0,0)[r]{
$j_2$
}}
\put(48,247){\makebox(0,0)[r]{
$j_2$
}}
\put(148,247){\makebox(0,0)[r]{
$j_2$
}}

\end{picture}
\ey
\ee
 }
}\end{figure*}
\en

\subsubsection{Validity of formulas at  nodes of transition graphs of distributed processes }

Let $ {\cal P} $ be a DP. 
$\forall\, at \in G_{\cal P} $,
$\forall\, \varphi \in Fm $, the notation $ at \models \varphi $ 
denotes the statement
\be{asfasdfaerg43g35}
\mbox{$ \varphi $ is true at the node $ at $}.
\ee

\re{asfasdfaerg43g35}
is valid 
iff $ \forall \, s \in \Sigma_{{\cal P} _{\dagger}} \; \;
at_s = at \Rightarrow s \models \varphi. $

It is easy to prove that TGs $ G_{{\cal P}_i} $, where $ {\cal P}_i \; (i = 1, \ldots, 4) $ are DPs defined in section \ref{pkergnmf43564654632} , have the following properties: 
\be{dfgdsgw34533y6dfg}
\left.\by
Init(G_{{\cal P}_1})\models 
\{\varphi_1, [{c_{AB}}]=\emptyset\}
\\
Init(G_{{\cal P}_2})\models 
\{\varphi_2, 
k_{AB}^{-1}[\circ]=\emptyset\}\\
Init(G_{{\cal P}_3})\models 
\{\varphi_3, 
[{c_{AJ}}]=\emptyset,\\\hspace{20mm}
[{c_{BJ}}]=\emptyset,
[\bar c]=
\emptyset\}
\\
Init(G_{{\cal P}_4})\models 
\{
\varphi_4,
k_{AJ}^{-1}[\circ]=\emptyset,
\\\hspace{20mm}
k_{BJ}^{-1}[\circ]=\emptyset,
\bar k^{-1}[\circ]=
\emptyset
\}
\ey\right\}\ee
where 
$$\by
\varphi_1=
\{c_{AB}\}\,\bot_{\bf C} \,P_{\dagger} \\
\varphi_2= 
\c{\{k_{AB}\} \,\bot_{\bf K}\,P_{\dagger} \\
k_{AB}^{-1}[P_{\dagger} ]\subseteq
k_{AB}^{-1}[\circ]}\\
\varphi_3=
 \{c_{AJ},c_{BJ},\bar c\}\,\bot_{\bf C} \,P_{\dagger} \\
\varphi_4=
\c{
\{k_{AJ},k_{BJ},\bar k\}\,\bot_{\bf K}\,P_{\dagger} \\
k_{AJ}^{-1}[P_{\dagger} ]\subseteq
k_{AJ}^{-1}[\circ]\\
k_{BJ}^{-1}[P_{\dagger} ]\subseteq
k_{BJ}^{-1}[\circ]\\
\bar k^{-1}[P_{\dagger} ]\subseteq
\bar k^{-1}[\circ]
}\ey
$$

Indeed, for any DP $ {\cal P} $, a state $ s \in \Sigma_{{\cal P} _{\dagger}} $ has the property $ at_s = Init (G_{{\cal P}}) $ if $ s = 0 $, or there is a path from 0 to $ s $ with edge labels
of the form 
$ \alpha_{P_{\dagger}} $, and for each formula $ \psi_i = \{\varphi_i, \ldots \} $ in \re{dfgdsgw34533y6dfg}
\bi
\i the truth of $ \psi_i $ in the initial state of DP $ ({\cal P}_i )_{\dagger} $ follows from the definitions of the concept of an initial state and 
the SP $ P_{\dagger} $, and
\i the truth of $ \psi_i $ in a state $ s $, to which there is a path from 0 with edge labels of the form $ \alpha_{P_{\dagger}} $, is substantiated by the statement
\be{dsfdasfaff3fwtg536h}\by
\forall \, s', s''\in \Sigma_{{\cal P} _{\dagger}}:
s' \ral {\alpha_{P_{\dagger}}} s''
\\
(s' \models \psi_i
\; \; \Rightarrow \; \;
s'' \models \psi_i) \ey\ee
which follows from theorem \ref{doredtuslna32b23322eq} ($ i = 1,3 $), or theorem \ref{doredtuslna32b322eq} ($ i = 2,4 $).
\ei

We will use statements \re{dfgdsgw34533y6dfg} 
in  solutions for the verification problem for some properties of DPs $ {\cal P}_i \; (i = 1, \ldots, 4) $
presented below. 

\section{Verification of cryptographic protocols}
\label{dsfgstrhyytegrdfgs}

Methods of verification of CPs presented in this section are based on a representation of  CPs in the form of DPs. 
To prove  properties of DPs, we use  theorems from  previous section. First method of verification of CPs described below is based on the concept of a TG, and the second method is based on  theorem \ref{doredtuslna32b322e3323q} and is most suitable for verification of authentication CPs.

In this section we 
assume that the symbol $ {\cal P} $ denotes a DP that does not contain   
$ P_{\dagger} $. 

\subsection{Verification method based on transition graphs }
\label{fdges5s5hysrhtgdf}

\subsubsection{Method description }

Some properties of DPs can be expressed by formulas,
 related to  reachable nodes of  corresponding TGs. 
 For example, one of  properties of DP $ {\cal P}_i$, where
 $i = 1, \ldots, 4$, 
 defined in section \ref{pkergnmf43564654632}, 
 has the following form:
\be{dfgdfbdbgfrrtyuu}
\by
\mbox{if a reachable node  $ at = (at_A, at_B) $ or}\\
\mbox{$ (at_A, at_J, at_B) $ of TG $ G_{{\cal P}_i} $ is such that }\\
\mbox{$ at_B = 1 $ ($ i = 1,2 $) or $ at_B = 2 $ ($ i = 3,4 $),}\\\mbox{then $ at \models x = y $.} \ey \ee

This property is called an {\bf integrity} and
has the following meaning:
\bi \i
if an execution of the DP $ ({\cal P}_i) _{\dagger} = \{A, B, P_{\dagger} \} $ or $ \{A, J, B, P_{\dagger} \} $ has reached a state where 
the receiver $ B $ ends a part of its execution, related to 
a receiving of a message from the sender $ A $,
\i then for any opposition of the adversary $ P_{\dagger} $ the transmitted message $ x $ in $ A $ is equal to the value that the variable $ y $ in $ B $ will be assigned. \ei

If a property of a DP $ {\cal P} $ has the form $ at \models \varphi $, where $ at $ is a reachable node of $ G_{{\cal P}} $, 
then a method for verification this property is the following:
\bi \i for each reachable node $ at '$ located on some path from $ Init (G_{{\cal P}}) $ to $ at $, a formula $ \varphi_{at'} $ 
which is true in $ at '$,
is calculated,  and
\i the property $ \varphi_{at} \leq \varphi $ is checked.
\ei

The need to calculate the above formulas for all  nodes on 
 paths from $ Init (G_{{\cal P}}) $ to $ at $ is due to the fact that to calculate $ \varphi_{at} $ you need to know the formulas $ \varphi_{at '} $ for each reachable node $ at' $ from which there is a realizable edge to $ at $, and so on.

The method for calculating  $ \varphi_{at} $ is as follows:
\bi
\i if $ \varphi_{at '} $ is calculated for some reachable node $ at' $ such that there is an edge 
$ at '\ra {\alpha} at $, then the formula $ \alpha (\varphi_{at '}) $ is calculated, the meaning of which is as follows: if $ \varphi_{at'} $ is true
in  state $ s $, and $ s \ra { \alpha} s '$, then $ \alpha (\varphi_{at'}) $ is true in each state to which there is a path from $ s' $ with edge labels of the form $ \alpha_{P_{\dagger}} $,
\i $ \varphi_{at} $ is defined as an analogue of a disjunction of  formulas from the set
$\{ \alpha (\varphi_{at '}) \mid at'\ra{\alpha} at\}.$
\ei

For an initial node $ at ^ 0 = Init (G_{{\cal P}}) $ the corresponding formula $ \varphi_{at ^ 0} $ is assumed to be given. For example, for DPs $ {\cal P}_i \; (i = 1, \ldots, 4) $ defined in  \ref{pkergnmf43564654632}, 
such formulas can be taken 
from the corresponding statements in \re{dfgdsgw34533y6dfg}.

Below we present examples of using this method 
for a verification of
DPs, in which secure channels ($ {\cal P}_1 $ and $ {\cal P}_3 $) or secure keys ($ {\cal P}_2 $ and $ {\cal P}_4 $) are used. 
Before verification of these DPs, 
we  state  theorems used for verification of these DPs.

\subsubsection{Theorems  for verification of 
 processes with secure channels}

In this section, we will use the following notation: 
\be{zfdgdgfdshsdgs}\by
\forall\,E\subseteq Tm,
\forall\,\alpha\in Act, \forall\,c\in Tm_{\bf C}
\\
E_{c,\alpha}=
E\cup\{e\}, \mbox{ if $\alpha=c!e$, and }
\\
E_{c,\alpha}=
E, \mbox{ otherwise.}\ey\ee

\refstepcounter{theorem}
{\bf Theorem \arabic{theorem}\label{dored32tuslna322b322e332323q}}.

Let  $ {\cal P} $ be a DP,
$ E$ be a subseteq of $\l {{\cal P}}_0 $,  
$ s, s'$ be states from $\Sigma_{{\cal P} _{\dagger}} $ 
such that $ s \ral {\! \alpha_{P}} s' $, where $ P \in {\cal P} $, and if $ \alpha = c! e $, then the  implication holds:
\be{12usl453211}
c^s\not\in E\;\;\Rightarrow\;\;
Var(e^s)\cap  E =\emptyset.
\ee

Then  $\forall\,E',E''\subseteq Tm$,
$\forall\,c\in E_{\bf C}$
the following implication holds:
$$\by
s\models 
\c{E\,\bot_{\bf C}\,P_{\dagger} \\
E'\subseteq [c]\subseteq E''}
\Rightarrow\\\Rightarrow
s'\models 
\c{E\,\bot_{\bf C}\,P_{\dagger} \\
E'_{c,\alpha}\subseteq [c]\subseteq E''_{c,\alpha}}\ey$$

{\bf Proof}.

According to \re{szdgdfhdg443e35gy6h5}, the value of the formula $ E \, \bot_{\bf C} \, P_{\dagger} $ in $ s $ depends only on the sets $ [P_{\dagger}] ^ s $ and $ [c] ^ s \; (\forall \, c \in Tm_{\bf C}) $, and
when passing from $ s $ to $ s' $
\bi \i
if $ \alpha = c? e $ or $  \alpha = (e: = e ')$, then 
these sets do not change,
\i if $ \alpha = c! e $, then  only  $ [c] ^ s $ can change by adding the term $ e ^ s $ to it,
\ei
therefore  \re{12usl453211} implies that 
$$ s \models E \, \bot_{\bf C} \, P_{\dagger} \Rightarrow s' \models E \, \bot_{\bf C} \, P_{\dagger} .$$

Implication $$ s \models E '\subseteq [c] \subseteq E''
\Rightarrow s '\models E'_{c, \alpha} \subseteq [c] \subseteq E''_{c, \alpha} $$ follows from definition \re{zfdgdgfdshsdgs}.
$ \blackbox $ \\

\refstepcounter{theorem}
{\bf Theorem \arabic{theorem}\label{dor23224563q}}.

Let there are given
\bi \i DP $ {\cal P} $, subset $ E \subseteq \l {{\cal P}}_0 $,  node $ at \in G_{{\cal P}} $,
\i  set $ \{at_i \ra {\alpha_i} at \mid i \in I \} $ of edges of 
TG $ G_{{\cal P}} $ (with a common end $ at $), and if $ G_{{ \cal P}} $ contains an edge of the form $ at '\ra {\alpha} at $, 
which does not belong this 
set, then  $ at' $ is unreachable,
\i set $ \{\varphi_i \mid i \in I \} $ of
formulas corresponding to the above edges, where $ \forall \, i \in I \; \; at_i \models \varphi_i $, and $ \varphi_i $ consists of the following EFs:
\bi\i
$ E \, \bot_{\bf C} \, P_{\dagger} $,
\i
$ E '_{i, c} \subseteq [c] \subseteq
E''_{i, c} $, where $c \in E_{\bf C}$, and $ E '_{i, c}, E''_{i, c} \subseteq Tm $, 
\i equalities $ e = e '$, where $ e, e' \in Tm $.\ei
\ei

Then $ at \models \varphi $, where $ \varphi $ consists of the  EFs
\bi\i
$ E \, \bot_{\bf C} \, P_{\dagger}$,
\i $ \bigcap_{i \in I} (E '_{i, c})_{c, \alpha_i}
\subseteq [c] \subseteq
\bigcup_{i \in I} (E''_{i, c})_{c, \alpha_i} $, and
\i equalities $ e = e '$,
occurred in each  $ \varphi_i \; (i \in I) $.
\ei

{\bf Proof}.

This theorem is a consequence of  theorems \ref{dored32tuslna322b322e332323q} and \ref{doredtuslna32b23322eq}. $ \blackbox $ \\

\refstepcounter {theorem}
{\bf Theorem \arabic {theorem} \label {dored231e332323323q}}.

Let  states $ s, s'\in \Sigma_{{\cal P} _{\dagger}} $ are such that $ s \ral {\! c? \hat x} s '$.

Then the following implication  holds:
$$ s \models \{[c '] = \{e \}, c = c' \}
\Rightarrow
s' \models
\{x = e \}.$$  

{\bf Proof}.

This theorem follows directly from the definition of an execution of an action of the form $ c? \hat x $ (see section \ref{hfdsklds}). $ \blackbox $

\subsubsection {Reduction of transition graphs}

If  analyzed property of a TG $ G_{{\cal P}} $ is related  with its reachable nodes, then  unreachable nodes and 
 edges associated with these nodes
 can be removed from this TG. 
Such  operation of a removing is called a {\bf reduction} of a TG. 
The resulting graph is called a {\bf reduced} TG, 
and is denoted by the same notation $ G_{{\cal P}} $.
If  the reduced TG has
unreachable nodes,
then this TG can be reduced again, and so on.

Unrealizable edges and unreachable
nodes of TGs can be detected with use of the following theorems (we omit proofs of these theorems). \\

\refstepcounter{theorem}
{\bf Theorem \arabic{theorem}\label{dor23224332563q}}.

Let ${\cal P} $ be a DP, and $ at \in G_{{\cal P}} $.
If 
\bi
\i
$ at \models \{[c] = \emptyset, c = c '\} $, where $ c, c' \in Tm_{\bf C} $, or
\i
$ at \models \{k ^ {- 1} [\circ] = \emptyset, k = k '\} $, where $ k, k' \in Tm_{\bf K} $,
\ei
and there is an edge $at\ra{\alpha}at'$, where
$\alpha$ has the form $c '? e $ or $? k' (e) $ respectively,
then this edge is unrealizable.
$ \blackbox $ \\

\refstepcounter {theorem}
{\bf Theorem \arabic {theorem} \label {doredtuslna322b322e332323q}}.

Let  $ {\cal P} $ be a DP, and $at \in G_{{\cal P}} $, then
\bi
\i if all edges ending in $ at $ are unrealizable, then $ at $ is unreachable, and
\i if $ at $ is unreachable, then all edges starting with $ at $ are unrealizable. $ \blackbox $
\ei

\subsubsection {Verification of   $ {\cal P}_1 $} \label {asfgdsgsrtr66yet3ttf}

Let us apply the theorems stated above to verification  property \re{dfgdfbdbgfrrtyuu} for
DP $ {\cal P}_1 $ described by  diagram \re{primerdfadsljfk1}. TG $ G_{{\cal P}_1} $ has the form \re{zfdsas3453y46354y325221}.

Theorem \ref{dor23224332563q} and  first statement in \re{dfgdsgw34533y6dfg}, which has the form
\be{fdgdsfdg45w6whh}
A ^ 0B ^ 0 \models
\{\varphi_1,
[{c_{AB}}] = \emptyset \} \ee
 justify the unrealizability of the edge marked with a black circle in TG \re{zfdsas3453y46354y325221}. By theorem \ref{doredtuslna322b322e332323q}, this implies 
  unreachability of  node $ A ^ 0B ^ 1 $.

After reduction of TG \re{zfdsas3453y46354y325221} by removing unreachable nodes and associated edges we get the graph \\
\be{zfdsas3453y4346354y325221}
\by
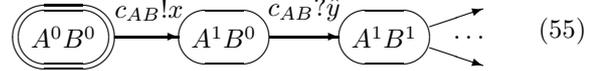
\begin{picture}(50,0)
\put(-60,0){\oval(34,20)}
\put(-60,0){\oval(38,24)}
\put(-60,0){\makebox(0,0)[c]{${ A^0B^0}$}}

\put(0,0){\oval(34,20)}
\put(0,0){\makebox(0,0)[c]{${A^1B^0}$}}

\put(60,0){\oval(34,20)}
\put(60,0){\makebox(0,0)[c]{${A^1B^1}$}}

\put(-41,0){\vector(1,0){24}}
\put(17,0){\vector(1,0){26}}

\put(30,6){\makebox(0,0)[b]{
$c_{AB}?\hat y$
}}
\put(-28,6){\makebox(0,0)[b]{
$c_{AB}!x$
}}
\put(77,4){\vector(3,1){20}}
\put(77,-4){\vector(3,-1){20}}
\put(90,0){\makebox(0,0){
$\ldots$}}
\end{picture}
\ey
\ee\\
In \re{zfdsas3453y4346354y325221} there is only one
node $ (A ^ 1B ^ 1) $ satisfying the condition in \re{dfgdfbdbgfrrtyuu}. Thus, it is required to prove that 
\be{fdgdsfdgfgergs}
 A ^ 1B ^ 1 \models x = y .\ee

From \re{fdgdsfdg45w6whh} by theorem \ref{dor23224563q} we get $$ A ^ 1B ^ 0 \models
\{\varphi_1,
[{c_{AB}}] = \{x \} \}, $$
whence, by theorems \ref{dor23224563q} and \ref{dored231e332323323q},
we get \re{fdgdsfdgfgergs}.

\subsubsection {Verification of  $ {\cal P}_3 $} \label {54y6ewewrteryeuyut}

Now consider the problem of proving   \re{dfgdfbdbgfrrtyuu} for DP $ {\cal P}_3 $, described by   \re{primerdfadsljfk3}, in which actions are defined according to \re{sdfgdsfgdsfgw3rt4ge46h5urjhyr}.
 $G_{{\cal P}_3} $ has the form \re{zfdsas3453y46354y51}.

Theorem \ref{dor23224332563q} and  third statement in \re{dfgdsgw34533y6dfg}, which has the form
\be{1fdgdsfdg45w6whh}
A ^ 0J ^ 0B ^ 0 \models
\{\varphi_3,
[{c_{AJ}}] = \emptyset,
[{c_{BJ}}] = \emptyset,
[{\bar c}] = \emptyset \}
\ee
 justify the unrealizability of the edges marked with black circles in  \re{zfdsas3453y46354y51}.
By theorem \ref{doredtuslna322b322e332323q}, this implies  unreachability of all nodes of the upper tier in TG \re{zfdsas3453y46354y51} except for the node $ A ^ 0J ^ 0B ^ 0 $.
 
After reduction of TG \re{zfdsas3453y46354y51} by removing the unreachable top tier nodes and associated edges we get the reduced TG
\re{zfdsas3453y46322354y51}.
\begin{figure*}
 {\def\arraystretch{1}
{\small
\be{zfdsas3453y46322354y51}
\by
\begin{picture}(120,170)

\put(-75,0){\circle*{4}}
\put(-75,100){\circle*{4}}
\put(-45,30){\circle*{4}}
\put(-45,130){\circle*{4}}

\put(117,4){\vector(3,1){20}}
\put(117,-4){\vector(3,-1){20}}
\put(130,0){\makebox(0,0){$\ldots$}}
\put(147,34){\vector(3,1){20}}
\put(147,26){\vector(3,-1){20}}
\put(160,30){\makebox(0,0){$\ldots$}}
\put(177,64){\vector(3,1){20}}
\put(177,56){\vector(3,-1){20}}
\put(190,60){\makebox(0,0){$\ldots$}}

\put(117,104){\vector(3,1){20}}
\put(117,96){\vector(3,-1){20}}
\put(130,100){\makebox(0,0){$\ldots$}}
\put(147,134){\vector(3,1){20}}
\put(147,126){\vector(3,-1){20}}
\put(160,130){\makebox(0,0){$\ldots$}}
\put(177,164){\vector(3,1){20}}
\put(177,156){\vector(3,-1){20}}
\put(190,160){\makebox(0,0){$\ldots$}}

\put(-100,100){\oval(34,20)}
\put(-100,100){\makebox(0,0)[c]{${\scriptstyle A^1J^0B^0}$}}
\put(0,100){\oval(34,20)}
\put(0,100){\makebox(0,0)[c]{${\scriptstyle A^1J^0B^1}$}}
\put(100,100){\oval(34,20)}
\put(100,100){\makebox(0,0)[c]{${\scriptstyle A^1J^0B^2}$}}

\put(-70,130){\oval(34,20)}
\put(-70,130){\makebox(0,0)[c]{${\scriptstyle A^1J^1B^0}$}}
\put(30,130){\oval(34,20)}
\put(30,130){\makebox(0,0)[c]{${\scriptstyle A^1J^1B^1}$}}
\put(130,130){\oval(34,20)}
\put(130,130){\makebox(0,0)[c]{${\scriptstyle A^1J^1B^2}$}}

\put(-100,160){\oval(34,20)}
\put(-100,160){\oval(38,24)}
\put(-100,160){\makebox(0,0)[c]{${\scriptstyle A^0J^0B^0}$}}

\put(-40,160){\oval(34,20)}
\put(-40,160){\makebox(0,0)[c]{${\scriptstyle A^1J^2B^0}$}}
\put(60,160){\oval(34,20)}
\put(60,160){\makebox(0,0)[c]{${\scriptstyle A^1J^2B^1}$}}
\put(160,160){\oval(34,20)}
\put(160,160){\makebox(0,0)[c]{${\scriptstyle A^1J^2B^2}$}}

\put(-100,0){\oval(34,20)}
\put(-100,0){\makebox(0,0)[c]{${\scriptstyle A^2J^0B^0}$}}
\put(0,0){\oval(34,20)}
\put(0,0){\makebox(0,0)[c]{${\scriptstyle A^2J^0B^1}$}}
\put(100,0){\oval(34,20)}
\put(100,0){\makebox(0,0)[c]{${\scriptstyle A^2J^0B^2}$}}

\put(-70,30){\oval(34,20)}
\put(-70,30){\makebox(0,0)[c]{${\scriptstyle A^2J^1B^0}$}}
\put(30,30){\oval(34,20)}
\put(30,30){\makebox(0,0)[c]{${\scriptstyle A^2J^1B^1}$}}
\put(130,30){\oval(34,20)}
\put(130,30){\makebox(0,0)[c]{${\scriptstyle A^2J^1B^2}$}}

\put(-40,60){\oval(34,20)}
\put(-40,60){\makebox(0,0)[c]{${\scriptstyle A^2J^2B^0}$}}
\put(60,60){\oval(34,20)}
\put(60,60){\makebox(0,0)[c]{${\scriptstyle A^2J^2B^1}$}}
\put(160,60){\oval(34,20)}
\put(160,60){\makebox(0,0)[c]{${\scriptstyle A^2J^2B^2}$}}

\put(-100,148){\vector(0,-1){38}}

\put(-100,90){\vector(0,-1){80}}
\put(0,90){\vector(0,-1){80}}
\put(100,90){\vector(0,-1){80}}

\put(-70,120){\vector(0,-1){80}}
\put(30,120){\vector(0,-1){80}}
\put(130,120){\vector(0,-1){80}}

\put(-40,150){\vector(0,-1){80}}
\put(60,150){\vector(0,-1){80}}
\put(160,150){\vector(0,-1){80}}

\put(-83,100){\vector(1,0){66}}
\put(17,100){\vector(1,0){66}}
\put(-83,0){\vector(1,0){66}}
\put(17,0){\vector(1,0){66}}

\put(-53,130){\vector(1,0){66}}
\put(47,130){\vector(1,0){66}}
\put(-53,30){\vector(1,0){66}}
\put(47,30){\vector(1,0){66}}

\put(-23,160){\vector(1,0){66}}
\put(77,160){\vector(1,0){66}}
\put(-23,60){\vector(1,0){66}}
\put(77,60){\vector(1,0){66}}

\put(-90,110){\vector(1,1){10}}
\put(10,110){\vector(1,1){10}}
\put(110,110){\vector(1,1){10}}
\put(-60,140){\vector(1,1){10}}
\put(40,140){\vector(1,1){10}}
\put(140,140){\vector(1,1){10}}

\put(-90,10){\vector(1,1){10}}
\put(10,10){\vector(1,1){10}}
\put(110,10){\vector(1,1){10}}
\put(-60,40){\vector(1,1){10}}
\put(40,40){\vector(1,1){10}}
\put(140,40){\vector(1,1){10}}

\put(-97,130){\makebox(0,0)[r]{
$\alpha_1$
}}
\put(3,150){\makebox(0,0)[r]{
$\alpha_1$
}}
\put(103,150){\makebox(0,0)[r]{
$\alpha_1$
}}

\put(-97,50){\makebox(0,0)[r]{
$\alpha_2$
}}
\put(3,50){\makebox(0,0)[r]{
$\alpha_2$
}}
\put(103,50){\makebox(0,0)[r]{
$\alpha_2$
}}

\put(-67,80){\makebox(0,0)[r]{
$\alpha_2$
}}
\put(33,80){\makebox(0,0)[r]{
$\alpha_2$
}}
\put(133,80){\makebox(0,0)[r]{
$\alpha_2$
}}

\put(-42,110){\makebox(0,0)[l]{
$\alpha_2$
}}
\put(58,110){\makebox(0,0)[l]{
$\alpha_2$
}}
\put(158,110){\makebox(0,0)[l]{
$\alpha_2$
}}


\put(-50,0){\makebox(0,0)[b]{
$\beta_1$
}}
\put(50,0){\makebox(0,0)[b]{
$\beta_2$
}}
\put(-20,30){\makebox(0,0)[b]{
$\beta_1$
}}
\put(70,30){\makebox(0,0)[b]{
$\beta_2$
}}
\put(10,60){\makebox(0,0)[b]{
$\beta_1$
}}
\put(110,60){\makebox(0,0)[b]{
$\beta_2$
}}

\put(-50,100){\makebox(0,0)[b]{
$\beta_1$
}}
\put(50,100){\makebox(0,0)[b]{
$\beta_2$
}}
\put(-20,130){\makebox(0,0)[b]{
$\beta_1$
}}
\put(70,130){\makebox(0,0)[b]{
$\beta_2$
}}
\put(10,160){\makebox(0,0)[b]{
$\beta_1$
}}
\put(110,160){\makebox(0,0)[b]{
$\beta_2$
}}


\put(-82,17){\makebox(0,0)[r]{
$j_1$
}}
\put(18,17){\makebox(0,0)[r]{
$j_1$
}}
\put(118,17){\makebox(0,0)[r]{
$j_1$
}}

\put(-82,117){\makebox(0,0)[r]{
$j_1$
}}
\put(18,117){\makebox(0,0)[r]{
$j_1$
}}
\put(118,117){\makebox(0,0)[r]{
$j_1$
}}

\put(-52,47){\makebox(0,0)[r]{
$j_2$
}}
\put(48,47){\makebox(0,0)[r]{
$j_2$
}}
\put(148,47){\makebox(0,0)[r]{
$j_2$
}}

\put(-52,147){\makebox(0,0)[r]{
$j_2$
}}
\put(48,147){\makebox(0,0)[r]{
$j_2$
}}
\put(148,147){\makebox(0,0)[r]{
$j_2$
}}

\end{picture}
\ey
\ee
 }
}\end{figure*}

Below we give a list of statements, each of which follows from the previous ones (the first follows from \re{1fdgdsfdg45w6whh}) according to  theorems 
 \ref{dor23224563q} and 
\ref{dored231e332323323q}:
\be{31fdgdsfdg45w6whh}
\by
A^1J^0B^0\models
\{\varphi_3,
[{c_{AJ}}]=\{\bar c\},
[{c_{BJ}}]=\emptyset,\\\hspace{20mm}
[{\bar c}]=\emptyset\}
\\
A^2J^0B^0\models
\{\varphi_3,
[{c_{AJ}}]=\{\bar c\},
[{c_{BJ}}]=\emptyset,\\\hspace{20mm}
[{\bar c}]=\{x\}\}\\
A^1J^1B^0\models
\{\varphi_3,
[{c_{AJ}}]=\{\bar c\},
[{c_{BJ}}]=\emptyset, \\\hspace{20mm}
[{\bar c}]=
\emptyset,
u=\bar c\}\\
A^2J^1B^0\models
\{\varphi_3,[{c_{AJ}}]=\{\bar c\},
[{c_{BJ}}]=\emptyset,\\\hspace{20mm}
[{\bar c}]=\{x\},
u=\bar c\}\ey
\ee

By theorem \ref{dor23224332563q}, these statements imply  unrealizability of the edges marked with black circles in TG \re{31fdgdsfdg45w6whh}.
After removing these edges and the corresponding unreachable nodes (using theorem \ref{doredtuslna322b322e332323q}), we get the reduced TG 
\re{zfd334sas3453y46322354y51}.
\begin{figure*}
{\def\arraystretch{1}
{\small
\be{zfd334sas3453y46322354y51}
\by
\begin{picture}(120,120)

\put(85,110){\circle*{4}}

\put(177,64){\vector(3,1){20}}
\put(177,56){\vector(3,-1){20}}
\put(190,60){\makebox(0,0){$\ldots$}}

\put(177,114){\vector(3,1){20}}
\put(177,106){\vector(3,-1){20}}
\put(190,110){\makebox(0,0){$\ldots$}}

\put(-100,50){\oval(34,20)}
\put(-100,50){\makebox(0,0)[c]{${\scriptstyle A^1J^0B^0}$}}

\put(-70,80){\oval(34,20)}
\put(-70,80){\makebox(0,0)[c]{${\scriptstyle A^1J^1B^0}$}}

\put(-100,110){\oval(34,20)}
\put(-100,110){\oval(38,24)}
\put(-100,110){\makebox(0,0)[c]{${\scriptstyle A^0J^0B^0}$}}

\put(-40,110){\oval(34,20)}
\put(-40,110){\makebox(0,0)[c]{${\scriptstyle A^1J^2B^0}$}}
\put(60,110){\oval(34,20)}
\put(60,110){\makebox(0,0)[c]{${\scriptstyle A^1J^2B^1}$}}
\put(160,110){\oval(34,20)}
\put(160,110){\makebox(0,0)[c]{${\scriptstyle A^1J^2B^2}$}}

\put(-100,0){\oval(34,20)}
\put(-100,0){\makebox(0,0)[c]{${\scriptstyle A^2J^0B^0}$}}

\put(-70,30){\oval(34,20)}
\put(-70,30){\makebox(0,0)[c]{${\scriptstyle A^2J^1B^0}$}}

\put(-40,60){\oval(34,20)}
\put(-40,60){\makebox(0,0)[c]{${\scriptstyle A^2J^2B^0}$}}
\put(60,60){\oval(34,20)}
\put(60,60){\makebox(0,0)[c]{${\scriptstyle A^2J^2B^1}$}}
\put(160,60){\oval(34,20)}
\put(160,60){\makebox(0,0)[c]{${\scriptstyle A^2J^2B^2}$}}

\put(-100,98){\vector(0,-1){38}}

\put(-100,40){\vector(0,-1){30}}

\put(-70,70){\vector(0,-1){30}}

\put(-40,100){\vector(0,-1){30}}
\put(60,100){\vector(0,-1){30}}
\put(160,100){\vector(0,-1){30}}

\put(-23,110){\vector(1,0){66}}
\put(77,110){\vector(1,0){66}}
\put(-23,60){\vector(1,0){66}}
\put(77,60){\vector(1,0){66}}

\put(-90,60){\vector(1,1){10}}
\put(-60,90){\vector(1,1){10}}

\put(-90,10){\vector(1,1){10}}
\put(-60,40){\vector(1,1){10}}

\put(-97,80){\makebox(0,0)[r]{
$\alpha_1$
}}
\put(-97,25){\makebox(0,0)[r]{
$\alpha_2$
}}
\put(-67,55){\makebox(0,0)[r]{
$\alpha_2$
}}

\put(-42,85){\makebox(0,0)[l]{
$\alpha_2$
}}
\put(58,85){\makebox(0,0)[l]{
$\alpha_2$
}}
\put(158,85){\makebox(0,0)[l]{
$\alpha_2$
}}


\put(10,60){\makebox(0,0)[b]{
$\beta_1$
}}
\put(110,60){\makebox(0,0)[b]{
$\beta_2$
}}

\put(10,110){\makebox(0,0)[b]{
$\beta_1$
}}
\put(110,110){\makebox(0,0)[b]{
$\beta_2$
}}


\put(-82,17){\makebox(0,0)[r]{
$j_1$
}}

\put(-82,67){\makebox(0,0)[r]{
$j_1$
}}

\put(-52,47){\makebox(0,0)[r]{
$j_2$
}}

\put(-52,97){\makebox(0,0)[r]{
$j_2$
}}

\end{picture}
\ey
\ee
 }
}\end{figure*}

From the last statement in \re{31fdgdsfdg45w6whh} using  theorems \ref{dor23224563q} and 
\ref{dored231e332323323q} we get:
\be{sfgfdsgds3w4563y34t4w}
\by
A^1J^2B^0\models 
\{\varphi_3,[{c_{AJ}}]=\{\bar c\},
[{c_{BJ}}]=\{u\},\\\hspace{20mm}
[{\bar c}]=
\emptyset,u=\bar c
\}\\
A^1J^2B^1\models
\{\varphi_3,[{c_{AJ}}]=\{\bar c\},
[{c_{BJ}}]=\{u\},
\\\hspace{20mm}
[{\bar c}]=
\emptyset,
u=\bar c,
v=u\}
\ey\ee

By theorem \ref{dor23224332563q}, from the last statement in \re{sfgfdsgds3w4563y34t4w} it follows that the edge 
in TG \re{zfd334sas3453y46322354y51}
marked with a black circle is unrealizable. Removing this edge and the corresponding unreachable nodes (to find which we use  theorem \ref{doredtuslna322b322e332323q}), we get the reduced TG 
\re{sdfgadsfsr4rwt35yt4}.
\begin{figure*}
{\def\arraystretch{1}
{\small
\be{sdfgadsfsr4rwt35yt4}
\by
\begin{picture}(50,130)

\put(147,4){\vector(3,1){20}}
\put(147,-4){\vector(3,-1){20}}
\put(160,0){\makebox(0,0){
$\ldots$}}

\put(-130,120){\oval(34,20)}
\put(-130,120){\oval(38,24)}
\put(-130,120){\makebox(0,0)[c]{${\scriptstyle A^0J^0B^0}$}}

\put(-50,120){\oval(34,20)}
\put(-50,120){\makebox(0,0)[c]{${\scriptstyle A^1J^0B^0}$}}

\put(-50,80){\oval(34,20)}
\put(-50,80){\makebox(0,0)[c]{${\scriptstyle A^1J^1B^0}$}}

\put(-50,40){\oval(34,20)}
\put(-50,40){\makebox(0,0)[c]{${\scriptstyle A^1J^2B^0}$}}

\put(50,120){\oval(34,20)}
\put(50,120){\makebox(0,0)[c]{${\scriptstyle A^2J^0B^0}$}}

\put(50,80){\oval(34,20)}
\put(50,80){\makebox(0,0)[c]{${\scriptstyle A^2J^1B^0}$}}

\put(50,40){\oval(34,20)}
\put(50,40){\makebox(0,0)[c]{${\scriptstyle A^2J^2B^0}$}}

\put(-50,100){\makebox(0,0)[r]{
$j_1$
}}

\put(50,100){\makebox(0,0)[l]{
$j_1$
}}

\put(-50,60){\makebox(0,0)[r]{
$j_2$
}}

\put(50,60){\makebox(0,0)[l]{
$j_2$
}}

\put(-90,122){\makebox(0,0)[b]{
$\alpha_1$
}}

\put(0,123){\makebox(0,0)[b]{
$\alpha_2$
}}

\put(0,83){\makebox(0,0)[b]{
$\alpha_2$
}}

\put(0,43){\makebox(0,0)[b]{
$\alpha_2$
}}

\put(-50,110){\vector(0,-1){20}}
\put(50,110){\vector(0,-1){20}}

\put(-50,70){\vector(0,-1){20}}
\put(50,70){\vector(0,-1){20}}

\put(-111,120){\vector(1,0){44}}

\put(-33,120){\vector(1,0){66}}
\put(-33,80){\vector(1,0){66}}
\put(-33,40){\vector(1,0){66}}

\put(-50,0){\oval(34,20)}
\put(-50,0){\makebox(0,0)[c]{${\scriptstyle A^1J^2B^1}$}}

\put(50,0){\oval(34,20)}
\put(50,0){\makebox(0,0)[c]{${\scriptstyle A^2J^2B^1}$}}

\put(0,3){\makebox(0,0)[b]{
$\alpha_2$
}}

\put(-33,0){\vector(1,0){66}}

\put(130,0){\oval(34,20)}
\put(130,0){\makebox(0,0)[c]{${\scriptstyle A^2J^2B^2}$}}

\put(-50,30){\vector(0,-1){20}}
\put(50,30){\vector(0,-1){20}}
\put(67,0){\vector(1,0){46}}

\put(-50,20){\makebox(0,0)[r]{
$\beta_1$
}}

\put(50,20){\makebox(0,0)[l]{
$\beta_1$
}}
\put(90,3){\makebox(0,0)[b]{
$\beta_2$
}}

\end{picture}
\ey
\ee
 }
}\end{figure*}

Applying  theorems \ref{dor23224563q} and \ref{dored231e332323323q}, we calculate the formulas corresponding to the remaining nodes: 
\be{sdfasfa43qt53tt}
\by
A^2J^2B^0\models
\{\varphi_3,
[{c_{AJ}}]=\{\bar c\},
[{c_{BJ}}]=\{u\},\\\hspace{20mm}
[{\bar c}]=\{x\},
u=\bar c\}
\\
A^2J^2B^1\models
\{\varphi_3,
[{c_{AJ}}]=\{\bar c\},
[{c_{BJ}}]=\{u\},\\\hspace{20mm}
[{\bar c}]=
\{x\},
u=\bar c,
v=u\}
\\
A^2J^2B^2\models
\{
x=y
\}
\ey\ee

Since \bi \i in TG \re{sdfgadsfsr4rwt35yt4} node $ A ^ 2J ^ 2B ^ 2 $ is the only node that satisfies the condition in \re{dfgdfbdbgfrrtyuu}, and \i according to the last statement in \re{sdfasfa43qt53tt},  this node satisfies
the statement in \re{dfgdfbdbgfrrtyuu}, \ei
then the problem of proving  property \re{dfgdfbdbgfrrtyuu} for DP $ {\cal P}_3 $ is solved. 

\subsubsection{Theorems 
for verification of distributed processes
with secure keys}

In this section, we will use the following notation: 
\be{sdfasgads43qg5}
\by
\forall\,E\subseteq Tm,
\forall\,\alpha\in Act,\forall\,k\in Tm_{\bf K}\\
E_{k,\alpha}=
E\cup\{e\}, \mbox{ if $\alpha=!k(e)$, and }
\\E_{k,\alpha}= E, \mbox{ otherwise.}
\ey\ee

\refstepcounter{theorem}
{\bf Theorem \arabic{theorem}\label{do1red344322tus323q}}.

Let $ {\cal P} $ be a DP, 
$ Var ({\cal P})_{\bf C} = \{\circ \} $, 
$P\in {\cal P}$,
$ E$ be a subset of $\l {{\cal P}}_0 $, 
$ s, s '$ be states from 
$\Sigma_{{\cal P} _{\dagger}} $, such that $ s \ral {\! \alpha_{P}} s' $,
 and if $ \alpha =! e $, then 
\be{2223usl45334321}
\forall\,x\in E_{\bf X}\;\;
x\,\bot_{{\bf K}, E}\,e^s.
\ee
Then $ \forall \, E ', E''\subseteq Tm $, $ \forall \, k \in E_{\bf K}$ 
the following implication holds:
$$\by
s\models \c{
E\,\bot_{\bf K}\,P_{\dagger} ,k^{-1}[P_{\dagger} ]\subseteq  k^{-1}[\circ]
\\
E'\subseteq  k^{-1}[\circ]
\subseteq E''}
\Rightarrow\\\Rightarrow
s'\models \c{
E\,\bot_{\bf K}\,P_{\dagger} , k^{-1}[P_{\dagger} ]\subseteq  k^{-1}[\circ]\\
E'_{k,\alpha}\subseteq  k^{-1}[\circ]
\subseteq E''_{k,\alpha}}\ey$$

{\bf Proof}.

Values of  formulas $ E \, \bot_{\bf K} \, P_{\dagger} $ and $ k ^ {- 1} [P_{\dagger}] \subseteq k ^ {- 1} [\circ] $ in $ s $ depend only on sets $ [P_{\dagger}] ^ s $ and $ [\circ]_s $ (for $ E \, \bot_{\bf K} \, P_{\dagger} $ this follows from \re{f443dgh1sdfsg4341}), and
\bi \i
if $ \alpha =? e $ or $\alpha = ( e: = e ')$, then 
$ [P_{\dagger}] ^ s $ and $ [\circ]_s $ do not change
when passing from $ s $ to $ s' $ ,
and
\i if $ \alpha =! e $, then 
only  $ [\circ]_s $ changes by adding $ e ^ s $
when passing from $ s $ to $ s' $, 
\ei
therefore, \re{2223usl45334321} implies the implication
$$\by
 s \models \{E \, \bot_{\bf K} \, P_{\dagger},
k ^ {- 1} [P_{\dagger}] \subseteq k ^ {- 1} [\circ] \} \Rightarrow\\\Rightarrow
s' \models \{E \, \bot_{\bf K} \, P_{\dagger},
k ^ {- 1} [P_{\dagger}] \subseteq k ^ {- 1} [\circ] \}.\ey $$

Implication $$\by
 s \models E '\subseteq k ^ {- 1} [\circ] \subseteq E''
\Rightarrow\\\Rightarrow s '\models E'_{k, \alpha} \subseteq k ^ {- 1} [\circ] \subseteq
E''_{k, \alpha}\ey $$ follows from \re{sdfasgads43qg5}.
$ \blackbox $ \\

Moreover, an analog of theorem \ref{dor23224563q} holds, 
with the replacement
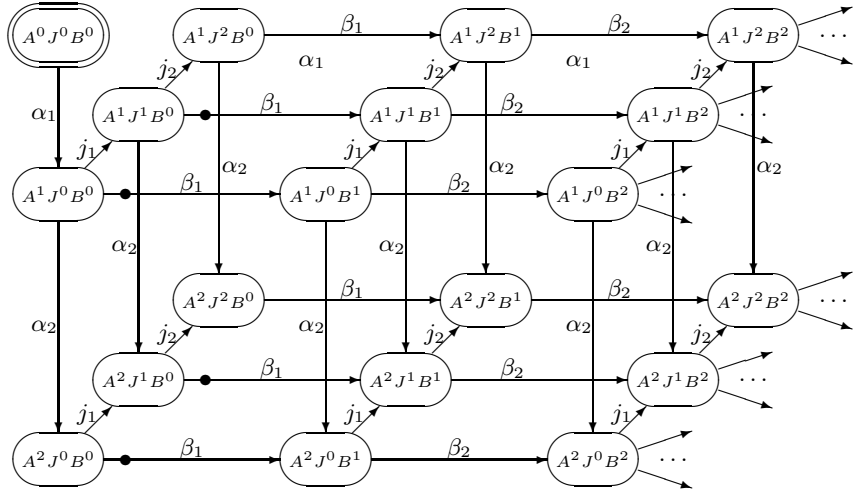
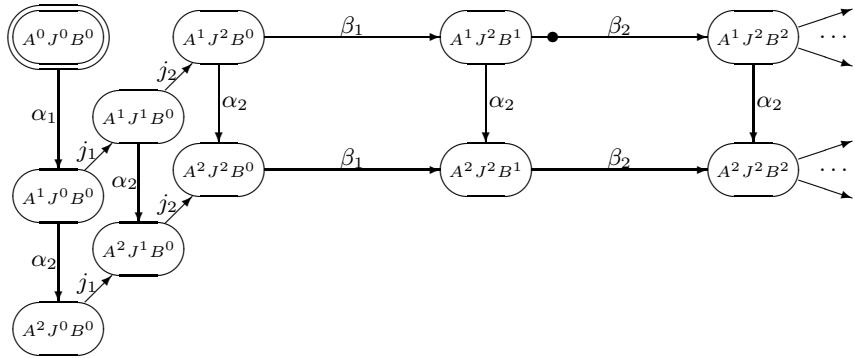
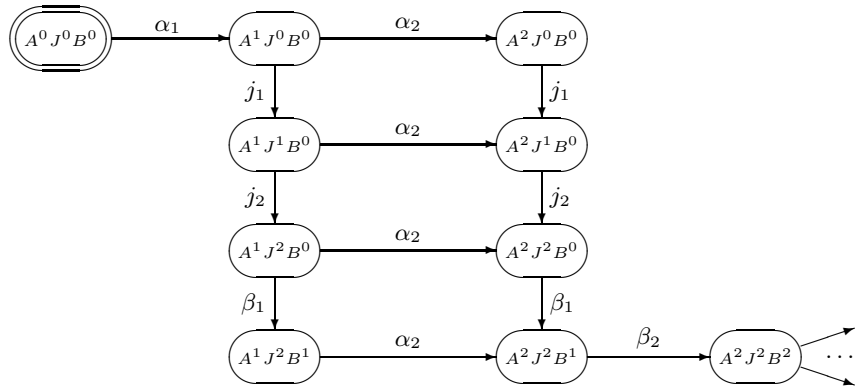
\bi \i
$ E \, \bot_{\bf C} \, P_{\dagger} $
on 
$ \{E \, \bot_{\bf K} \, P_{\dagger},
k ^ {- 1} [P_{\dagger}] \subseteq k ^ {- 1} [\circ] \} $, 
\i $ [c] $ on $ k ^ {- 1} [\circ] $,
\i $ E_{i, c} $ on $ E_{i, k} $,
$ (E_{i, c})_{c, \alpha_i} $ on
$ (E_{i, k})_{k, \alpha_i} $,
\ei
and an analog of 
theorem \ref{dored231e332323323q} holds, 
with the replacement
$ c? \hat x $ on $? k (\hat x) $, 
$ [c '] $ on $ (k') ^ {- 1} [\circ] $,  
$ c = c '$ on $ k = k' $. 

\subsubsection{Verification of 
${\cal P}_2$}
\label{dfgfrw54h6yhytg}

A proof of  property \re{dfgdfbdbgfrrtyuu} for DP $ {\cal P}_2 $ described by diagram \re{primerdfadsljfk2} is carried out similarly to the proof of this property for DP $ {\cal P}_1 $ in section
\ref{asfgdsgsrtr66yet3ttf}. TG $ G_{{\cal P}_2} $ 
has the form \re{zfdsas3453y46354y3252211}.
An unrealizability of the edge marked with a black circle in 
$ G_{{\cal P}_2} $
is substantiated with use of theorem
\ref{dor23224332563q} and second statement in
\re{dfgdsgw34533y6dfg}.
 By theorem \ref{doredtuslna322b322e332323q}, 
 we get an unreachability of the node $ A ^ 0B ^ 1 $. After reduction of TG  $G_{{\cal P}_2} $ we get the TG 
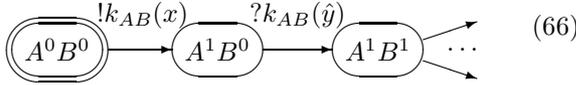
\be{zfdsas3453y43463522354y325221}
\by
\begin{picture}(50,20)
\put(-60,0){\oval(34,20)}
\put(-60,0){\oval(38,24)}
\put(-60,0){\makebox(0,0)[c]{${ A^0B^0}$}}

\put(0,0){\oval(34,20)}
\put(0,0){\makebox(0,0)[c]{${A^1B^0}$}}

\put(60,0){\oval(34,20)}
\put(60,0){\makebox(0,0)[c]{${A^1B^1}$}}

\put(-41,0){\vector(1,0){24}}
\put(17,0){\vector(1,0){26}}

\put(30,8){\makebox(0,0)[b]{
$?  k_{AB}(\hat y)$
}}
\put(-28,8){\makebox(0,0)[b]{
$! k_{AB}(x)$
}}
\put(77,4){\vector(3,1){20}}
\put(77,-4){\vector(3,-1){20}}
\put(90,0){\makebox(0,0){
$\ldots$}}

\end{picture}
\ey
\ee\\
In \re{zfdsas3453y43463522354y325221} there is a single node $ (A ^ 1B ^ 1) $ that satisfies the condition in \re{dfgdfbdbgfrrtyuu}. Thus, it is required to prove that $ A ^ 1B ^ 1 \models x = y $. This property follows from the statement $ A ^ 1B ^ 0 \models \{\varphi_2, k_{AB} ^ {- 1} [{\circ}] = \{x \} \}. $

\subsubsection {Verification of 
 $ {\cal P}_4 $} \label {dfgfrw54h6yhytg1}

 Proof of property \re{dfgdfbdbgfrrtyuu} for DP $ {\cal P}_4 $ described by diagram \re{primerdfadsljfk3}, where actions $ \alpha_i, \beta_i, j_i \; (i = 1,2) $ 
 are defined according to 
 \re{sdfgdsfgdsfgw3rt4ge46h5urjhyr1}, is carried out similarly to the proof of this property for DP $ {\cal P}_3 $ in section
 \ref{54y6ewewrteryeuyut}. TG $ G_{{\cal P}_4} $ 
 has the form \re{zfdsas3453y46354y51}. 
After reduction of TG 
 \re{zfdsas3453y46354y51} we get the same TGs
 \re{zfdsas3453y46322354y51}, \re{zfd334sas3453y46322354y51}, \re{sdfgadsfsr4rwt35yt4}, 
as in the case of verification of DP ${\cal P}_3$.
We will not describe in detail a solution of the verification problem
for DP $ {\cal P}_4 $, we will only present  statements associated with  nodes of TG \re{sdfgadsfsr4rwt35yt4} for this case. 
$${\def\arraystretch{1.2}
\by
A^1J^0B^0\models
\{\varphi_4,
k_{AJ}^{-1}[\circ]=\{\bar k\},
k_{BJ}^{-1}[\circ]=
\emptyset,\\\hspace{20mm}
\bar k^{-1}[\circ]=
\emptyset
\}
\\
A^2J^0B^0\models
\{\varphi_4,
k_{AJ}^{-1}[\circ]=\{\bar k\},
k_{BJ}^{-1}[\circ]=\emptyset,\\\hspace{20mm}
\bar k^{-1}[\circ]=\{x\}
\}
\\
A^1J^1B^0\models
\{\varphi_4,
k_{AJ}^{-1}[\circ]=\{\bar k\},
k_{BJ}^{-1}[\circ]=\emptyset,\\\hspace{20mm}
\bar k^{-1}[\circ]=\emptyset,
u=\bar k\}
\\
A^2J^1B^0\models
\{\varphi_4,
k_{AJ}^{-1}[\circ]=\{\bar k\},
k_{BJ}^{-1}[\circ]=\emptyset,\\\hspace{20mm}
\bar k^{-1}[\circ]=\{x\},
u=\bar k\}
\\
A^1J^2B^0\models
\{\varphi_4,
k_{AJ}^{-1}[\circ]=\{\bar k\},
k_{BJ}^{-1}[\circ]=\{u\},\\\hspace{20mm}
\bar k^{-1}[\circ]=
\emptyset,
u=\bar k
\}
\\
A^1J^2B^1\models
\{\varphi_4,
k_{AJ}^{-1}[\circ]=\{\bar k\},
k_{BJ}^{-1}[\circ]=\{u\},\\\hspace{20mm}
\bar k^{-1}[\circ]=
\emptyset,
u=\bar k,
v=u\}
\\
A^2J^2B^0\models
\{\varphi_4,
k_{AJ}^{-1}[\circ]=\{\bar k\},
k_{BJ}^{-1}[\circ]=\{u\},\\\hspace{20mm}
\bar k^{-1}[\circ]=\{x\},
u=\bar k\}
\\
A^2J^2B^1\models
\{\varphi_4,
k_{AJ}^{-1}[\circ]=\{\bar k\},
k_{BJ}^{-1}[\circ]=\{u\},\\\hspace{20mm}
\bar k^{-1}[\circ]=
\{x\}, u=\bar k,
v=u\}
\\
A^2J^2B^2\models\{x=y\}.
\ey
}$$

\subsection{Yahalom protocol verification}
\label{sdfdsgerre434343456}

In this and  next section, we consider a method for verifying CPs, based on theorem \ref{doredtuslna32b322e3323q}. This method is not explicitly described, since it can be 
understood by  examples 
of verification of Yahalom CP (in this section) and 
a CP of message transmission with unlimited number of participants 
(in section \ref{sdfgstrghw54y46ywh}).

\subsubsection {Description of  Yahalom protocol}
\label {zfgvds3w456w3}

Yahalom protocol is designed to authenticate agents communicating 
over the open  channel $ \circ $.
It is assumed that
\bi
\i there are given a 
set $ Ag \subseteq Var_{\bf A}$, and a
{\bf trusted intermediary} $ J \in 
Var_{\bf A}$, 
these agents can communicate 
 through  channel $ \circ $,
\i each  $ A \in Ag $ has a shared secret key $ k_{AJ} $ with 
$ J $, on which $ A $ and $ J $ can encrypt and decrypt messages using a symmetric encryption system, and only $ A $ and $ J $ know the key $ k_{AJ} $.
\ei
 
The following agents participate in each Yahalom session: 
an {\bf initiator} $ A \in Ag $, a trusted intermediary $ J $, and 
a {\bf responder} $ B \in Ag $. Each agent from $ Ag $ can be an initiator in some sessions, and a responder in others. The same agent may be both an initiator and a responder in the same session (i.e. it is possible that $ A = B $). A Yahalom  session with an initiator $ A $, a responder $ B $ and 
a trusted intermediary $ J $ is a set of four message 
transfers: 
\be{dasfasdfgfdss}
\by 1.&
A\to B&:&
A,n_A\\2.&
B\to J&:&B, 
k_{BJ}(A, n_A, n_B)\\3.&
J\to A&:&
k_{AJ}(B, k,n_A,n_B), k_{BJ}
(A,
k)\\4.&
A\to B&:&k_{BJ}
(A,
k),k(n_B)
\ey\ee

Transfers in \re{dasfasdfgfdss} have the following meaning:
\bn
\i $ A $ sends $ B $ a request for an authentication and a generation of a session key $ k $, this request consists of 
the agent name $ A $ and  nonce $ n_A $,
\i $ B $ sends $ J $ a request to generate a session key $ k $, in its request it includes its name, the name of the agent $ A $, for communication with which this key is needed, the received nonce $ n_A $, and its nonce $ n_B $,
\i $ J $ generates session key $ k $ and sends $ A $ a pair of messages,
\bi \i from  first message $ A $ can extract  $ k $, \i
and  second message is intended for $ A $ to forward it to $ B $, \ei
\i $ A $ sends $ B $ a pair of messages,
\bi \i
 first of which it received from $ J $,  $ B $ can extract
 a session key $ k $ from this message, 
\i using $ k $,  $ B $ decrypts  second message, if a result of 
the decryption  matches its nonce $ n_B $, then this is a proof for him that a sender of this message is exactly $ A $.
\ei
\en

A Yahalom session is described 
by  diagram \re{sdfg1dsgwer33r333}.
\begin{figure*}
\be{sdfg1dsgwer33r333}
\!\!\!\!\!\!\!\!\!\!\!\!\!\!
\by
\begin{picture}(0,95)
\put(-40,92){\makebox(0,0){$I_A$}}
\put(40,92){\makebox(0,0){$R_B$}}
\put(0,92){\makebox(0,0){$J$}}

\put(-40,80){\circle*{4}}
\put(-37,80){\makebox(0,0)[l]{$0$}}
\put(-40,60){\circle*{4}}
\put(-37,62){\makebox(0,0)[l]{$1$}}
\put(-40,0){\circle*{4}}
\put(-37,20){\makebox(0,0)[l]{$2$}}
\put(-40,20){\circle*{4}}
\put(-37,0){\makebox(0,0)[l]{$3$}}
\put(-40,10){\vector(1,0){80}}
\put(-40,0){\line(0,1){80}}
\put(0,0){\line(0,1){80}}
\put(-43,130){\makebox(0,0)[r]{$\;$}}
\put(3,130){\makebox(0,0)[l]{$\;$}}
\put(-43,110){\makebox(0,0)[r]{$\;$}}
\put(3,110){\makebox(0,0)[l]{$\;$}}
\put(-43,90){\makebox(0,0)[r]{$\;$}}
\put(3,92){\makebox(0,0)[l]{$\;$}}
\put(-2,68){\makebox(0,0)[r]{$\;$}}
\put(43,70){\makebox(0,0)[l]{$\;$}}
\put(-3,50){\makebox(0,0)[r]{$\;$}}
\put(43,50){\makebox(0,0)[l]{$\;$}}
\put(-43,70){\makebox(0,0)[r]{$
!(A,\bar n^i_A)$}}
\put(43,70){\makebox(0,0)[l]{$?
(\hat a^i_B, \hat n^i_B)$}}
\put(43,50){\makebox(0,0)[l]{$!
(B, k_{BJ}(a^i_B,{n^i_B},\bar n^r_B))$}}
\put(0,48){\makebox(0,0)[r]{$
?(\hat
a^r_J, k_{\hat
a^r_JJ}(\hat
a^i_J,\hat n^i_J,
\hat n^r_J)\!)$}}
\put(-43,30){\makebox(0,0)[r]{$
?(k_{AJ}( 
a^r_A,\hat k^i_A,n^i_A,\hat n^r_A), \hat x)$}}
\put(2,28){\makebox(0,0)[l]{$
!(k_{a^i_J J}(a^r_J,\bar k_J,n^i_J,n^r_J), k_{a^r_JJ}
(a^i_J,\bar
k_J))$}}
\put(-43,10){\makebox(0,0)[r]{$
!(x,k^i_A(n^r_A))$}}
\put(43,10){\makebox(0,0)[l]{$?
(k_{BJ}(a^i_B,\hat k^r_B), \hat k^r_B(n^r_B))$}}

\put(0,80){\circle*{4}}
\put(3,80){\makebox(0,0)[l]{$0$}}
\put(0,42){\circle*{4}}
\put(3,42){\makebox(0,0)[l]{$1$}}
\put(0,0){\circle*{4}}
\put(3,0){\makebox(0,0)[l]{$2$}}
\put(40,80){\circle*{4}}
\put(37,80){\makebox(0,0)[r]{$0$}}
\put(40,60){\circle*{4}}
\put(37,60){\makebox(0,0)[r]{$1$}}
\put(40,42){\circle*{4}}
\put(37,42){\makebox(0,0)[r]{$2$}}
\put(40,0){\circle*{4}}
\put(37,0){\makebox(0,0)[r]{$3$}}
\put(-40,70){\vector(1,0){80}}
\put(40,50){\vector(-1,0){40}}
\put(0,30){\vector(-1,0){40}}
\put(40,0){\line(0,1){80}}
\end{picture}\ey
\ee\end{figure*}
In this diagram,  \bi\i
left and right threads correspond to SPs $ I_A $ and $ R_B $, describing a behavior of the initiator $ A $ and the responder $ B $, respectively, \i middle thread corresponds to a SP, describing a behavior of the intermediary $ J $, this SP is denoted by the same symbol $ J $. \ei

The meaning of  variables in these SPs is seen from the comparison of actions in these SPs with the corresponding actions in \re{dasfasdfgfdss}. Superscripts $ i $ 
and $ r $ on variables mean that these variables 
presumably contain an information about an initiator ($ i $) or a responder ($ r $) of this session.

We assume that $$ Agent (I_A) = A ,\;
 Agent (R_B) = B ,\;
 Agent (J) = J .$$

A DP $ {\cal P} $ corresponding to Yahalom  has the form
\be{sdgffdsghdsfe334w}
\by {\cal P} =
\{
\{I_A ^{*} \mid A \in Ag \},
\{R_B ^{*} \mid B \in Ag \}, J ^{*} \}. \ey \ee

We will use the following notations:
\bi \i
if $ {\cal P} $ is a DP, and $ \pi $ is a path in $ \Sigma_{{\cal P} _{\dagger}} $, then  $ \pi \ni P ^ {i, i '} : s \ra {\alpha} s '$ means that $ \pi $ contains the edge $ s \ra {\alpha_P} s' $, and
$ at_{s_P} = i $,
$ at_{s'_P} = i '$,
\i  $ s \models E \, \bot_{\bf K} \, e $ denotes the
statement $$ \forall \, x \in E_{\bf X} \; \; x \, \bot_{{\bf K}, E} \, e ^ s .$$
\ei

It is not hard to prove that 
\be{fgfhgfhhrt54}
s\models E\,\bot_{\bf K}\,(e,e')\Leftrightarrow\left\{\by
s\models E\,\bot_{\bf K}\,e\;\;\mbox{ and }\\
s\models E\,\bot_{\bf K}\,e'.\ey\right.\ee

\subsubsection{Properties of Yahalom protocol}

The following properties of 
DP \re{sdgffdsghdsfe334w} will be verified:
\bi
\i {\bf secrecy} of keys and nonces $ n ^ r_B $: 
\be{dfgdsf3241325hg}\by
\forall\,s\in \Sigma_{{\cal P}_{\dagger} }\quad
s\models E
\,\bot_{\bf K}\,
P_{\dagger} ,\\ \mbox{ where }
E=\{k_{BJ}, k_J, n^r_B\mid B\in Ag\}
\ey\ee
\i {\bf authentication of the initiator to the responder}: $
\forall\, R_B \in {\cal P} $, $\forall\, s \in \Sigma_{{\cal P} _{\dagger}} $,\\ if $ s \models at_{R_B} = 3 $, then $ \exists $ $ I_A \in {\cal P} $: 
\be{sdfgad32sgfdsgsdfds}\by
s\models \{at_{I_A}=3,
a^r_A=B, a^i_B=A, \\\hspace{10mm}
n^i_A=n^i_B,
n^r_A=n^r_B,k^i_A=k^r_B\},\ey
\ee

\i {\bf authentication of the responder to the initiator}: 
$\forall\,I_{A}\in {\cal P}$,
$\forall\,s\in \Sigma_{{\cal P}_{\dagger} }$, 
\\if  $s\models at_{I_A}=2$, 
then  $\exists\,R_B\in {\cal P}$:
\be{afdgdsfgw454y264uhtrr}\by
s\models \{at_{R_B}=2,
a^r_A=B,
a^i_B =A, \\\hspace{10mm}
n^i_A=n^i_B,
n^r_A=n^r_B\}.\ey
\ee
\ei

\subsubsection{Secrecy of keys and nonces $ n ^ r_B $ }
\label{ratrete5y334rrrrre}

Prove \re{dfgdsf3241325hg} by contradiction.

Suppose $ S = \{s \in \Sigma_{{\cal P} _{\dagger}} \mid s \not \models \varphi \} \neq \emptyset $, where $ \varphi $ is a formula in \re{dfgdsf3241325hg}.

$ \forall \, s \in S $ denote by $ \pi_s $ a path of minimum length from $ 0 $ to $ s $. Let $ s $ be a state from $ S $ with the least length of $ \pi_s $. Since $ 0 \models \varphi $, then $ s \neq 0 $.

Let $ s' \ra {\! \alpha_{P}} s $ be an edge from $ \pi_s $ ending in $ s $.

From the definition of $ s $ it follows that $ s' \models \varphi $, $ s \not \models \varphi $. If $ P = P_{\dagger} $, then from  theorem \ref{doredtusl2nab33222eq} it follows that $ s \models \varphi $, i.e. we have a contradiction.

Therefore, $ P \in \{I_A, R_B, J \mid
A, B \in Ag \} $,
and
\be{fdgdsghsdhsghsr54}\by
\alpha_P =! e, \;
[\circ]_{s} = [\circ]_{s'} \cup \{e \}, \\
\exists \, y \in E_{\bf X}:
\neg (y \, \bot_{{\bf K}, E} \, e ^ s). \ey\ee

There is only  justification
of the existence of the edge $ s' \ra {\! \alpha_{P}} s $ with  properties \re{fdgdsghsdhsghsr54}:
\be{zdfgdg54w6yhutrhrs}
\pi_s \ni I_A ^ {2,3}: s' \ra {! e} s, \; \;
\mbox{where }\; e = (x, k ^ i_A (n ^ r_A)). \ee

Since $ s' \models at_{I_A} = 2 $, then
$ \exists \, s_1 \leq_{\pi_s} s' $:
\be{1fgdsgsde45strhsrhst}
\by
\pi_s \ni
I_A ^ {1, 2}:
s'_1 \ra {? e_1} s_1, \mbox{where }\\
e_1 = (k_{AJ} (a ^ r_A, \hat k ^ i_A, n ^ i_A, \hat n ^ r_A),
\hat x
). \ey\ee

Since $ s_1 \leq_{\pi_s} s '$ and $ s' \models \varphi $, then $ s_1 \models \varphi $. In particular, $ s_1 \models E \, \bot_{\bf K} \, e_1 $. By \re{fgfhgfhhrt54}, this implies $ s_1 \models E \, \bot_{\bf K} \, x $.

By theorem \ref{doredtuslna32b322e3323q},  $ s_1 \models \varphi $, $ e_1 ^ s \in [\circ]_{s_1} $, and $ k_{AJ} \in E $  imply: $ \exists \, s_2 \leq_{\pi_s} s'_1 $:
\bi\i $ \pi_s $ contains the edge $ s'_2 \ral {(! e_2)_{P}} s_2 $, where $ P \in {\cal P} $,  \i first component $ k_{AJ} (\ldots) $ of the term $ e_1 ^ s $ is a subterm of $ e_2 ^ s $.\ei

There is only  justification
the existence of an edge with such properties:
\be{1sfgfasgdsfh3gsw5445hwrtgs}
\left\{\by\pi_s\ni
 J^{1, 2}: s'_2\ra{!e_2}s_2,
\mbox{ where }\\
e_2=(k_{a^i_{ J} { J}}(a^r_{ J},\bar k_{ J},n^i_{ J},n^r_{ J}), \ldots
),\\
k_{(a^i_{ J})^s { J}}((a^r_{ J})^s,\bar k_{ J},\ldots
)=\\
=k_{AJ}(a^r_A,(k^i_A)^s,\ldots
).
\ey\right.\ee
(ellipsis in \re{1sfgfasgdsfh3gsw5445hwrtgs}
and below denotes a component of a pair that is not of interest for consideration). 

 \re{1sfgfasgdsfh3gsw5445hwrtgs} implies $ (k ^ i_A) ^ s = \bar k_J $, so $ s \models E \, \bot_{\bf K} \, k ^ i_A (n ^ r_A) $.
From the above property $ s_1 \models E \, \bot_{\bf K} \, x $
and \re{fgfhgfhhrt54} we get: $ s \models E \, \bot_{\bf K} \, ( x, k ^ i_A (n ^ r_A)) $, i.e. $ s \models E \, \bot_{\bf K} \, e $, which contradicts the assumption $ s \not \models E \, \bot_{\bf K} \, e $. $ \blackbox $ \\

Proven property $ \forall \, s \in \Sigma_{{\cal P} _{\dagger}} \; \; s \models \varphi $ will be used below.
In  proofs presented below, for each application of theorem \ref{doredtuslna32b322e3323q}, there is only one way to justify the existence of the edge \re{fgfdsgdsgsdfgfdsgh} in the graph $ \Sigma_{{\cal P} _{\dagger}} $, and we will 
not
mention the uniqueness of such a justification. This uniqueness is ensured by a suitable definition of actions of the form $!e $ in SPs occurred 
in DPs under consideration. 

\subsubsection{Authentication of the initiator to the responder}
\label{sfgdfsgsefds1}

Let a SP $ R_B \in {\cal P} $ and a state $ s \in \Sigma_{{\cal P} _{\dagger}} $ are such that $ s \models at_{R_B} = 3 $.
Prove that $ \exists \, I_A \in {\cal P} $: \re{sdfgad32sgfdsgsdfds} holds.

Let $ \pi $ be a path from $ 0 $ to $ s $.
The statement
$ s \models at_{R_B} = 3 $ implies that $ \exists \, s_1 \leq_{\pi} s $:
$$\by \pi \ni
R_B ^ {2, 3}: s'_1 \ra {
? e_1}
s_1, \mbox{where }\\
e_1 = (k_{BJ} (a ^ i_B, \hat k_B ^ r), \hat k_B ^ r (n ^ r_B)). \ey$$

By theorem \ref{doredtuslna32b322e3323q}, 
from 
$ s_1 \models \varphi $, $ e_1 ^ s \in [\circ]_{s_1} $, $ k_{BJ} \in E $ it follows that
$ \exists \, s_2 \leq_{\pi} s'_1 $:
\be{fgfdsgdsfgdsfgdsfgdsfgdshsg}
\left \{
\by
\pi \ni
J ^ {1, 2}: s'_2 \ra {!e_2}
s_2,  \mbox{ where }\\
e_2 =
(\ldots,
k_{a ^ r_JJ}
(a ^ i_J, \bar
k_J)), \\
k_{(a ^ r_J) ^ sJ}
((a ^ i_J) ^ s, \bar
k_J) =
k_{BJ} ((a ^ i_B) ^ s, (k_B ^ r) ^ s).
\ey \right. \ee

Second equality in \re{fgfdsgdsfgdsfgdsfgdsfgdshsg} implies 
that \be{fdgfdsfgds0} (a ^ r_J) ^ s = B, (a ^ i_J) ^ s = (a ^ i_B) ^ s, \bar k_J = (k ^ r_B) ^ s. \ee

From $ s'_2 \models at_{J} = 1 $ it follows that $ \exists \, s_3 \leq_{\pi} s'_2 $:
\be{dfgdshgsdhfh354t4}\by
\pi \ni
J ^ {0, 1}: s'_3 \ra {? E_3}
s_3, \mbox{ where }\\
e_3 =
(\ldots,
k_{\hat
a ^ r_JJ} (\hat
a ^ i_J, \hat n ^ i_J,
\hat n ^ r_J)). \ey\ee 

From \re{fdgfdsfgds0} and \re{dfgdshgsdhfh354t4}
it follows that
$$k_{BJ}(\mbox{3 terms})\subseteq
e_3^s\in [\circ]_{s_3},$$ whence by theorem 
\ref{doredtuslna32b322e3323q}, with considering 
$s_3 \models \varphi$ and $k_{BJ}\in E$
we get:
$\exists\,s_4\leq_{\pi} s'_3$:
\be{efasfae33w44rere}
\left\{\by\pi\ni
R_{\grave B}^{1, 2}: s'_4\ra{!e_4}s_4,
\mbox{ where }\\e_4=(\ldots, 
k_{{\grave B}J}(a^i_{\grave B},{n^i_{\grave B}},\bar n^r_{\grave B}))\\
k_{{\grave B}J}((a^i_{\grave B})^s,(n^i_{\grave B})^s,\bar n^r_{\grave B})=\\=k_{BJ}(
(a^i_B)^s,(n^i_J)^s,
(n^r_J)^s)
\ey\right.\ee

Second equality in \re{efasfae33w44rere}
implies that
\be{fdgfdsfgds1}
\grave B=B, (n^i_B)^s=(n^i_J)^s, \bar n^r_B=(n^r_J)^s.\ee

By theorem \ref{doredtuslna32b322e3323q},
from $$s_1\models\varphi,
(k^r_B(n^r_B))^s
\subseteq e_1^s
\in
[\circ]_{s_1}, 
(k^r_B)^s=\bar k_J \in E$$
it follows that
$\exists\,s_5\leq_{\pi} s'_1$:
\be{asfgsg54hsttrsrhgdsfs}
\left\{\by\pi\ni
I_A^{2, 3}: s'_5\ra{!e_5}s_5,
\mbox{ where }\\
e_5=(\ldots,
k^i_A(n^r_A))\\
(k^i_A(n^r_A))^s=\bar k_J(n^r_B)
\ey\right.\ee

From  second equality in \re{asfgsg54hsttrsrhgdsfs}
it follows that
\be{fdgfdsfgds2}
(k^i_A)^s=\bar k_J, (n^r_A)^s = n^r_B.\ee

From $s_5 \models at_{I_A}=2$ it follows that 
$\exists\,s_6\leq_{\pi} s'_5$: 
\be{fgdsgsde45strhsrhst}\by\pi\ni
I_A^{1, 2}:
s'_6\ra{?e_6}s_6,\mbox{ where }\\
e_6=(k_{AJ}( a^r_A,\hat k^i_A,n^i_A,\hat n^r_A), 
\ldots
).\ey\ee

From \re{fdgfdsfgds2} and 
\re{fgdsgsde45strhsrhst}
it follows that 
\be{sdfgdsgrg3w43w33t3}\by
k_{AJ}(a^r_A,(k^i_A)^s,n^i_A,( n^r_A)^s) = \\= k_{AJ}(a^r_A,\bar k_J,n^i_A, n^r_B)
\subseteq e_6^s\in [\circ]_{s_6}.\ey\ee

By theorem \ref{doredtuslna32b322e3323q},
from
$s_6 \models \varphi$, $k_{AJ}\in E$, and
 \re{sdfgdsgrg3w43w33t3} it follows that
$\exists\,s_7\leq_{\pi} s'_6$:
\be{sfgfasgdsfhgsw5445hwrtgs}
\left\{\by\pi\ni
\grave J^{1, 2}: s'_7\ra{!e_7}s_7,
\mbox{ where }\\
e_7=(k_{a^i_{\grave J} {\grave J}}(a^r_{\grave J},\bar k_{\grave J},n^i_{\grave J},n^r_{\grave J}), \ldots
)\\
k_{(a^i_{\grave J})^s { J}}((a^r_{\grave J})^s,\bar k_{\grave J},(n^i_{\grave J})^s,(n^r_{\grave J})^s)=\\
=k_{AJ}(a^r_A,\bar k_J,n^i_A, n^r_B)
\ey\right.\ee

From  second equality in  \re{sfgfasgdsfhgsw5445hwrtgs}
it follows that
\be{fdgfdsfgd3343s1}\by
(a^i_{\grave J})^s=A,
(a^r_{\grave  J})^s=a^r_A,
\grave J=J, \\
(n^i_J)^s=n^i_A, (n^r_J)^s=
\bar n^r_B.\ey\ee

 \re{sdfgad32sgfdsgsdfds}
follows from 
\re{fdgfdsfgds0},
\re{fdgfdsfgds1},
\re{fdgfdsfgds2},
\re{fdgfdsfgd3343s1}. $\blackbox$

\subsubsection{Authentication of the responder  to the initiator }
\label{fdsfgdstrhyh5tfddc}

Let a SP $I_A\in {\cal P}$ and a state
$s\in \Sigma_{{\cal P}_{\dagger} }$
are such that $s\models at_{I_A}=2$.
Prove that
$\exists\,R_B\in {\cal P}$:
\re{afdgdsfgw454y264uhtrr} holds.

Let $\pi$ be a path from 0 to $s$.
From $s\models at_{I_A}=2$ it follows that 
$\exists\,s_1\leq_{\pi} s$:
\be{fgdsgsde45strhsrh332st}\by
\pi\ni I_A^{1, 2}:
s'_1\ra{?e_1}s_1,\mbox{ where }\\
e_1=(k_{AJ}(a^r_A,\hat k^i_A,n^i_A,\hat n^r_A), 
\ldots
).\ey\ee

By theorem \ref{doredtuslna32b322e3323q},
from $$s_1\models \varphi,\;
k_{AJ}(\mbox{4 terms})
\subseteq e_1^s
\in
[\circ]_{s_1},\;
k_{AJ}\in E$$
it follows that 
$\exists\,s_2\leq_{\pi} s'_1$:
\be{fgf32dsgdsfgdsfg4dsfgdsfgds4rhsg}
\left\{
\by\pi\ni
J^{1, 2}:s'_2\ra{!e_2}
s_2,\mbox{ where }\\
e_2=
(
k_{a^i_J J}(a^r_J,\bar k_J,n^i_J,n^r_J), \ldots
)\\
k_{(a^i_J)^s J}((a^r_J)^s,\bar k_J,(n^i_J)^s,(n^r_J)^s)=\\=
k_{AJ}(a^r_A,(k^i_A)^s,
n^i_A,(n^r_A)^s)
\ey\right.\ee

From  second equality in \re{fgf32dsgdsfgdsfg4dsfgdsfgds4rhsg}
it follows that
\be{fdsgdshfdf456yhy}\by
(a^i_J)^s =A,
(a^r_J)^s=a^r_A,
\bar k_J=(k^i_A)^s,\\
(n^i_J)^s=n^i_A,
(n^r_J)^s=(n^r_A)^s.
\ey\ee

From $s_2\models at_{J}=1$ 
it follows that 
$\exists\,s_3\leq_{\pi} s'_2$:
\be{dfgd23shgsdhfh354t4}\by\pi\ni
J^{0, 1}:s'_3\ra{?e_3}
s_3,\mbox{ where }\\
e_3=
(\ldots, 
k_{\hat
a^r_JJ}(\hat
a^i_J,\hat n^i_J,
\hat n^r_J)).\ey\ee

From \re{fdsgdshfdf456yhy}
 and \re{dfgd23shgsdhfh354t4}
it follows that 
$$k_{a^r_AJ}(
A,n^i_A,
(n^r_A)^s) \subseteq e_3^s\in [\circ]_{s_3},$$ whence by theorem  \ref{doredtuslna32b322e3323q},
considering
$s_3\models \varphi$,
 and 
$k_{a^r_AJ}\in E$,
we get:
$\exists\,s_4\leq_{\pi} s'_3$:
\be{efasfae3233w44rere}
\left\{\by\pi\ni
R_{ B}^{1, 2}: s'_4\ra{!e_4}s_4,
\mbox{ where }\\e_4=(\ldots, 
k_{{B}J}(a^i_{ B},{n^i_{ B}},\bar n^r_{ B}))\\
k_{{ B}J}((a^i_{ B})^s,(n^i_{ B})^s,\bar n^r_{ B})=\\=k_{a^r_AJ}(
A,n^i_A,
(n^r_A)^s).
\ey\right.\ee

Second equality in \re{efasfae3233w44rere}
implies equalities from which \re{afdgdsfgw454y264uhtrr} 
follows:
$$
B=a^r_A,
(a^i_{ B})^s=A,
(n^i_{ B})^s=n^i_A,
\bar n^r_{ B}=(n^r_A)^s. \;\;\blackbox
$$

\subsection{Verification of the protocol 
of message transmission with unlimited number of participants}
\label{sdfgstrghw54y46ywh}

In this section we 
consider an example of verification of a CP intended for 
EM transmission with unlimited number of participants.
This CP is a generalization of the Wide-Mouth Frog CP 
$ {\cal P}_4 $, considered in section \ref{pkergnmf43564654632}. 

\subsubsection{Protocol Description}

Participants of this CP are agents from the set $ Ag \subseteq Var_{\bf A}$ and a trusted intermediary $ J \in Var_{\bf A}$.
Each  $ A \in Ag $ uses the key $ k_{AJ} $ to communicate with $ J $, which is available only to $ A $ and $ J $.
The encrypted transmission of the message $ x $ from  $ A \in Ag $ to $ B \in Ag $ consists of the following actions:
\bi \i exchange of messages between $ A $ and $ J $, as a result of which $ J $ learns the name $ A $ of the sender, the name of $ B $ of the recipient, and the key $ k $, on which $ x $ 
will be encrypted,
\i exchange of messages between $ J $ and $ B $, as a result of which $ B $ learns the name $ A $ of the sender of the message that $ B $ will receive from $ A $, and the key $ k $ on which this message will be encrypted,
\i transfer of EM $ k (x) $ from $ A $ to $ B $.
\ei

An execution of a session of this CP with the initiator $ A $, the responder $ B $ and the trusted intermediary $ J $ is the following set of message transfers: 
\be{dasfasdfg43rtss}
\by 
1.&A\to J&:& k_{AJ}(A,n_A)\\
2.&J\to A&:& k_{AJ}(n_A,n_J)\\
3.&A\to J&:& k_{AJ}(n_J,k)\\
4.& J\to B&:&k_{BJ}(n_A)\\
5.& B\to J&:&k_{BJ}(n_A, n_B, B)\\
6.& J\to B&:&k_{BJ}(A, n_B, k)\\
7.&A\to B&:&k(x)
\ey\ee

This session is represented by the  diagram
\re{sdfgdsgwer33r333}.
\begin{figure*}
\be{sdfgdsgwer33r333}\hspace{5mm}\by
\begin{picture}(0,152)
\put(-60,152){\makebox(0,0){$I_A$}}
\put(60,152){\makebox(0,0){$R_B$}}
\put(0,152){\makebox(0,0){$J$}}
\put(-60,140){\circle*{4}}
\put(-57,140){\makebox(0,0)[l]{$0$}}
\put(-60,120){\circle*{4}}
\put(-57,120){\makebox(0,0)[l]{$1$}}
\put(-60,100){\circle*{4}}
\put(-57,100){\makebox(0,0)[l]{$2$}}
\put(-60,80){\circle*{4}}
\put(-57,81){\makebox(0,0)[l]{$3$}}
\put(-60,0){\circle*{4}}
\put(-57,0){\makebox(0,0)[l]{$4$}}
\put(0,140){\circle*{4}}
\put(-3,140){\makebox(0,0)[r]{$0$}}
\put(0,120){\circle*{4}}
\put(-3,120){\makebox(0,0)[r]{$1$}}
\put(0,100){\circle*{4}}
\put(-3,100){\makebox(0,0)[r]{$2$}}
\put(-60,130){\vector(1,0){60}}
\put(0,110){\vector(-1,0){60}}
\put(-60,90){\vector(1,0){60}}
\put(-60,10){\vector(1,0){120}}
\put(-60,0){\line(0,1){140}}
\put(-63,130){\makebox(0,0)[r]{$!k_{AJ}(a^r_A,\bar n^i_A)$}}
\put(3,130){\makebox(0,0)[l]{$? k_{\hat a^i_JJ}(\hat a^r_J,\hat n^i_J)$}}
\put(-63,110){\makebox(0,0)[r]{$? k_{AJ}(\bar n^i_A,\hat n^j_A)$}}
\put(3,110){\makebox(0,0)[l]{$! k_{a^i_JJ}(n^i_J,\bar n^j_J)$}}
\put(-63,90){\makebox(0,0)[r]{$! k_{AJ}(n^j_A, \bar k^i_A)$}}
\put(3,92){\makebox(0,0)[l]{$? k_{a^i_JJ}(\bar n^j_J,\hat k_J)$}}
\put(-2,68){\makebox(0,0)[r]{$! k_{a^r_JJ}(n^i_J)$}}
\put(63,70){\makebox(0,0)[l]{$? k_{BJ}(\hat n^i_B)$}}
\put(-3,50){\makebox(0,0)[r]{$? k_{a^r_JJ}(n^i_J, \hat n^r_J,a^r_J)$}}
\put(63,50){\makebox(0,0)[l]{$! k_{BJ}(n^i_B,\bar n^r_B,B)$}}
\put(-3,30){\makebox(0,0)[r]{$! k_{a^r_JJ}(a^i_J,n^r_J,
 k_J)$}}
\put(63,30){\makebox(0,0)[l]{$? k_{BJ}(
\hat a^i_B,\bar  n^r_B,\hat k^i_B)$}}
\put(-63,10){\makebox(0,0)[r]{$
!\bar k^i_{A}( x^i_A)$}}
\put(63,10){\makebox(0,0)[l]{$?k^i_B(\hat x^i_B)$}}

\put(0,80){\circle*{4}}
\put(3,79){\makebox(0,0)[l]{$3$}}
\put(0,60){\circle*{4}}
\put(3,60){\makebox(0,0)[l]{$4$}}
\put(0,40){\circle*{4}}
\put(3,40){\makebox(0,0)[l]{$5$}}
\put(0,20){\circle*{4}}
\put(3,20){\makebox(0,0)[l]{$6$}}
\put(60,80){\circle*{4}}
\put(57,79){\makebox(0,0)[r]{$0$}}
\put(60,60){\circle*{4}}
\put(57,60){\makebox(0,0)[r]{$1$}}
\put(60,40){\circle*{4}}
\put(57,40){\makebox(0,0)[r]{$2$}}
\put(60,20){\circle*{4}}
\put(57,20){\makebox(0,0)[r]{$3$}}
\put(60,0){\circle*{4}}
\put(57,0){\makebox(0,0)[r]{$4$}}
\put(65,-5){\makebox(0,0)[l]{$P_B$}}
\put(0,70){\vector(1,0){60}}
\put(60,50){\vector(-1,0){60}}
\put(0,30){\vector(1,0){60}}
\put(60,0){\line(0,1){80}}
\put(0,20){\line(0,1){120}}
\end{picture}\ey
\ee\end{figure*}

A DP $ {\cal P} $ corresponding to this CP has the form 
\re{sdgffdsghdsfe334w}.
Properties of this CP that must be verified:
\bi
\i {\bf secrecy} of keys, transmitted messages and nonces: 
\be{dfgdsf3241325h3g}\by
\forall\,s\in \Sigma_{{\cal P}_{\dagger} }\;\;
s\models E
\,\bot_{\bf K}\,
P_{\dagger} ,\;\mbox{ where }\\
E=\{k_{AJ}, k^i_A, x^i_A, n^i_A\mid A\in Ag\}\ey
\ee
\i {\bf integrity} of transmitted messages: 
\be{sdfgad32sgfd2sgsdf332ds}\by
\mbox{$\forall\,R_B\in {\cal P}$, 
$\forall\,s\in \Sigma_{{\cal P}_{\dagger} }$,}\\ 
\mbox{if
$s\models at_{R_B}=4$,
then $\exists$ 
$I_A\in {\cal P}$:
}
\\
s\models \{at_{I_A}=4,
a^r_A=B, a^i_B=A, \\
\hspace{5mm} n^i_A=n^i_B,
k^i_A=k^i_B,
x^i_A=x^i_B
\}\ey\ee
\ei

\subsubsection{Verification of the protocol}
\label{fgsrh654sey5tg}

The proof of  secrecy property \re{dfgdsf3241325h3g}  coincides with the beginning of the reasoning in section \ref{ratrete5y334rrrrre}, with the only difference that there is no way to justify the existence of  edge $ s' \ra {\! \alpha_{ P}} s $ with properties \re{fdgdsghsdhsghsr54}.

Prove the integrity property \re{sdfgad32sgfd2sgsdf332ds}.
We will use in this proof  property \re{dfgdsf3241325h3g}
proven above.

Let a SP $ R_B \in {\cal P} $ and a state $ s \in \Sigma_{{\cal P} _{\dagger}} $ are such that $ s \models at_{R_B} = 4 $.
Prove that $ \exists \, I_A \in {\cal P} $: the statement in  third and fourth  lines \re{sdfgad32sgfd2sgsdf332ds} holds.

Let $ \pi $ be a path from $ 0 $ to $ s $.
From $ s \models at_{R_B} = 4 $ it follows that 
\be{zdfgdfgdsghsrth6srje34}
\by
\exists\,s_1\leq_{\pi} s: \;
\pi\ni
R_B^{3, 4}:s'_1\ra{
?e_1}
s_1,\\\mbox{ where }
e_1=k^i_B(\hat x^i_B),
\\
\exists\,s_2\leq_{\pi} s'_1: \;
\pi\ni
R_B^{2, 3}:s'_2\ra{
?e_2}
s_2,\\\mbox{ where }
e_2=k_{BJ}(\hat a^i_B, \bar n^r_B,\hat k^i_B).
\ey\ee

By theorem \ref{doredtuslna32b322e3323q},
from
second statement in \re{zdfgdfgdsghsrth6srje34},
$e_2^{s}\in [\circ]_{s_2}$,
and $k_{BJ}\in E$,
it follows  
\be{fgf32dsgdsfg3dsfg4dsfgdsfgds4rhsg}
\left\{
\by\exists\,s_3\leq_{\pi} s'_2:
\pi\ni
J^{5, 6}:s'_3\ra{!e_3}
s_3,\\\mbox{ where }
e_3=k_{a^r_JJ}(a^i_J,n^r_J,
 k_J),\\
k_{(a^r_J)^sJ}((a^i_J)^s,(n^r_J)^s,
 (k_J)^s)=\\=
k_{BJ}(\hat a^i_B, \bar n^r_B,\hat k^i_B).
\ey\right.\ee

From second equality in 
\re{fgf32dsgdsfg3dsfg4dsfgdsfgds4rhsg}
it follows that
\be{afdgsr4ew56yh5eagre}\by
(a^r_J)^s = B,\;
(a^i_J)^s=(a^i_B)^s,\\
(n^r_J)^s=\bar n^r_B,\;
(k_J)^s= (k^i_B)^s.\ey
\ee

From first statement in
\re{fgf32dsgdsfg3dsfg4dsfgdsfgds4rhsg}
and \re{afdgsr4ew56yh5eagre}
we get
\be{arfgste3w5wty6}\by
\exists\,s_4\leq_{\pi} s'_3:
\pi\ni
J^{4, 5}:s'_4\ra{?e_4}
s_4,\\\mbox{ where }
e_4=k_{BJ}(n^i_J, n^r_B,B).\ey\ee

By theorem \ref{doredtuslna32b322e3323q},
from \re{arfgste3w5wty6},
 $e_4^{s}\in [\circ]_{s_4}$, 
$k_{BJ}\in E$,
it follows that
\be{21fgf32dsgdsfg433dsfg4dsfgdsfgds4rhsg}
\left\{
\by\exists\,s_5\leq_{\pi} s'_4:
\pi\ni
\grave B^{1, 2}:s'_5\ra{!e_5}
s_5,\\\mbox{ where }
e_5=k_{\grave BJ}(n^i_{\grave B},\bar n^r_{\grave B},
{\grave B}),
\\
k_{{\grave B}J}((n^i_{{\grave B}})^s,
\bar n^r_{{\grave B}},
{{\grave B}})
=\\=
k_{BJ}((n^i_J)^s, \bar n^r_B,B).
\ey\right.\ee

From second equality in 
\re{21fgf32dsgdsfg433dsfg4dsfgdsfgds4rhsg}
it follows that
\be{fdsgsr546uh76yrhtg}
\bar n^r_{\grave B}= \bar n^r_{B},\; 
\grave B = B, \;
(n^i_{{ B}})^s=(n^i_J)^s.\;
\ee 

From \re{arfgste3w5wty6}
it follows that
\be{arfgste3w5wt443y6}
\by
\exists\,s_6\leq_{\pi} s'_4:
\pi\ni
 J^{2, 3}:s'_6\ra{?e_6}
s_6,\\\mbox{ where }
e_6=k_{a^i_{ J}J}(n^j_{ J},  k_{ J}).
\ey\ee

By theorem \ref{doredtuslna32b322e3323q},
from \re{arfgste3w5wt443y6},
$e_6^{s}\in [\circ]_{s_6},
k_{(a^i_{ J})^sJ}\in E$
it follows that
\be{2132fg4f3244dsfgdsfgds4rhsg}
\left\{
\by\exists\,s_7\leq_{\pi} s'_6:
\pi\ni
A^{2, 3}:s'_7\ra{!e_7}
s_7,\\\mbox{ where }
e_7=k_{AJ}(n^j_A, \bar k^i_A),
\\
k_{AJ}((n^j_A)^s, \bar k^i_A)
=
k_{(a^i_{ J})^sJ}(\bar n^j_{ J},  (k_{ J})^s).
\ey\right.\ee

From second equality in 
\re{2132fg4f3244dsfgdsfgds4rhsg}
we get:
\be{fdgdsfdsgrege1}
A=(a^i_{ J})^s,\;
(n^j_A)^s=\bar n^j_{ J},\;
\bar k^i_A=(k_{ J})^s.
\ee

From first statement in 
\re{2132fg4f3244dsfgdsfgds4rhsg}
we get:
\be{arfgste3w325wt443y6}\by
\exists\,s_{8}\leq_{\pi} s'_7:
\pi\ni
I_A^{1, 2}:s'_{8}\ra{?e_{8}}
s_{8},\\\mbox{ where }
e_{8}=k_{AJ}(\bar n^i_A,\hat n^j_A).\ey\ee

By theorem \ref{doredtuslna32b322e3323q},
from \re{arfgste3w325wt443y6},
$e_{8}^{s}\in [\circ]_{s_{8}},
k_{AJ}\in E$
it follows that
\be{2132fg4f324324dsfgdsfgds4rhsg}
\left\{
\by\exists\,s_{9}\leq_{\pi} s'_{8}:
\pi\ni
{\grave J}^{1, 2}:s'_{9}\ra{!e_{9}}
s_{9},\\\mbox{ where }
e_{9}=k_{a^i_{{\grave J}} J}(n^i_{{\grave J}}, \bar n^j_{{\grave J}}),
\\
k_{(a^i_{{\grave J}})^s J}((n^i_{{\grave J}})^s, \bar n^j_{{\grave J}})
=
k_{AJ}(\bar n^i_A,(n^j_A)^s).
\ey\right.\ee

From second equality in  
\re{2132fg4f324324dsfgdsfgds4rhsg}
we get:
\be{fdgdsfdsgrege2}
(a^i_{{\grave J}})^s =A,\;
(n^i_{{\grave J}})^s=\bar n^i_A,\;
\bar n^j_{{\grave J}}=
(n^j_A)^s
\ee

From \re{fdgdsfdsgrege1} and
\re{fdgdsfdsgrege2}
we get: 
\be{hrdttrhtrshstrdryhtrse}\by
\bar n^j_{\grave J}=\bar n^j_{{ J}}=(n^j_A)^s,\;
 {{\grave J}} = { J},\\
 (a^i_{{ J}})^s =A,\;
(n^i_{{J}})^s=\bar n^i_A.
\ey\ee

From \re{afdgsr4ew56yh5eagre}
and  \re{fdgdsfdsgrege1}
we get:
\be{dfgdsgw45g46h4y}
(k^i_B)^s =(k_J)^s= \bar k^i_A\in E,\ee
so by theorem \ref{doredtuslna32b322e3323q},
from first statement in
\re{zdfgdfgdsghsrth6srje34}
and $e_{1}^{s}\in [\circ]_{s_{1}}$
it follows that
\be{23232132fg4f324324dsfgdsfgds4rhsg}
\left\{
\by\exists\,s_{11}\leq_{\pi} s'_{1}:
\pi\ni
{\grave A}^{3,4}:s'_{11}\ra{!e_{11}}
s_{11},\\
\mbox{ where }
e_{11}=\bar k^i_{\grave A}(x^i_{\grave A}),
\\
\bar k^i_{\grave A}(x^i_{\grave A})=
(k^i_B)^s((x^i_B)^s).
\ey\right.\ee

From  second equality in 
\re{23232132fg4f324324dsfgdsfgds4rhsg}
and
\re{dfgdsgw45g46h4y}
we get:
\be{sgfasdfaga43w35hyafgd}
\bar k^i_{\grave A}=(k^i_B)^s=\bar k^i_A,\;
\grave A=A, \;
x^i_{A}=(x^i_B)^s.
\ee

From  first statement in 
\re{2132fg4f324324dsfgdsfgds4rhsg}
and \re{hrdttrhtrshstrdryhtrse}
we get:
\be{hdsfjldskj53438w}\by
\exists\,s_{10}\leq_{\pi} s'_{9}:
\pi\ni
 J^{0, 1}:s'_{10}\ra{?e_{10}}
s_{10},\\\mbox{ where }
e_{10}=k_{\hat a^i_{ J}J}(\hat a^r_{ J},
\hat n^i_{ J}).\ey\ee

From \re{afdgsr4ew56yh5eagre} and \re{hrdttrhtrshstrdryhtrse}
it follows that
$$e^s_{10}=k_{AJ}(B, \bar n^i_A).$$

By theorem  \ref{doredtuslna32b322e3323q},
from \re{hdsfjldskj53438w},
 $e_{10}^{s}\in [\circ]_{s_{10}}$,
$k_{AJ}\in E$,
it follows that
\be{213322fg4f324324dsfgdsfgds4rhsg}
\left\{
\by\exists\,s_{12}\leq_{\pi} s'_{10}:
\pi\ni
{\grave A}^{0, 1}:s'_{12}\ra{!e_{12}}
s_{12},\\\mbox{ where }
e_{12}=k_{{\grave A} J}(a^r_{{\grave A}}, \bar n^i_{{\grave A}}),
\\
k_{\grave A J}(a^r_{\grave A},\bar n^i_{\grave A})=
k_{AJ}(B, \bar n^i_A).
\ey\right.\ee
From  second equality in 
\re{213322fg4f324324dsfgdsfgds4rhsg}
we get: 
\be{fd32gdsfdsgrege2}
\bar n^i_{{\grave A}}=\bar n^i_A,\;
\grave A = A,\;
a^r_{{ A}} = B.
\ee

\re{sdfgad32sgfd2sgsdf332ds}
is justified as follows: 
\bi
\i $s\models at_{I_A}=4$ follows from  
\re{23232132fg4f324324dsfgdsfgds4rhsg} and
\re{sgfasdfaga43w35hyafgd}:
$s_{11}\models at_{\grave A}=4$,
 $\grave A=A$,  $s_{11}\leq_\pi s$,
\i $s\models a^r_A=B$ follows from 
\re{fd32gdsfdsgrege2},
\i $s\models a^i_B=A$ follows from  
\re{afdgsr4ew56yh5eagre} and \re{fdgdsfdsgrege1},
\i $s\models n^i_A=n^i_B$ follows from 
\re{fdsgsr546uh76yrhtg} and
\re{hrdttrhtrshstrdryhtrse},
\i $s\models k^i_A=k^i_B$ follows from 
\re{dfgdsgw45g46h4y},
\i $s\models x^i_A=x^i_B$ follows from 
\re{sgfasdfaga43w35hyafgd}. $\blackbox$
\ei

\section {Conclusion}

In this work, a new  model 
of cryptographic protocols
was built, and examples of its use for solving 
verification problems of properties of integrity, secrecy and correspondence are presented.

For further activities on the development of this model and verification methods based on it, the following research directions  can be named:
\bi
\i development of specification languages  for description of 
CP properties, allowing to express, for example, properties 
of zero knowledge in authentication CPs,  properties of non-traceability in CPs of electronic payments, 
properties of an anonymity and a correctness of vote counting in  CPs of electronic voting,
and  development of methods for verification properties expressed in these languages,
\i construction of methods for automated synthesis of CPs by describing  properties that they must satisfy.
\ei 

\section{Compliance with Ethical Standards}

This study was funded by 
the Ministry of Digital Development, Communications and Mass Media of the Russian Federation and Russian Venture Company (Agreement No.004/20 dated 20.03.2020, IGK 0000000007119P190002).

The author declares that he has no conflict of interest. 

This article does not contain any studies with human participants performed by the author.

\end{document}